\documentclass[epjST]{svjour}
\usepackage{graphicx}
\usepackage{amsmath}
\usepackage{amssymb}
\usepackage{hyperref}
\usepackage{braket}

\newcommand{\barQ}{{\bar{Q}}}
\newcommand{\qqb}{{Q\bar{Q}}}
\newcommand{\calC}{{\cal{C}}}
\newcommand{\calD}{{\cal{D}}}

\newcommand{\calO}{{\cal{O}}}
\newcommand{\calP}{{\cal{P}}}
\newcommand{\bfk}{{\bf{k}}}

\newcommand{\bfq}{{\bf{q}}}
\newcommand{\bfr}{{\bf{r}}}
\newcommand{\bfx}{{\bf{x}}}
\newcommand{\bfD}{{\bf{D}}}
\newcommand{\bfE}{{\bf{E}}}
\newcommand{\bfR}{{\bf{R}}}
\newcommand{\tr}{{\rm{tr}}}

\begin{document}
\title{Quarkonium propagation in the quark gluon plasma}
\author{Rishi Sharma\inst{1}\fnmsep\thanks{\email{rishi@theory.tifr.res.in}}}
\institute{Tata Institute for Fundamental Research}
 
\abstract{
In relativistic heavy ion collisions at RHIC and the LHC, a quark gluon plasma
(QGP) is created for a short duration of about $10$fm/c. Quarkonia (bound
states of $c\bar{c}$ and $b\bar{b}$) are sensitive probes of this phase on
length scales comparable to the size of the bound states which are less than
$1$fm. Observations of quarkonia in these collisions provide us with a lot of
information about how the presence of a QGP affects various quarkonium states.
This has motivated the development of the theory of heavy quarks and their
bound states in a thermal medium, and its application to the phenomenology of
quarkonia in heavy ion collisions. We review some of these developments here.
}
\maketitle
 
\section{Introduction}
\label{intro}

We can understand the dynamical properties of the QGP with the help of various
probes that test its response on different length and time scales. For example,
the long wavelength (comparable to the system size) dynamics of the energy
momentum tensor of high temperature quantum chromodynamics (QCD) is given by
hydrodynamics
(see~\cite{Romatschke:2009im,Teaney:2009qa,Hirano:2012qz,Song:2013gia,Gale:2013da,Heinz:2013th,Shen:2014vra,Jeon:2015dfa,Jaiswal:2016hex}
for reviews). The hydrodynamic properties of the thermal medium are
experimentally studied by detecting particles with low (less than roughly
$1$GeV) momenta that are expected to be connected to the collective flow of the
medium. As a concrete example, a very well studied observable is the
distribution of particles as a function of the azimuthal angle. The Fourier
coefficients of this distribution are called elliptic flow coefficients and the
observed values of these coefficients constrain the values of the viscosity of
the QGP. (For example
see~\cite{Romatschke:2007mq,Heinz:2009cv,Schenke:2010nt}.) Indeed, the low
value of the shear viscosity to the entropy ratio [$\eta/s\sim 1/(4\pi)$] needed to
match the observed elliptic flow in heavy ion collisions at RHIC and the LHC is
one of the striking results from the study of the QGP. 

Another family of probes, hard probes, are characterized by an energy scale
(for example the momentum or the mass of the particle propagating through the medium)
which is much larger than the temperature of the QGP. 

The propagation of highly energetic low-mass partons is sensitive to the medium
correlations on the light cone directions along the direction of
motion~\cite{Wiedemann:2000ez,CaronHuot:2008ni,Majumder:2012sh}.

For slowly moving heavy quarks [bottom ($b$) and charm ($c$) quarks], the
dynamics are sensitive to the chromo-electric field correlator calculated at low
frequencies as we will discuss in more detail. 

Quarkonia are bound states of heavy quarks and anti-quarks. The masses of the
quarks ($m_c\sim 1.5$GeV, $m_b\sim 4.9$GeV~\cite{Eichten:2007qx}) are much
larger than the QCD scale ($\Lambda_{QCD}\sim 0.2$GeV). Their binding energies
are significantly smaller than the quark masses and therefore it is fruitful to
think of them as non-relativistic bound states. They feature multiple energy
scales: the heavy quark mass $M$, the inverse size $1/r$, and the binding
energy $E_b$. 

A rough understanding of these energy scales can be obtained at leading order
in $\alpha_s$ in vacuum. At this order the interaction in the singlet channel is
Coulombic attraction, and $1/r\sim Mv$ and $E_b\sim Mv^2$ where $v\sim
\alpha_s(Mv)$ is the relative velocity between the $Q$ and
$\bar{Q}$~\cite{Bodwin:1994jh}.  Realistically, $v\sim 0.5$ for $c$ and $v\sim
0.3$ for $b$~\cite{Bodwin:1994jh,Quigg:1979vr,Eichten:1979ms,Eichten:2007qx} is
not very small, and perturbative estimates are not quantitatively reliable,
especially for the charm quark.  Still, the system may be amenable to a
non-relativistic treatment, with potentials that are not perturbative but
obtained non-perturbatively using lattice QCD~\cite{Otto:1984qr}. 

Multiple observables for the quarkonium systems in vacuum have been computed in the
potential model and compared with experimental results to test how well the
potential description works for these systems. Several studies have computed the
energy levels for charmonia and bottomonia in the framework of potential
models. These potentials typically feature a short distance perturbative piece
added to a long-distance non-perturbative piece, for example, the Cornell
potential.  For a detailed early calculation of the charmonium and bottomonium
spectra, leptonic decay rates, and transition rates using phenomenological 
potentials, see Ref.~\cite{Eichten:1979ms}.

The non-relativistic treatment for the study of quarkonia has since been put in
an effective field theory (EFT) framework called potential NRQCD
(pNRQCD)~\cite{Brambilla:1999xf}. (See Ref.~\cite{Brambilla:2004jw} for a
comprehensive review.) The $Q\bar{Q}$ potential is part of the leading order
term in the pNRQCD lagrangian (see Eq.~\ref{eq:LpNRQCD} below), and terms
suppressed by $M$ appear systematically as higher dimensional terms.

For the lowest energy quarkonium states, the short distance part of the
potential is the most important, and this part can be calculated perturbatively
using pNRQCD, although one needs to calculate the potential to high order in
$\alpha_s$~\cite{Brambilla:1999xf,Brambilla:2004jw,Brambilla:2009bi} for
quantitative accuracy. The spectra using these potentials evaluated to
$\alpha^5$ were calculated in Refs.~\cite{Brambilla:1999xj,Kniehl:2002br}. At
this order spin-spin, spin-orbit, and tensor couplings are present. For excited
quarkonia which are larger in size, the long-distance non-perturbative part of
the potential is important. This part is constrained by lattice QCD. 

The phenomenology addressed using these non-relativistic techniques is quite
extensive. We refer the reader to Refs.~\cite{Eichten:2007qx,Voloshin:2007dx,Brambilla:2010cs}
for broad overviews. The potential model calculations of quarkonia give a very
good description of the experimentally observed spectra of charmonia as well as
bottomonia, at least below the thresholds for open heavy flavor production.
More recent results on bottomonia~\cite{Patrignani:2012an} reiterate the point
that the potential models describe the energy levels of quarkonium states below
the threshold, but describing states near or above the threshold is challenging. For
other recent comparisons of quarkonium spectra with experimental results see
Refs.~\cite{Repko:2012rk}. 

Another set of observables that are widely studied are the electromagnetic
transition rates and leptonic decay widths. The leptonic decay widths are
sensitive to the wavefunction near the origin. For the lowest few states, the
agreement with experimental results for the leptonic decay widths is very
good~\cite{Radford:2007vd,Brambilla:2020xod}, and becomes worse for higher excited states. The
electromagnetic transitions rates, on the other hand, are sensitive to the
dipole matrix elements. Several of these rates have been measured for
bottomonia and the potential model does describe the results
well~\cite{Radford:2007vd}. 

From potential models for the singlet wavefunction, one can estimate $1/r\sim
1$GeV for $\Upsilon(1S)$ and $1/r\sim 0.5$GeV for $J/\psi$ with $E_b\sim
0.4-0.5$GeV for both~\cite{Quigg:1979vr}. The spatial extent will naturally
be larger for higher excited states and the binding energies smaller. 

In all these calculations, one considers only the singlet state wavefunction
for the quarkonia. If the coupling $\alpha_s$ on the scale of the inverse
length scale of the quarkonia is weak, then the system is in the weak coupling
regime. In vacuum, in the weak coupling regime, both the singlet and the octet
wavefunction feature in the non-relativistic theory describing quarkonia.
However, in the strong coupling regime, the octet wavefunction has a sufficient energy gap compared to
the singlet wavefunction and can be integrated out~\cite{Brambilla:1999xf,Brambilla:2004jw}.
This is the regime considered in most phenomenological applications.

In the thermal medium, the vacuum scales $1/r$ and $E_b$ need to be compared to the thermal scales. The
temperature $T$ in the QGP phase ranges from about $0.2-0.5$GeV. Another scale
of interest is the Debye screening mass $m_D$ for static chromo-electric
fields. In perturbation theory at leading order, $m_D\sim gT$, but at this
scale $g$ is not small~\cite{Kaczmarek:2002mc,Digal:2003jc,Kaczmarek:2003ph,Kaczmarek:2005ui,Doring:2007uh} and
non-perturbative effects are important.

It is intuitively clear that if $T\gtrsim E_b$, absorption of thermal
gluons in the medium can lead to the dissociation of quarkonia (gluo-dissociation).
The rate for this process in the weak coupling limit was first calculated in
Ref.~\cite{Peskin:1979,Bhanot:1979vb} where it was assumed that the octet state
resulting from the absorption of the gluon by the singlet state can be taken to
be free. The process involves the transfer of energy from the medium to quarkonia.

It was also realized very early~\cite{Matsui:1986} that screening of the
$Q\bar{Q}$ interaction in the QGP could lead to the weakening of the $Q\bar{Q}$
binding.  The authors proposed that this effect can be probed experimentally by
looking at the suppression of $J/\psi$ yields in heavy ion ($AA$)
collisions relative to their appropriately normalized yields in $pp$ collisions
($R_{AA}$). The proposal was soon extended to excited states of quarkonia
~\cite{Karsch:1987pv,Karsch:1991,Digal:2001ue}. 

It is useful to note here that the connection of the final yields of quarkonia
to processes like dissociation and screening mentioned above is subtle because
there are additional effects in play. As a specific example, we note that
regeneration (also known in the literature as recombination or coalescence) of
quarkonia from $Q$ and $\bar{Q}$ generated in different hard collisions can be
important and may even lead to an enhancement of the quarkonium yields in $AA$
collisions compared to (normalized) yields in $pp$ collisions. We will make a
few comments about this process below and refer the reader
to~\cite{Rapp:2008tf,Rapp:2009my} for detailed reviews on the topic. Cold
nuclear matter effects, associated with the fact that the distribution of
partons in nuclei is different from their distribution in nucleons, also
affect $R_{AA}$~\cite{ConesadelValle:2011fw}. Despite these caveats, it is
important to understand the processes involving screening and dissociation of a
$Q$, $\bar{Q}$ moving together in the QGP to understand $R_{AA}$.

A very rough estimate of the importance of the screening of the $Q\bar{Q}$
interaction by the medium can be obtained by comparing the screening scale
$m_D$ and the inverse size $1/r$ for various quarkonium states. Significant
suppression in the QGP for a given quarkonium state is expected for $m_D>1/r$.
Screening is especially important when $E_b$ is much less than the medium
scales like $T$ and $m_D$. In this case, the frequency of the gluons exchanged
between the $Q$ and $\bar{Q}$ can be neglected, and the screened interaction
can be treated as a potential as is familiar from non-relativistic quantum
systems. However, even in this zero frequency limit, the problem of $Q\bar{Q}$
propagation in the medium does not quite reduce to a static problem of
$Q\bar{Q}$ interacting via a real screened potential, and evolution is
sensitive to scatterings with gluons in the medium.

This point was clarified in Ref.~\cite{Laine:2006ns} where the authors showed using finite $T$
perturbation theory that in addition to getting screened, the potential between
the $Q$ and $\bar{Q}$ develops an imaginary part. In weak coupling, the real
part is the usual screened Coulomb potential. The imaginary part is related to
the spectral function of gluons at zero frequency. Because of the importance of
this result, we will review it further in Sec.~\ref{sec:screening}. 

From the above discussion, it is clear that this problem involves multiple
scales. Depending on the relative order between $1/r$, $E_b$, $T$, $m_D$,
different processes play a role. The scale $1/r$ is larger than $E_b$, $T$,
$m_D$ with a clear hierarchy for the lowest energy states (see
Tables~\ref{tab:tablec},~\ref{tab:tableb}) of quarkonia. These values of $r$
are quoted from Ref.~\cite{Burnier:2015tda} and $r$ estimates in older
potential model calculations are somewhat smaller and show a larger hierarchy
between $1/r$ and $E_b$. (For example see Ref.~\cite{Eichten:1979ms}.) They were
computed using the $T=0$ potentials calculated on the
lattice~\cite{Burnier:2015tda}. (We note as an aside that in the paper the same
lattice set up was then also used to calculate the real and the imaginary parts
of the potential at finite $T$.) $E_b$ and $T$ are both of the order of a few
$100$MeV. (The weakly bound excited states do not have a clear hierarchy
between $1/r$ and the other scales, and need more care.)  

\begin{table}[h!]
  \begin{center}
    \begin{tabular}{|c|c|c|}
      \hline
      Meson & $r$(fm) & $Mv\sim 1/r$(GeV)\\
      \hline
      $J/\psi$ & 0.56 & 0.35 \\
      $\psi(2S)$ & 1.15 & 0.17 \\
      $\chi_c(1P)$ & 0.81 & 0.24 \\
      %$J/\psi$ & 0.47 & 0.43 \\
      %$\chi_c(1P)$ & 0.74 & 0.27 \\
      %$\psi(2S)$ & 0.96 & 0.21 \\
      \hline
    \end{tabular}
    \caption{$r$ for charmonia\cite{Burnier:2015tda}
    \label{tab:tablec}}
  \end{center}
\end{table}

\begin{table}[h!]
  \begin{center}
    \begin{tabular}{|c|c|c|}
      \hline
      Meson & $r$(fm) & $Mv\sim 1/r$(GeV)\\
      \hline
      $\Upsilon(1S)$ & 0.29 & 0.67\\
      $\Upsilon(2S)$ & 0.59 & 0.34\\
      $\Upsilon(3S)$ & 0.86 & 0.23\\
      $\chi_b(1P)$ & 0.48 & 0.41 \\
      $\chi_b(2P)$ & 0.77 & 0.26\\
      $\chi_b(3P)$ & 1.1 & 0.18\\
      %$\Upsilon(1S)$ & 0.20 & 1.0\\
      %$\chi_b(1P)$ & 0.39 & 0.51 \\
      %$\Upsilon(2S)$ & 0.48 & 0.42\\
      %$\Upsilon(1D)$ & 0.53 & 0.38\\
      %$\chi_b(2P)$ & 0.64 & 0.31\\
      %$\Upsilon(3S)$ & 0.72 & 0.28\\
      %$\Upsilon(2D)$ & 0.75 & 0.27\\
      %$\Upsilon(4S)$ & 0.92 & 0.22\\
      \hline
    \end{tabular}
    \caption{$r$ for bottomonia\cite{Burnier:2015tda}
    \label{tab:tableb}}
  \end{center}
\end{table}

However, the relative order between $E_b$, $T$, $m_D$ is not as obvious and
depends on the state being considered and the value of $T$. For static
quantities, $2\pi T$ is the relevant scale as it separates the lowest Matsubara
frequency from the next. For dynamic processes, $T$ defines the typical energy
carried by the typical degrees of freedom.  If the hierarchy between $E_b$,
$T$, and $m_D$ on one hand, and $\Lambda_{QCD}$ on the other is not very
strong then a quantitatively controlled calculation of these effects requires
non-perturbative calculations using lattice QCD. 

One way to non-perturbatively analyze how quarkonia are affected by the thermal
medium is by calculating the spectral function corresponding to the
current-current correlation function for heavy
quarks~\cite{Datta:2003ww,Umeda:2002vr,Asakawa:2003re}. Lattice QCD provides a
quantitatively controlled method to calculate correlation functions of
operators separated in imaginary time. The extraction of the spectral function 
from imaginary time data is challenging as this requires an analytic
continuation to real time, but much progress has occurred since the early
efforts~\cite{Datta:2003ww,Jakovac:2006sf,Mocsy:2007yj,Mocsy:2007jz,Bazavov:2009us,Petreczky:2010tk,Aarts:2011sm,Karsch:2012na}.
We will not review this progress here but refer the readers to excellent
reviews on this topic~\cite{Petreczky:2012rq,Datta:2014wga,Rothkopf:2019ipj}.

The difficulty in calculating real time quantities on the lattice
behooves us to consider alternative methods to calculate quarkonium
properties theoretically. Effective field theories (EFTs) provide a framework to systematically integrate
out degrees of freedom to focus on the relevant ones at the scale of interest.
This allows us to write observables in factorized forms as matching
coefficients multiplied by matrix elements that can be found within the EFT.
The coefficients can be computed non-perturbatively by matching. The relevant
EFT for quarkonia is pNRQCD~\cite{Brambilla:1999xf,Brambilla:2004jw}. We will briefly review the
development of this theory at finite $T$ in Sec.~\ref{sec:pNRQCD}.  

The connection of the calculations of quarkonium properties in a uniform and time
independent medium at temperature $T$ to phenomenology requires stitching
together the effects of locally equilibrated media as the quarkonia move
through the evolving QGP. 

One approach to this is as follows. At any instant, each quarkonium state has a
decay rate $\Gamma$ which depends on $T$. Schematically, the decay rate has the
form given in Eq.\ref{eq:Gamma}. The background evolution of the QGP is
modelled via hydrodynamics and gives the evolution of $T$ with time at any
location. By solving the rate equations with a $T$ dependent $\Gamma$ gives the
suppression of the direct production of that state due to the QGP. By adding
all the feed-down contributions, we can obtain the net
suppression of quarkonia in heavy ion collisions. We will review some of the
studies that have used this approach in Sec.~\ref{sec:lowpT}. 

In Sec.~\ref{sec:highpT} we will discuss one specific example of such a
framework used for the calculation of quarkonia with a large transverse
momentum ($p_T$) in Ref.~\cite{Aronson:2017ymv}.  

One point to note here is that many phenomenological studies use the decay rates
calculated to quarkonia at rest in the thermal medium. For small relative
velocities between quarkonia and the medium, the rates are not substantially
changed but the differences can be substantial for large relative velocities.
For the modification of gluo-dissociation due to Doppler shifts see
Ref.~\cite{Hoelck:2016tqf}. For an attempt to calculate these effects in an EFT
see~\cite{Escobedo:2011ie}.

The microscopic calculation of $\Gamma$ at a given time requires the knowledge
of the state of $Q\bar{Q}$ at that time (Eq.\ref{eq:Gamma}). In most studies
it is assumed that the state is an eigenstate of the Hamiltonian at time $T$:
the adiabatic approximation. (For recent work analyzing deviations from the
adiabatic approximation see Ref.~\cite{Boyd:2019arx}.) However, if the $T$
evolution of the background is rapid, this approximation is invalid and one
needs to follow the coherent quantum evolution of the state with time and the
rate equation is not sufficient to describe these dynamics. Since the evolution
involves exchange of energy and momentum with the medium (as evinced by the
fact that the potential is complex), the $\qqb$ system is an open quantum
system~\cite{Akamatsu:2011se,Akamatsu:2012vt,Akamatsu:2013,Akamatsu:2015kaa,Kajimoto:2017rel,Akamatsu:2018xim,Miura:2019ssi,Brambilla:2016wgg,Brambilla:2017zei,Brambilla:2019tpt,Brambilla:2019oaa,Blaizot:2015hya,Blaizot:2017ypk,Blaizot:2018oev,Yao:2017fuc,Yao:2018zze,Yao:2018nmy,Yao:2018sgn,Yao:2020eqy,Islam:2020bnp}.
Its description involves solving the evolution equation for the density matrix
$\rho$. We will review progress on the quantum description of the system in
Sec.~\ref{sec:open}. Then we will review in some detail one recent work by the
author~\cite{Sharma:2019xum} on including correlated noise in the quantum
evolution of the $\qqb$ system.

The list of the topics in this review is quite biased. We will not be able to
cover many areas of this vast and active research field. In particular,
important topics left out include,
\begin{enumerate}
\item{
Regeneration is the process of coalescence of a $Q$ and a $\bar{Q}$ to form
quarkonia from $Q$ and $\bar{Q}$ generated in different hard collisions.  This
process is important especially for charmonia with low transverse momentum, if
$c$, $\bar{c}$ are abundant in the heavy ion environment.  The observed
relative yields and spectra of $J/\psi$ with low transverse momenta in PbPb
collisions at the LHC with $\sqrt{s}=2.76$TeV do suggest a significant
contribution from recombination processes (see Ref.~\cite{Andronic:2015wma} for a
discussion). We refer the reader to~\cite{Rapp:2008tf,Rapp:2009my} for detailed
reviews on the topic.
}
\item{AdS/CFT is a technique which is very useful for obtaining qualitative
insight into dynamical properties of theories which share some crucial
similarities with QCD. We refer the interested reader to Ref.~\cite{CasalderreySolana:2011us} for
a comprehensive review of the application of AdS/CFT to understand high
temperature QCD.}
\item{Cold nuclear matter (CNM) effects contribute to the yields of quarkonia. These
are especially important at forward rapidity. We refer to articles in the review
for this phenomenology~\cite{ConesadelValle:2011fw}.}
\item{For broad reviews of the phenomenology of quarkonia (and open heavy quark) in heavy
ion collisions we refer the reader to the following review articles
Refs.~\cite{Mocsy:2013syh,Aarts:2016hap,Andronic:2015wma}.
For a broad overview of quarkonium physics in general see
Ref.~\cite{Brambilla:2010cs}.}
\end{enumerate}

\section{Screening, damping, and dissociation in perturbation theory
~\label{sec:screening}}
Consider a $Q\bar{Q}$ state in a thermal medium at a temperature
higher than the crossover temperature. The state is affected due to the
screening of the interaction between the $Q$ and $\bar{Q}$. Furthermore,
absorption of a gluon from the medium or the emission of a gluon converts the
singlet state to an octet state and vice-versa.  A description of the evolution
of the state requires a consistent description of all these processes.   

While weak coupling analyses are not expected to be quantitatively accurate at
temperatures of interest for LHC and RHIC phenomenology, key insight into the problem came from the weak coupling analysis by
Ref.~\cite{Laine:2006ns}: the potential between the $Q$ and $\barQ$ in a
thermal medium is complex at finite $T$. The authors calculated the 
following correlator at finite $T$
\begin{equation}
\calC(t,r) = \langle 
\bar\psi(t,\frac{\bfr}{2})W[(t,\frac{\bfr}{2});(t,-\frac{\bfr}{2})]\gamma^\mu\psi(t,-\frac{\bfr}{2})
\bar\psi(0,-\frac{\bfr}{2})\gamma_\mu W[(t,-\frac{\bfr}{2});(t,\frac{\bfr}{2})]\psi(0,\frac{\bfr}{2})
\rangle\;.
~\label{eq:Ctr}
\end{equation}
The correlator~\cite{Philipsen:2002az} is a correlator of ``point split''
currents, where the fields composing the currents are split by a spatial
separation $\bfr$. $\psi$ is the heavy quark field operator. $W$ are Wilson lines connecting the 
$\psi$'s, needed for gauge invariance. 

For infinitely massive quarks, the location of the heavy quark does not change
with time and the field operator, $\psi$, only evolves with time by a Wilson line
in the time direction,
\begin{equation}
\psi(t,\bfx)= W[(t,\bfx);(0,\bfx)]\psi(0,\bfx)
\end{equation}
The equal time $\psi$ operators can be contracted using the standard
commutation rules. (For example see Ref.~\cite{McLerran:1981pb} for evolution
in Euclidean time.) Therefore, in this case, $\calC(t,r)$ is simply the
expectation value of the Wilson loop in a thermal medium of gluons and light quarks.

The computation in Ref.~\cite{Laine:2006ns} (see also Ref.~\cite{Burnier:2007qm}) was done in a weakly coupled plasma to the lowest order in $g$.
The authors showed that at late times, $\calC$ satisfies the Schr\"{o}dinger
equation ($\nabla^2/M$ drops out in the large $M$ limit)
\begin{equation}
i \partial_t \calC(t, r) = V(r) \calC(t, r)\;,
\end{equation}
where
\begin{equation}
V(r) =  - \frac{g^2 C_F}{4\pi}[m_D + \frac{\exp(-m_D r)}{r}] 
- i\frac{g^2 C_F T}{4\pi}\phi(m_D r)\;.
~\label{eq:Vr}
\end{equation} 
$C_F = (N_c^2-1)/(2N_c)$ and $N_c=3$ is the number of colors.

The function $\phi(x)$ has the form,
\begin{equation}
\phi(x) = 2 \int _0^\infty \frac{dz z}{(z^2+1)^3}[1 - \frac{\sin(zx)}{zx}]\;,
~\label{eq:phix}
\end{equation}
which increases monotonically from $\phi(0)=0$ to $\phi(\infty)=1$.

A nice intuition for the presence of an imaginary piece in the $Q\bar{Q}$
potential is given as follows~\cite{Beraudo:2007ky}. It arises due to the
scattering of the heavy quark and the anti-quark with the light constituents
(light quarks and gluons) of the medium, mediated via the exchange of
$T$-channel longitudinal gluons. (Longitudinal gluon exchanges dominate in the large $M$
regime, since the transverse modes are suppressed by the velocity of the $Q$ or
$\bar{Q}$.)  

More concretely, the propagator for longitudinal gluons have the form
\begin{equation}
\calD_L^{\mu\nu}(q^\mu) =
P_{\mu\nu}^L\frac{1}{(q^0)^2-\bfq^2-\Pi_L(q^\mu)}=P_{\mu\nu}^L D(q^{\mu})\;,
~\label{eq:D_L}
\end{equation}
where $P_L$ is the longitudinal projector
\begin{equation}
P_L^{\mu\nu}(q^\mu) = \delta{\mu\nu} - \frac{q^\mu q^\nu}{q^2} -
P_T^{\mu\nu}(q^\mu)\;.
\end{equation}
$P_T^{\mu\nu}(q^\mu)$ is the transverse projector with only spatial components
\begin{equation}
P_T^{ij}(q^\mu) = \delta^{ij} - \frac{\bfq^i\bfq^j}{\bfq^2}\;.
\end{equation}

The self energy for the longitudinal gluons is~\cite{Kapusta:2006pm}
\begin{equation}
\Pi_L(q^\mu) = m_D^2[1-\frac{(q^0)^2}{\bfq^2}]
[1-\frac{q^0}{2|\bfq|}\log(\frac{q^0+|\bfq|}{q^0-|\bfq|})]
~\label{eq:PiL}
\end{equation}
The Debye screening mass $m_D$ is given by
\begin{equation}
m_D^2 = g^2 T^2(\frac{N_c}{3}+\frac{N_f}{6})\;,~\label{eq:mD}
\end{equation}
where $N_f$ is the number of flavors.

From Eq.~\ref{eq:D_L} the retarded propagator $D^R$ can be obtained from the
analytic continuation of $D(q^\mu)$ in Eq.~\ref{eq:D_L} to slightly imaginary
frequencies, $q^0\rightarrow q^0+i\epsilon$. The spectral function
has the usual form
\begin{equation}
\begin{split}
\rho_L(q^\mu) &= \frac{1}{i}\Bigl\{D^R(q^\mu) - [D^{R}(q^\mu)]^*\Bigr\}\\
&=
\frac{2 \Im m[\Pi_L(q^\mu)]}
{\Bigl[(q^0)^2-\bfq^2-\Re e[\Pi_L(q^\mu)]\Bigr]^2+\Bigl[\Im
m[\Pi_L(q^\mu)]\Bigr]^2}\;.
\end{split}
~\label{eq:rho}
\end{equation}
For $q^0<|\bfq|$, i.e. for space-like momenta, the $\log$ term 
(Eq.~\ref{eq:PiL}) is imaginary and the spectral function is non-zero. This
reflects the well known phenomenon called Landau damping, associated with the
absorption of space-like gluons by particles in the plasma. For time-like
momenta this process is kinematically not allowed at leading order. 

The key realization is that for large $M$, the kinetic energy change of the
heavy quark due to gluon exchanges with the light quarks and the gluons in the
medium is small, and the damping of a single heavy quark is governed by the
gluonic spectral function for $q^0\rightarrow 0$. For small $q^0$ 
($q^0\ll |\bfq|$, well in the space-like regime) the spectral function is given by
\begin{equation}
\rho_{L}(q^\mu)|_{q^0\ll |\bfq|} \approx \frac{\pi m_D^2 q^0}{|\bfq|(\bfq^2+m_D^2)^2}\;,
~\label{eq:rho0HTL}
\end{equation}

One can show that for a single $Q$~\cite{Beraudo:2007ky} the collision rate is
\begin{equation}
\Gamma_Q = g^2 C_F\int \frac{d^3 \bfq}{(2\pi)^3} 
[n_B(q^0) \rho_L(q^\mu)]|_{q^0\ll |\bfq|}\;,~\label{eq:GammaQFinal}
\end{equation}
where
\begin{equation}
n_B(q^0) = \frac{1}{\exp(q^0/T)-1}\approx \frac{T}{q^0}
\label{eq:smallq0}
\end{equation}
for small $q^0$. Therefore, for $q^0\ll T$, $q^0$ in $\rho_L$
(Eq.~\ref{eq:rho0HTL}) cancels out and the collision rate is finite in the
static limit. 

The final form of the collision rate, Eq.~\ref{eq:GammaQFinal}, is intuitively
easy to understand. It is simply given by the rate of scattering of $Q$ with the
gluons in the medium with $q^0\rightarrow 0$. This rate is proportional to
the gluonic occupation number and their spectral density and does not refer
explicitly to the light particles with which $Q$ scatters. All the information
about the light particles goes into the calculation of the gluonic spectral
density at $q^0\rightarrow 0$.

Now coming back to the problem of a $Q\bar{Q}$ separated by $r$, we note that
at large separation, the $Q$ and $\bar{Q}$ experience collisions with the
medium partons independently. At finite $r$, the two sources interfere
destructively and the net collision rate for the $Q\bar{Q}$ is smaller.
Substituting  $r\rightarrow 0$ in  Eq.~\ref{eq:phix} gives $0$ for this rate, 
which corresponds to completely destructive interference from a $Q$ and a
$\bar{Q}$ in the color singlet channel at the same location.

The damping term is clearly related to the decay of the
quarkonium states. As a consequence of taking the static limit, the decay
appears naturally as an imaginary part of the potential in Eq.~\ref{eq:Vr}.   

One can go a step further and attempt to connect the damping term to
phenomenology. For this purpose, one can interpret the imaginary part of the
potential as a manifestation of processes that take $Q\bar{Q}$ out of the
Hilbert space of two bound particles. 

Suppose that the $Q\bar{Q}$ prepared in a singlet state with a
wavefunction $\Psi(\bfr)$ which is an eigenstate of the non-hermitian
hamiltonian
\begin{equation}
[- \frac{\nabla^2_{\bfr}}{M} + V(r)]\Psi(\bfr) = (E_R-iE_I)\Psi(\bfr)
\end{equation}
where $M/2$ is the reduced mass of the $Q\bar{Q}$ system. Then
the decay rate is for the state is simply given by
\begin{equation}
\Gamma_{Q\bar{Q}} = 2E_I \;.~\label{eq:GammaEI}
\end{equation}

Alternatively, if the $Q\bar{Q}$ is prepared in a singlet state with a 
wavefunction $\Psi'(\bfr)$ which is an eigenstate of the real part of the
potential $\Re e[V(r)]$, then the leading order decay rate from the imaginary
potential is 
\begin{equation}
\Gamma_{Q\bar{Q}} = 2\langle \Psi'|\Im m[V(r)]|\Psi'\rangle
\;.~\label{eq:GammaImV}
\end{equation}

The survival probability for quarkonia after spending time $[t_0,t]$ in the
medium can be estimated from,
\begin{eqnarray}
    \label{eq:classical_RAA}
    && \frac{d N_{\psi}(t)}{dt} = - \Gamma_{\psi}(t) N_{\psi}(t) ,\nonumber \\
    && N_{\psi}(t) = e^{-\int_{t_{0}}^{t} dt' \Gamma_{\psi}(t')}N_{\psi}(t_{0}).
\end{eqnarray}
The exponential pre-factor gives $R_{AA}$.

These relations (Eqs.~\ref{eq:GammaEI},~\ref{eq:GammaImV} with
Eq.~\ref{eq:classical_RAA}) have been used in phenomenological applications to
estimate the suppressions of quarkonium states in the QGP.

The calculation of $\Im m[V]$ above ignores the kinetic energy of the heavy quarks and in the
next section we will discuss how this restriction can be relaxed. In
particular, we will find that the absorption of light-like (on-shell) gluons
with finite $q^0>E_b$ leads to a process known as gluo-dissociation~\cite{Peskin:1979,Bhanot:1979vb,Brambilla:2011sg},
which is important if $E_b$ is not much smaller than $T$. 

Finally, we note here that it is not immediately clear that in the QGP, Eq.~\ref{eq:GammaEI} is
the correct way to interpret $\Im m[V]$. The $Q\bar{Q}$ system is an open
quantum system that exchanges energy and momentum with the thermal medium. It
was shown in Refs.~\cite{Akamatsu:2011se,Kajimoto:2017rel} that the evolution
equation for the $Q\bar{Q}$ density matrix can be written in terms of a
stochastic Schr\"{o}dinger equation where the noise term in the stochastic
potential is related to $\Im m[V]$. This provides a rigorous interpretation of
the complex potential. This is discussed in more detail in
Sec.~\ref{sec:open}. For a recent, detailed review of this treatment of
quarkonia, we refer the reader to Ref.~\cite{Akamatsu:2020ypb}.

\subsection{pNRQCD~\label{sec:pNRQCD}}
One way to go beyond the static limit considered above and systematically
include terms $\calO(1/M)$ and above is to use a non-relativistic EFT. Given
the substantial hierarchy between the scale $M$ and the scales $1/r$, $E_b$,
$T$, and $\Lambda_{QCD}$, it is natural to integrate out modes from the
ultra-violet scale to a momentum scale (the largest of the remaining four).
This gives the theory called non-relativistic QCD (NRQCD). (See
Ref.~\cite{Bodwin:1994jh} and references therein for a review.) This theory
features two component non-relativistic fields ($\chi$ and $\psi$ for the $Q$
and the $\barQ$ respectively)
\begin{equation}
\chi^\dagger[D_0 - \frac{\bfD^2}{2M}]\chi
+\psi^\dagger[D_0 - \frac{\bfD^2}{2M}]\psi + \cdot\cdot
\end{equation}
$D_0$ and $\bfD$ are covariant derivatives. The dots correspond to operators
featuring four $\chi$'s, $\psi$'s, and gluonic fields. Importantly, the
lagrangian takes the kinetic energy of the quarks into account in a consistent
manner. 

The next steps depend on the hierarchy between the scales $1/r$, $E_b$, $T$,
and $\Lambda_{QCD}$. In the following, let us take the well motivated choice
that the scale $1/r$ is the next largest. (See
Tables~\ref{tab:tablec},~\ref{tab:tableb}). In this case the energy $E_b$ of the
$Q$, $\barQ$ (in the $\qqb$ rest frame), and the temperature, are both less than its momentum. 
Consequently, the exchanged gluons carry small energy, and at the energy scale
these exchanges can be replaced by a
potential~\cite{Brambilla:1999xf,Brambilla:2004jw}. 

The $Q$ and $\bar{Q}$ are non-relativistic and their state can be described by
a wavefunction, $\Phi(\bfR,\bfr)$. $\bfR$ refers to the centre of the $Q$ and
$\barQ$ and $\bfr$ to their relative separation. Schematically, the state is of the
form ~\cite{Brambilla:2004jw},
\begin{equation}
\int d^3 \bfr d^3 \bfR \Psi_{\alpha\beta}(\bfR,\bfr)
\psi^\dagger_\alpha(\bfR+\bfr/2) \chi_\beta(\bfR-\bfr/2)|{\rm{long\; wavelength\;
Gluons}}\rangle\;,
\end{equation}
where we have explicitly written the color indices $\alpha$, $\beta$ for
clarity.

The long wavelength gluons carry momenta and energies of the order of $E_b$ in
vacuum. In the thermal medium, gluons with energies of the order of $T$
are part of the medium and also participate in the EFT. 

The wavefunction $\Psi$ can be decomposed into singlet and octet wavefunctions 
$S(\bfR,\bfr,t)$, $O^a(\bfR,\bfr,t)$ as follows,
\begin{equation}
\Psi(\bfR,\bfr) = 
P[e^{ig \int_{\bfR-\bfr/2}^{\bfR+\bfr/2} \tau^b {\bf{A}}^{b}\cdot d \bfr}]S(\bfR,\bfr)
+
P[e^{ig \int_{\bfR}^{\bfR+\bfr/2} \tau^b {\bf{A}}^{b}\cdot  d \bfr}][O^a(\bfR,\bfr)\tau^a]
P[e^{ig \int_{\bfR-\bfr/2}^{\bfR} \tau^b {\bf{A}}^{b}\cdot  d \bfr}]
\end{equation}
The gluonic fields with wavelength $1/E_b$, $1/T$ can be expanded in a
multipole expansion. At leading order in the expansion, 
\begin{equation}
\begin{split}
S(\bfR,\bfr)&=\frac{1}{N_c}\tr[\Psi(\bfR,\bfr)]\\
O^a(\bfR,\bfr)&=\frac{1}{T_F}\tr[\tau^a\Psi(\bfR,\bfr)]\
\end{split}
\end{equation}

Finally, the low energy theory includes any other light degrees of freedom
(eg. other light quarks). The structure of the lagrangian is
\begin{equation}
\begin{split}
L =& S^\dagger [i\partial_0 - \frac{\nabla^2}{M} + V_s(\bfr)] S
+S^\dagger [i\partial_0 - \frac{D^2}{M} + V_o(\bfr)] S\\
&
+gS^\dagger(\bfr\cdot \bfE) O
+gO^\dagger(\bfr\cdot \bfE) S
+g\{O^\dagger,\bfr\cdot \bfE\}O+\cdot\cdot
\end{split}
~\label{eq:LpNRQCD}
\end{equation}
$\bfE$ is the electric field operator that arises when one does the multipole
expansion of the long wavelength gluons. Dots represent terms further 
suppressed by $E_br$.

The structure of the theory just follows from the symmetries. The low energy
constants depend on the microscopic details and in particular the relative
order between $E_b$, $m_D$, $T$, and $\Lambda_{QCD}$. In a
series of
papers~\cite{Brambilla:2008cx,Escobedo:2008sy,Brambilla:2010vq,Escobedo:2011ie,Brambilla:2011sg,Brambilla:2013dpa,Brambilla:2011mk,Brambilla:2016wgg,Biondini:2017qjh,Brambilla:2017zei,Brambilla:2019tpt,Brambilla:2019oaa}
the finite $T$ version of pNRQCD was developed, considering various choices for
the relative order of the different scales. 

\begin{figure}[!htb]
\begin{center}
\includegraphics[width=0.47\linewidth]{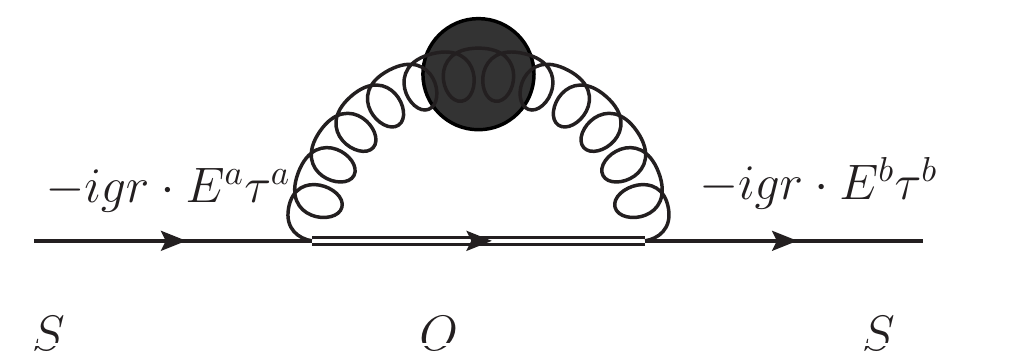}
\end{center}
\caption{Self energy correction to the singlet propagator~\cite{Brambilla:2008cx}. The singlet state is
represented by a single line. The octet by a double line. The vertex is
$-igr^i$ from the singlet-octet operator $gS^\dagger(\bfr\cdot \bfE) O$. At
leading order in $g$, the chromo-electric field is 
$E^a=-\partial_i A^{a\;0}-\partial_0A^{a\;i}$. The blob represent self energy
corrections for the gluon.}
\label{fig:SOS}
\end{figure}

The simplest analysis arises when weak coupling methods can be applied.  For a
concrete example, let us review a case considered in
Ref.~\cite{Brambilla:2008cx}. We start from NRQCD at the scale $1/r$. The low
energy coefficients (LECs) and the propagators at this scale do not have
thermal corrections since $1/r\gg T$ and appear at scales comparable to or
lower than $T$. 

For concreteness let us also assume that the next three scales are $T$,
$m_D\sim gT$ ($g\ll1$), and $E_b$ in decreasing order.  Comparing with the
discussion in Sec.~\ref{sec:screening}, one quantity of interest is the (complex)
singlet potential. Diagrammatically, this can be calculated by calculating the
self energy correction to the singlet propagator from the modes that are integrated
out.  

The leading (in the multipole expansion) thermal correction is obtained by the
dipole terms $gS^\dagger(\bfr\cdot \bfE) O +gO^\dagger(\bfr\cdot \bfE) S$ in
the Lagrangian (Eq.~\ref{eq:LpNRQCD}). Diagrammatically the correction is shown
explicitly in Fig.~\ref{fig:SOS}. 

The self-energy correction in the static limit (dropping the kinetic energies
of the heavy quarks) can simply be written as
\begin{equation}
\delta V_s(r) = - i r^2 T_F \int \frac{d^4k}{(2\pi)^4} 
\frac{i}{-k^0 - V_o(r) + V_s(r)+i\epsilon} [2\tilde{G}_{EE}(k^0, \bfk)]
\label{eq:deltaVs}
\end{equation}
Here, $T_F=1/2$ is the color Casimir in the fundamental representation
$\tilde{G}_{EE}$ is the thermal expectation value of the time ordered chromo-electric 
correlator
\begin{equation}
\tilde{G}_{EE}(k^0, \bfk) = \frac{1}{6N_c} g^2\langle 
\int d^4 x e^{i(k^0t-\bfk\cdot\bfr)}
{\cal{T}}\{\bfE^a(t, \bfr) W^{ab}(t, \bfr;0,{\bf{0}})\cdot\bfE^b(0,{\bf{0}})\}
\rangle_T\;.~\label{eq:GEE}
\end{equation}
$W^{ab}$ is the Wilson line in adjoint representation. The $2$ in
Eq.~\ref{eq:deltaVs} arises from the choice of the normalization in
Eq.~\ref{eq:GEE}.

At the scale $1/r$ $(\gg m_D)$ the gluonic propagators at the scale $1/r$ can
simply be taken to be the vacuum propagator. Therefore at leading order in the
coupling, $G_{EE}$ can be obtained from the free
propagators~\cite{Brambilla:2008cx}. This gives a real part and also an
imaginary part associated with the absorption of on-shell gluons
(gluo-dissociation)~\cite{Peskin:1979,Bhanot:1979vb,Brambilla:2011sg}. 

At scales lower than $T$, the appropriate propagators for the gluons are the
Hard-thermal-loop (HTL)~\cite{Kapusta:2006pm} resummed propagators. The
imaginary part of the gluon self energy [associated with the Landau damping of
the gluons via interaction with the medium (see Eq.~\ref{eq:PiL} for
$q^0<q$)] leads to an imaginary contribution to the self energy correction to
the singlet propagator. The process is related to the calculation shown above
(Sec.~\ref{sec:screening}), where all the fields are expanded in small $r$. This is
consistent with the assumed hierarchy ($r E_b\ll 1$) used in writing pNRQCD.

Thus we obtain the real and the imaginary parts of the singlet potential, which
can be used to calculate the decay rates as described in
Sec.~\ref{sec:screening}. The contributions to the complex potential in a
framework which includes both the gluo-dissociation and the Landau damping
contributions were calculated for the first time in~\cite{Brambilla:2008cx}.

As the above example illustrates, pNRQCD can be used to include all thermal
corrections (Landau damping and gluo-dissociation) systematically. Within
pNRQCD one can consider different hierarchies between the energy scales $1/r$,
$E_b$, $T$, and $m_D$ and pinpoint the dominant effects in each regime.

For example, it was clarified in Ref.~\cite{Brambilla:2008cx}, that formally the
result for the singlet complex potential Eq.~\ref{eq:Vr} is obtained in the
limit $1/r\ll T$. It is interesting to note, however, that the form
Eq.~\ref{eq:Vr} gives a good description of the singlet potential extracted
from the lattice over a wide range of $r$~\cite{Burnier:2016mxc}.

Assuming $1/r\gg T$ on the other hand, the form of the complex potential
depends on the relative order between $T$, $m_D$, and $E_b$. In the case $T\gg
m_D\gg E_b$ discussed above, there are two contributions to the singlet complex
potential: Landau damping and gluo-dissociation (singlet-octet transformation).
The Landau damping contribution dominates for $m_D\gg
E_b$~\cite{Brambilla:2008cx}. The hierarchy $E_b\gg m_D$ was studied in further
detail in Ref.~\cite{Brambilla:2010vq}, and an important result that was found
was that the gluo-dissociation contribution was dominant compared to the Landau
damping contribution for $E_b\gg m_D$. Another important point to note is that
if the relative order between $T$ and $E_b$ is reversed, namely, $T<E_b$, then
thermal effects do not affect the potential, but arise only when the energy of
the state are calculated.

A systematic calculation of gluo-dissociation including $1/N_c$ corrections was
done in Ref.~\cite{Brambilla:2011sg} and this was extended to dissociation via
scattering off light partons in the medium in Ref.~\cite{Brambilla:2013dpa}.

For a detailed discussion of various energy hierarchies that can potentially
arise and the relevant physical processes under these different conditions, we
refer the reader to the original
papers~\cite{Brambilla:2008cx,Escobedo:2008sy,Brambilla:2010vq,Escobedo:2011ie,Brambilla:2011sg,Brambilla:2013dpa,Brambilla:2011mk}  

More recently, pNRQCD has been used to derive master equations for the quantum
evolution of the density matrix of a $\qqb$ system in the
QGP~\cite{Brambilla:2016wgg,Brambilla:2017zei,Brambilla:2019tpt,Brambilla:2019oaa}.
We will discuss these further in Sec.~\ref{sec:open}.

Another power of the EFT approach is that it provides a tool to go beyond
perturbation theory. EFTs often allow us to write observables in a factorized
form; a product (or convolution) of a short distance matching coefficient with
a long distance matrix element. By matching the correlation functions in the 
EFT to those in the underlying theory, the matching coefficient can be fixed.
This matching can be done perturbatively (which was described above) 
or non-perturbatively (we will describe an example below). The matrix element
can computed within the EFT.

As an illustrative example, consider the case discussed in
Ref.~\cite{Brambilla:2016wgg,Brambilla:2017zei,Brambilla:2019tpt,Brambilla:2019oaa}.
In the strong coupling limit, the hierarchy $m_D\ll T$ is not satisfied and
hence the analysis discussed above is not valid. Furthermore, the gluonic
propagator is non-perturbative. 

However progress can be made in some cases. If we assume that all thermal
scales are larger than $E_b$, then both the potentials and the kinetic energies
of the singlet and octet states can be ignored in Eq.~\ref{eq:deltaVs}, the
energy transfer to the medium gluons can be ignored, and the
correction to the singlet and octet potentials is determined by
$\tilde{G}_{EE}(k^0\rightarrow 0, \bfk)$,
\begin{equation}
\delta V_s(r) =  r^2 T_F \int \frac{d^3k}{(2\pi)^3} \tilde{G}_{EE}(0, \bfk)\;.
\label{eq:deltaVsStatic}
\end{equation}

The relevant integral can be simply written in terms of the correlator
\begin{equation}
\begin{split}
{G}_{EE}(k^0=0, \bfr={\bf{0}})&=
\int \frac{d^3k}{(2\pi)^3} \tilde{G}_{EE}(0, \bfk) \\
&= \frac{1}{6N_c} g^2\langle 
\int dt 
{\cal{T}}\{\bfE^a(t, {\bf{0}}) W^{ab}(t, {\bf{0}}; 0,{\bf{0}})\cdot\bfE^b(0,{\bf{0}})\}
\rangle_T\;.~\label{eq:GEEStaticLocal}
\end{split}
\end{equation}

The heavy quark momentum diffusion constant of a single heavy quark is
proportional to the real part of the chromo-electric correlator $G_{EE}(k^0=0,
\bfr={\bf{0}})$ and because of its importance for the phenomenology of open
heavy flavor, substantial effort has been put towards its non-perturbative
determination. In particular, it has been computed on the
lattice~\cite{Banerjee:2011ra,Ding:2011hr,Francis:2015daa,Brambilla:2019oaa}.
Recently, an effort has been made to also estimate the imaginary part of the
correlator using lattice data~\cite{Brambilla:2019tpt,Brambilla:2019oaa}.

As another example, there are increasingly sophisticated non-perturbative
calculations of the complex singlet potential from the
lattice~\cite{Rothkopf:2011db,Burnier:2013fca,Burnier:2014ssa,Burnier:2015tda,Burnier:2016mxc,Bala:2019cqu,Bala:2019boe}. 

Non-perturbative calculations of the internal energies and the free energies of
a $Q$ and $\bar{Q}$ separated by a distance $r$ at finite temperature were
available much
earlier~\cite{Kaczmarek:2002mc,Kaczmarek:2003ph,Digal:2003jc,Kaczmarek:2005ui,Doring:2007uh}.
The latter is related to the Polyakov loop correlator~\cite{McLerran:1981pb}.
The connection between the Polyakov loop correlators and the singlet and octet
potentials were clarified in
Refs.~\cite{Brambilla:2010xn,Berwein:2012mw,Berwein:2013xza,Berwein:2015ayt,Berwein:2017thy}.
In this series of papers, it was shown using perturbation theory and pNRQCD that
the real part of the singlet potential is well approximated by the singlet free
energy. The Polyakov loop and the Polyakov loop correlators were calculated
up to $g^7$. These correlators were computed on the lattice in
Refs.~\cite{Bazavov:2016uvm,Bazavov:2018wmo} and the results show remarkable
agreement with the pNRQCD calculation up to $rT\sim 0.3$.

\section{Phenomenology~\label{sec:lowpT}}
Many of these advances in theory have been incorporated in the treatments
developed to address the phenomenology of quarkonium states in heavy ion
collisions at RHIC and the LHC. 

Both experiments have collected data on a number of observables including
$R_{AA}$. Another observable that has attracted attention is the collective
flow ($v_2$) of these states at these facilities. For some results on
charmonium at RHIC see\linebreak
Refs.~\cite{Adler:2003rc,Adare:2006ns,Adare:2008sh,Abelev:2009qaa,Adare:2011yf,Adamczyk:2012pw,Adamczyk:2012ey,Adare:2012wf,Adamczyk:2013tvk,Aidala:2014bqx,Adare:2015hva,Adamczyk:2016srz,STAR:2019yox,Adam:2019rbk}
and at the LHC see\linebreak
Refs.~\cite{Chatrchyan:2012np,Khachatryan:2014bva,Khachatryan:2016ypw,Sirunyan:2016znt,Aad:2010px,Aad:2010aa,Aaboud:2018quy,Aaboud:2018ttm,Abelev:2012rv,ALICE:2013xna,Abelev:2013ila,Adam:2015rba,Adam:2015isa,Adam:2015gba,Adam:2016rdg,Acharya:2017tgv,Acharya:2018jvc,Acharya:2018pjd,Acharya:2019iur,Acharya:2019lkh}.
For bottomonium states at RHIC see
Refs.~\cite{Adamczyk:2013poh,Adamczyk:2016dzv,Adare:2014hje} and
at the LHC see 
Refs.~\cite{Chatrchyan:2011pe,Khachatryan:2016xxp,Sirunyan:2017lzi}). 
These observables have been measured for low as well as high transverse
momentum ($p_T$) for various centralities and provide important constraints on
the models of $Q\bar{Q}$ propagation in the QGP. For a comprehensive review of
the observations and phenomenology of quarkonia (and open heavy flavor),
see~\cite{Andronic:2015wma}.

The connection from the theoretical calculations of the properties of
$Q\bar{Q}$ in a thermal medium to the phenomenology of quarkonia in heavy ion
collisions is quite non-trivial. 

First, one needs a model for the initial production of the $Q\bar{Q}$ state.
The heavy quark pairs are created early in the collisions, on a time scale
$\sim 1/(2M)$. It is often assumed that a singlet bound state are formed soon
after, on a time scale $1/E_b$ in the rest frame of the $\qqb$. However, one
can begin the evolution from a distribution of singlet and octet
states~\cite{Brambilla:2016wgg,Brambilla:2017zei}, which is a natural
consequence of the NRQCD formalism~\cite{Bodwin:1994jh}.

The state propagates in a thermal medium that evolves in time, is
anisotropic, and features spatial gradients. We assume that a local density
approximation still describes the physics: the dissociation rate is given by
the instantaneous temperature in the neighbourhood of the $\qqb$ state.

Let us consider the evolution of $\qqb$ states moving slowly compared to the
ambient medium. Then the formalism described above can be used to calculate the
rate. For a pure singlet state $\Psi$, the rate can be obtained from
Eq.~\ref{eq:GammaEI} or Eq.~\ref{eq:GammaImV}. Its calculation involves the
knowledge of the wavefunction $\Psi$ at a given instant.  Finally, with the
dissociation rate in hand, $R_{AA}$ is simply given the probability of survival
of the state.

$\Psi$ can be found from the initial state in two extreme approximations, the
adiabatic approximation or the sudden approximation. Both approaches have been
used in phenomenological applications.  We make a note here that neither
approximation is expected to work throughout the evolution and it might be more
appropriate to describe the $\qqb$ system as a continuously evolving quantum
state using the language of open quantum systems, as we shall discuss below in
Sec.~\ref{sec:open}.

Connection with phenomenology requires taking care of additional effects.
Excited states of quarkonia can feed-down to the lower energy states after
formation. CNM effects (see 
\cite{Adare:2007gn,Adler:2005ph,Adare:2010fn,Adamczyk:2016dhc,Abelev:2013yxa,Abelev:2014zpa,Abelev:2014oea,Adam:2015iga,Adam:2015jsa,Adamova:2017uhu,Acharya:2018yud,Acharya:2018kxc,Sirunyan:2017mzd,Aad:2015ddl,TheATLAScollaboration:2015zdl,Aaij:2013zxa}) 
for experimental constraints) modify the initial production rates in $AA$ collisions.

Additionally, as mentioned above, in collisions with high multiplicities of
heavy quarks, $Q$ and $\bar{Q}$ from different hard collisions may combine
(this is sometimes called coalescence, recombination, or regeneration) to form quarkonia. 

Due to the complexity of the problem, there have been various attempts to
address the phenomenology, with emphasis on a subset of processes affecting
$\qqb$ propagation in the QGP. We will provide a broad overview of some of
these developments below. 

%Before we give a broad overview of quarkonium
%phenomenology, it is useful to get a feel for the numbers. For $\Upsilon$,
%assuming that. Taking $E_b\sim300$MeV, a rough value of the temperature
%$T\sim400$MeV, $g=2$, the . The damping contribution
%(Eq.~\ref{}) gives
%\begin{equation}
%\Gamma\sim
%\end{equation}
%For a medium of length $5$fm, this gives $R_{AA}\sim$, which is in the ball
%park of what is obtained.

\begin{enumerate}
\item{In Refs.~\cite{Blaizot:2015hya,Blaizot:2017ypk,Blaizot:2018oev} the
quantum evolution equations for the open $\qqb$ system were simplified to derive classical transport
equations for quarkonia under certain assumptions. The advantage of this
approach is that it is not computationally difficult to include the process of
coalescence of quarkonia from $Q$, $\bar{Q}$ created in two distinct hard
collisions. (This is studied in detail Ref.~\cite{Blaizot:2017ypk})}
\item{The possible role of regeneration in quarkonium dynamics was realized
early\linebreak
~\cite{Grandchamp:2001pf,Grandchamp:2002wp,Grandchamp:2003,Grandchamp:2005yw}
and this effect has been included at leading order in several studies
~\cite{Zhao:2007,Zhao:2008vu}. 

Later, an ``in-medium T-matrix'' formalism was
developed~\cite{Mannarelli:2005pz,Cabrera:2006wh} for including
non-perturbative effects in the interactions between the heavy quarks between
each other and between the heavy quarks and the medium particles.  (See
Ref.~\cite{vanHees:2007me} for the formalism and its application to open heavy
flavor.) This was subsequently applied to
phenomenology~\cite{Riek:2010fk,Zhao:2010nk,Emerick:2011xu,Zhao:2012gc,Liu:2015ypa,Du:2017qkv,Liu:2017qah,Du:2019tjf}.

For other studies including regeneration of quarkonia from coalescence see\linebreak
Refs.~\cite{Greco:2003vf,Zhang:2002ug,Thews:2006ia,Bass:2006vu,Yan:2006ve,Capella:2007jv,Bravina:2008su,Yan:2006ve,Peng:2010zza,Ferreiro:2012rq}
(See~\cite{BraunMunzinger:2000ep,Thews:2000rj,Yan:2006,Kostyuk:2003kt,Thews:2006ia,Andronic:2011yq,Gupta:2014ova} and references therein
for statistical approaches.) 
}
\item{In a series of
papers~\cite{Yao:2017fuc,Yao:2018zze,Yao:2018nmy,Yao:2018sgn,Yao:2020eqy}
Boltzmann transport equations describing $\qqb$ evolution were derived and
applied to address the phenomenology of bottomonium states. The authors started
from the Lindblad equations for $\qqb$ system treating it as an open quantum
system, and derived the Boltzmann equations using some simplifying assumptions 
}
\item{In Ref.~\cite{Hoelck:2016tqf}, the gluo-dissociation and the Landau damping
contributions to the width were added incoherently and an effect important at
large momentum was included: the Doppler shift of the background gluons.
In Ref.~\cite{Hoelck:2017dby} the effect of the background electromagnetic field was 
considered.
}
\item{As discussed above, the QGP is anisotropic. Therefore the formalism
developed for screening and damping in homogeneous, isotropic media is not
directly applicable there. In a series of
papers~\cite{Dumitru:2007hy,Dumitru:2009ni,Dumitru:2009fy,Strickland:2011mw,Strickland:2011aa,Margotta:2011ta,Machado:2013rta,Alford:2013jva,Krouppa:2015yoa,Krouppa:2016jcl,Krouppa:2017jlg,Bhaduri:2018iwr,Boyd:2019arx},
a formalism to include this effect and see its effect on the phenomenology was
developed. In Refs.~\cite{Dumitru:2007hy,Dumitru:2009ni,Dumitru:2009fy}, the
complex color singlet potential in an anisotropic thermal medium was
calculated in weak coupling. In
Refs.~\cite{Strickland:2011mw,Strickland:2011aa,Margotta:2011ta,Krouppa:2015yoa,Krouppa:2016jcl},
Eq.~\ref{eq:GammaEI} was used with the color singlet potential and the
adiabatic approximation in an 
anisotropic thermal medium to calculate $R_{AA}$. Effects due to the magnetic
field were considered in Refs.~\cite{Machado:2013rta,Alford:2013jva}.
In Ref.~\cite{Krouppa:2017jlg} the complex singlet potential calculated on the
lattice used was the potential at the temperature $T$.  In
Ref.~\cite{Bhaduri:2018iwr} the formalism was used to calculate $v_2$.
Deviations from adiabatic evolution were discussed in Ref.~\cite{Boyd:2019arx}.

The gluo-dissociation contribution to the $T$ dependent potential in an
anisotropic medium was computed in pNRQCD in Ref.~\cite{Biondini:2017qjh}.
}
\item{In \cite{Song:2005yd,Park:2007zza,Song:2007gm} (also see
\cite{Liu:2013kkg}) the leading order calculation of the gluo-dissociation rate
by \cite{Peskin:1979,Bhanot:1979vb} was extended to next to leading order. In
particular, in Ref.~\cite{Park:2007zza}, dissociation via scattering off light
quarks and gluons in the medium were treated in a unified framework. [This
calculation was
further extended using the EFT framework in Ref.~\cite{Brambilla:2013dpa} (see
Sec.~\ref{sec:pNRQCD}).]

Assuming that gluo-dissociation is the dominant decay process for quarkonia,
quarkonium observables were calculated in
Refs.~\cite{Song:2010ix,Song:2010er,Song:2011nu,Song:2011xi}. Additional
effects on the formation dynamics of quarkonia were discussed in
Refs.~\cite{Song:2013lov,Lee:2013dca,Song:2014qoa,Song:2015bja}
}
\item{More recently, the theory of open quantum systems in the framework of 
pNRQCD has been applied to the phenomenology of bottomonia in 
Refs.~\cite{Brambilla:2016wgg,Brambilla:2017zei}.}
\end{enumerate}

After this broad overview of various approaches used for quarkonium
phenomenology, we now describe in some more detail a calculation of $R_{AA}$
done in Ref.~\cite{Aronson:2017ymv} for high $p_T$ quarkonia and comparison
with the observations at the LHC.

\subsection{Quarkonia at high $p_T$~\label{sec:highpT}}
For high $p_T$ ($p_T\gtrsim 2M$) quarkonia, it may be more appropriate to start
from a formalism used to study the propagation of highly energetic partons in
the quark gluon plasma. In this section, we will review results from
Ref.~\cite{Aronson:2017ymv} which applies this formalism to the calculation of
$R_{AA}$ for quarkonia with large $p_T$. 

The formalism is based on solving the rate equations for $Q\bar{Q}$ pairs. The
physical picture is simple. $Q\bar{Q}$ pairs with large transverse momenta, 
correlated over short distances are produced in hard processes. These form
quarkonium states over a time scale $t_{\rm{form.}}$. Collisions with the
medium gluons lead to the dissociation of quarkonium states over a time scale
$t_{\rm{diss.}}$. Mathematically, the differential cross-sections for the
$Q\bar{Q}$ states and the quarkonia evolve according to,
\begin{equation}
\begin{split}
\frac{d}{dt} \left( \frac{d\sigma^{Q\bar{Q}}(t;p_T)}{dp_T} \right) 
&= - \frac{1}{t_{\rm form.} } \frac{d\sigma^{Q\bar{Q}}(p_T)}{dp_T}  \, , \\
\frac{d}{dt} \left(  \frac{d\sigma^{\rm meson}(t;p_T)}{dp_T} \right) 
&=  \frac{1}{t_{\rm form.}} \frac{d\sigma^{Q\bar{Q}}(t;p_T)}{dp_T} \\
&- \frac{1}{t_{\rm diss.}}  \frac{d\sigma^{\rm meson}(t;p_T)}{dp_T}   \, . 
\label{rateq}
\end{split}
\end{equation}

Solving Eq.~\ref{rateq} for each $p_T$ requires the initial 
conditions of the differential yields and the knowledge of $t_{\rm diss.}$ and
$t_{\rm form.}$. We discuss each of these below.  

The production of high $p_T$ $Q\bar{Q}$ states is given by the NRQCD
formalism~\cite{Bodwin:1994jh}. The cross-sections for quarkonium production in
$pp$ collisions can be formally written as
\begin{equation}
d\sigma(ij\rightarrow {\rm{meson}}+X) (p_T) = 
\sum_n d\sigma(ij\rightarrow Q\bar{Q}[n]+X') (p_T) 
\langle {\cal{O}}[n]\rangle\;,~\label{factorization}
\end{equation}
where $i,\;j$ refer to initial partons (primarily $gg$ at the LHC).
$d\sigma(ij\rightarrow Q\bar{Q}[n]+X') (p_T)$ refer to short distance coefficients
for the production of $Q\bar{Q}$ in a particular color (for example singlet or
octet) and angular momentum ($^{2S+1} L_J$) configuration which we collectively
label as $[n]$ for short here. These coefficients can be computed in 
perturbation theory (See
Refs.~\cite{Baier:1983,Humpert:1987,Cho:1995ce,Cho:1995vh,Braaten:2000cm,Sridhar:1996vd}
for the computation to the NLO. See
Refs.~\cite{Butenschoen:2010rq,Butenschoen:Long,Butenschoen:polarised,Wang:2012is,Shao:2014yta}
for NNLO computations.)
$\langle {\cal{O}}[n]\rangle$ refer to long distance matrix elements (LDMEs)
which correspond to the probability of forming a particular mesonic state (for
example $J/\psi$, $\psi(2S)$, $\chi_{c J}^1$, or $\chi_{c J}^2$ states for
$c\bar{c}$) from the short distance $Q\bar{Q}$ states. These have to be fitted
to experimentally observed $p_T$ differential yields for the various mesons.
We used the fit in Ref.~\cite{Sharma:2012dy}.

In $AA$ collisions, Eq.~\ref{factorization} does not give the final meson
yields due to two reasons. First, the formed states can be dissociated by the
medium (final state effects). These are handled by Eq.~\ref{rateq}. Thus the left hand side of
Eq.~\ref{factorization} only gives the initial state for
$\frac{d\sigma^{Q\bar{Q}}(t;p_T)}{dp_T}$ in Eq.~\ref{rateq}~\footnote{Within
the ambit of final state interactions there is another effect that can be
included. The color octet state $Q\bar{Q}$ undergoes energy loss before giving
rise to the initial mesonic state. This effect was included in
Ref.~\cite{Aronson:2017ymv}}. The initial value of $\frac{d\sigma^{\rm
{meson}}(t;p_T)}{dp_T}$ is $0$. Second, the production cross-sections
$d\sigma(ij\rightarrow Q\bar{Q}[n]+X') (p_T)$ are themselves modified due to
initial state (CNM) effects. These can be estimated by measuring the
modification factor $pA$ collisions. For charmonia at low $p_T$ CNM effects are
substantial for forward and backward
rapidity~\cite{Abelev:2013yxa,Abelev:2014zpa,Adam:2015iga,Adam:2015jsa}, they
seem to be consistent with a small ($\sim 10\%$ for prompt $J/\psi$ and even
smaller for $\psi(2S)$) enhancement for $p_T>5$~GeV at central rapidities in
$p{\rm{Pb}}$ collisions~\cite{Sirunyan:2017mzd}. For
$\Upsilon(1s)$ at central rapidity, results are consistent with absence of CNM
effects~\cite{TheATLAScollaboration:2015zdl}. In our calculation we will ignore
these effects. 

The second ingredient in Eq.~\ref{rateq} is the dissociation time. Intuition
from the study of high $p_T$ partons suggests that an important process is
transverse momentum broadening: the energetic parton picks up momentum
perpendicular to its motion due to random kicks from thermal gluons in the
medium. Detailed formalisms to calculate this effect have been developed (see
Refs.~\cite{Baier:1996sk,Zakharov:1996fv,Baier:1998kq,Gyulassy:1999zd,Gyulassy:2000er,Wiedemann:2000za}
and references therein).  

For a time independent medium at the temperature $T$, an energetic parton
traversing a distance $L$ accumulates momentum transverse to its motion. In
weak coupling, assuming that scattering in the medium can be effectively
modelled by $T$ channel scattering off static colored sources, the opacity
$\chi=L/\lambda$ ($\lambda$ is the mean free path of the parton) is
substantially larger than $1$, the distribution of the net momentum transfer
($q_T$) to a single energetic quark transverse to its motion can be
approximated~\cite{Gyulassy:2002yv,Wiedemann:2000za,Adil:2006ra} by
\begin{equation}
\frac{d \calP(\bfq_T)}{d \bfq_T} \sim e^{- \frac{\bfq_T^2}{2\chi m_D^2\xi} }
~\label{broadening}
\end{equation}
$\xi\gtrsim 1$ is a parameter that roughly accounts for the fact in transverse
impact parameter space (the Fourier conjugate of $\bfq_T$), the small impact
parameter expansion receives logarithmic corrections. (If the exchanges are
strictly very soft, $\xi=1$). $m_D$ is the screening mass of the exchanged
gluons with the medium. Eq.~\ref{broadening} describes momentum broadening with
a broadening parameter $\hat{q}={2\chi m_D^2\xi}$.   

For an energetic $Q\bar{Q}$ pair moving together in the medium, the transverse
momentum transfers lead to a distortion of the state in transverse momentum
space and lead to the dissociation of bound states. This idea was first used to
propose a new energy loss mechanism for $D$ and $B$ mesons~\cite{Adil:2006ra}.
(See also~\cite{Sharma:2009hn}.)

In Refs.~\cite{Sharma:2012dy,Aronson:2017ymv} this was applied to the study of
quarkonium propagation in the medium. Here, we follow the effect of broadening
on a state of of the $Q$ and $\bar{Q}$~\cite{Aronson:2017ymv}. 

The analysis of transverse broadening of energetic $Q\bar{Q}$ states can be
nicely done by writing the states as light cone wavefunctions. The lowest Fock states
consist of $Q\bar{Q}$ in the color singlet configuration. In this
approximation the heavy meson state of momentum $\vec{P^+} = (P^+,{\bf P})$
can be approximated as:
\begin{eqnarray}       
 |\vec{P}^+ \rangle 
&=& \int \frac{d^2{\bf k}}{(2\pi)^{3}} \frac{dx}{ 2\sqrt{ x(1-x)}}
  \frac{\delta_{c_1c_2}}{\sqrt{3}} \, 
\psi(x,{\bf k})  \nonumber \\
&& \times a_Q^{\dagger\;  c_1 }(x\vec{P}^++{\bf k})  b_{\bar{Q}}^{\dagger \;  c_2  }
((1-x)\vec{P}^+-{\bf k})  |0 \rangle  \; , 
\label{Mp1}
\end{eqnarray}       
where  $a^\dagger$ ($b^\dagger$) represent an ``effective'' heavy quark
(anti-quark) in the $3$ ($\bar{3}$) state, $c_1,c_2$ being the color
indices~\cite{Sharma:2009hn,Sharma:2012dy}.

The light cone wavefunction in momentum space has the form,
\begin{eqnarray}
&& \psi(x,{\bf k}) = {\rm Norm} \times \exp\left(-\frac{{\bf k}^2 + m_Q^2 }{2 \Lambda^2(T) x(1-x) }   
  \right) \;   ,
\label{lowithnorm}
\end{eqnarray}
where the normalization is given by the relation,
\begin{eqnarray}
&& \frac{1}{2 (2\pi)^{3} } \int dx d^2{\bf k}  \;
| \psi(x,{\bf k}) |^2 = 1 \; .
\label{lonorm}
\end{eqnarray}
$x$ refers to the longitudinal fraction of the momentum. ${\bf k}$ refers to
the transverse momentum.

The transverse extent of the states
\begin{equation}
\frac{1}{2 (2\pi)^{3} } \int dx d^2{\bf k}  \;
{\Delta \bf k}^2  | \psi(x,{\bf k}) |^2 =  4\langle {\bf k^2} \rangle =  \frac{2}{3} \kappa^2  \; .
\label{lonorm2}
\end{equation}

To be concrete, let us follow the evolution of the $Q\bar{Q}$ state created in a hard collision
(early time) at a transverse location ${\bf x}_0$ (we will focus on central
rapidity) for a state that propagates with velocity ${\bf \beta}$,
such that  ${\bf x}(\tau) = {\bf x}_0 + {\bf \beta} (\tau-t_0)$.  The
temperature changes as the system evolves and the state moves and we call the
instantaneous value as $T$. The evolution is started at $t=t_0$ which is taken 
to be $t_0=0.6$fm. By this time the medium is likely to be thermalized and a
hydrodynamic description is possible. (For reviews of hydrodynamic
simulations see~\cite{Romatschke:2009im,Teaney:2009qa,Hirano:2012qz,Song:2013gia,Gale:2013da,Shen:2014vra,Jeon:2015dfa,Jaiswal:2016hex}.) 

After traversing the medium for a time $t$, the cumulative relative transverse 
dimensional momentum transfer  reads
\begin{eqnarray}
&&\chi m_D^2 \xi  = \int_{t_0}^t  d\tau \frac{m_D^2(\bf{x}(\tau),\tau) }{\lambda_q({\bf x}(\tau),\tau) } \xi\, , 
\end{eqnarray}
Here we use $N_f=2$.  The scattering inverse length of the quark is
${1}/{\lambda_q}  = \sigma_{qq}\rho_q + \sigma_{qg} \rho_g$, where $\rho_q$ and
$\rho_g$ are the partial densities of light quarks and gluons in the QGP.  The
elastic scattering cross sections are given by 
\begin{equation}
\sigma_{qq} = \frac{1}{18 \pi} \frac{g^4}{m_D^2} \, ,  \quad \sigma_{qg} = \frac{1}{8 \pi} \frac{g^4}{m_D^2} \, .
\end{equation}

We initialize the wavefunction $\psi_{i}(\Delta {\bf k}, x)$ of the
proto-quarkonium $Q\bar{Q}$  state with a width $\Lambda_0 \equiv \Lambda(T=0)
$.  This is a natural choice since in the absence of a medium it will evolve on
the time-scale of  ${\cal O}(1{\rm fm})$ into the observed heavy meson. By
propagating in the medium this initial wavefunction  accumulates transverse
momentum broadening  $\chi m_D^2 \xi$.  The probability that this  $Q\bar{Q}$
configuration will transition into a final-state heavy meson with thermal
wavefunction $\psi_{f}(\Delta {\bf k}, x)$ with $\Lambda(T)$ is given by 
\begin{equation}
\begin{split}
  P_{f\leftarrow i} (\chi m_D^2 \xi,T) & =  \left|  \frac{1}{2 (2\pi)^{3} }  \int d^{2}{\bf k} dx \,
\psi_{f}^* (\Delta {\bf k},x)\psi_{i}(\Delta {\bf k}, x) \right|^{2}  
\\
& \hspace*{-1.in}= \left| \frac{1}{2 (2\pi)^{3} }  \int dx \; {\rm Norm}_f {\rm Norm}_i \, \pi  
  \,  e^{-\frac{ m_{Q}^{2} }{ x(1-x)\Lambda(T)^{2} } }  e^{-\frac{ m_{Q}^{2} }{ x(1-x)\Lambda_0^{2} } }  \right.  \\
&  \hspace*{-.9in}  \times \left .  \,   \frac{ 2 [ x(1-x)\Lambda(T)^{2}]
[\chi m_D^{2}\xi+x(1-x)\Lambda_0^{2}] }
{   [ x(1-x)\Lambda(T)^{2}] + [\chi m_D^{2}\xi+x(1-x)\Lambda_0^{2}]  }  \; \right|^2 \, . \;\; \quad 
\label{sprob}
\end{split}
\end{equation}
In Eq.~(\ref{sprob}) ${\rm Norm}_i$  is the normalization of the initial state,
including the transverse momentum broadening 
from collisional interactions,  and  ${\rm Norm}_f$ 
if the normalization of the final state. 

The dissociation rate for the specific quarkonium state  can then be expressed
as 
\begin{equation}
t_{\rm diss.}  =  -  \frac{1}{P_{f\leftarrow i} (\chi m_D^2 \xi,T)}  
 \frac{d P_{f\leftarrow i} (\chi m_D^2 \xi,T) }{dt}  \, .
\end{equation}

Finally, the third ingredient we need is the formation time $t_{\rm form.}$. In
various studies this was taken to be $\sim\gamma \frac{1}{E_b}$ (where $\gamma$
takes care of the time dilation). Here we consider it as a parameter which we
vary between $1$fm to $1.5$fm. The intuition behind this choice is that for the
thermal wavefunction, $\tau_f$ refers to the time scale on which a
proto-quarkonuim  $\qqb$ state is affected more by the medium and is
dependent on the medium properties like $T$ than the boost of the $\qqb$
state. 

With all the ingredients in place, we evolve using Eq.~\ref{rateq} in a medium which is given by a
solution of the hydrodynamic equations. This is taken from the publicly
available iBNE-VISHNU implementation of hydrodynamics~\cite{Shen:2014vra} with
Glauber initial states. The distribution of the hard collisions is given by the
distribution of binary collisions in the Glauber model. In the azimuthal
direction, the $p_T$ momentum direction is isotropic. 

\begin{figure}
\includegraphics[width=0.47\linewidth]{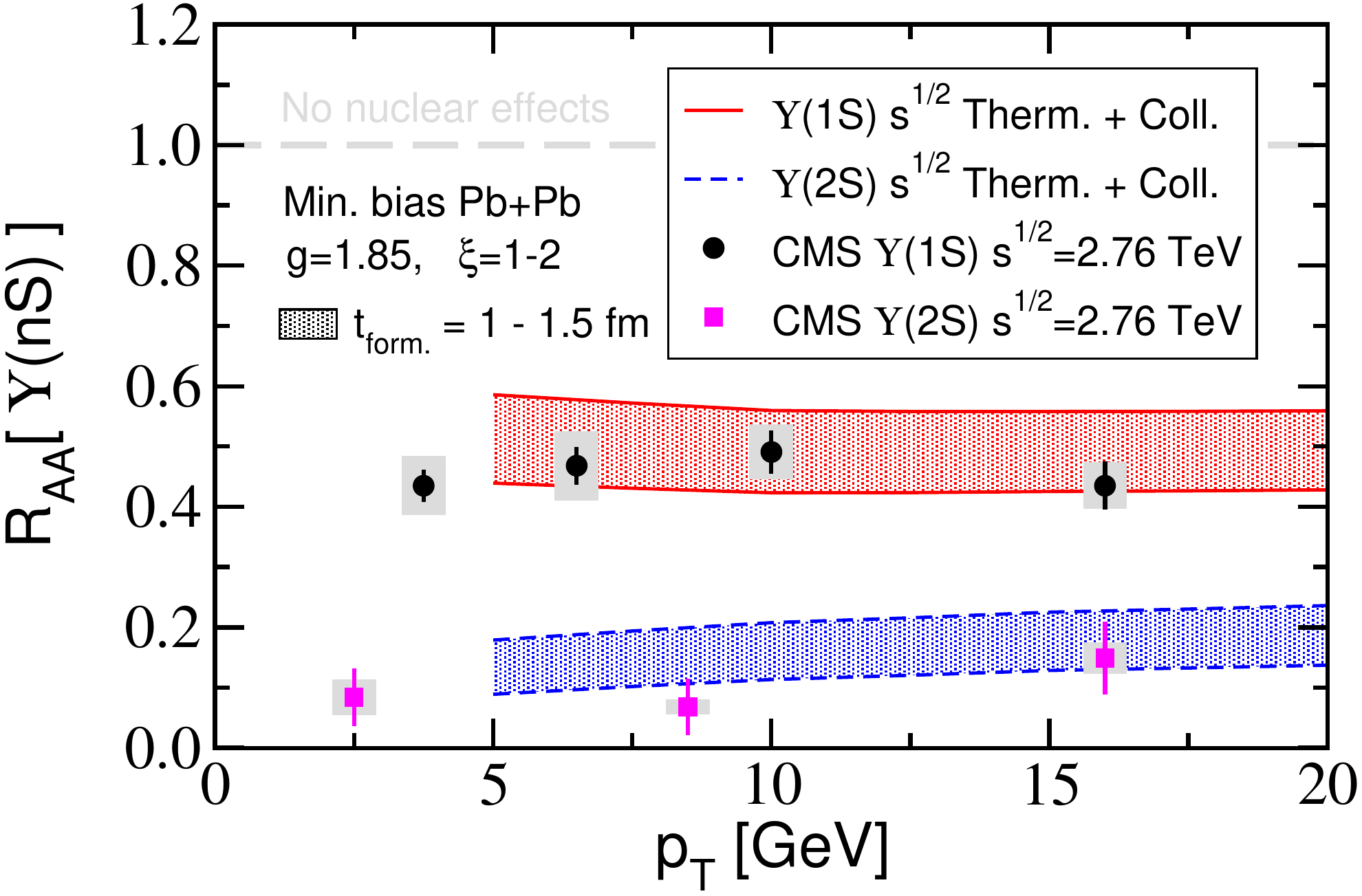}
\includegraphics[width=0.5\linewidth]{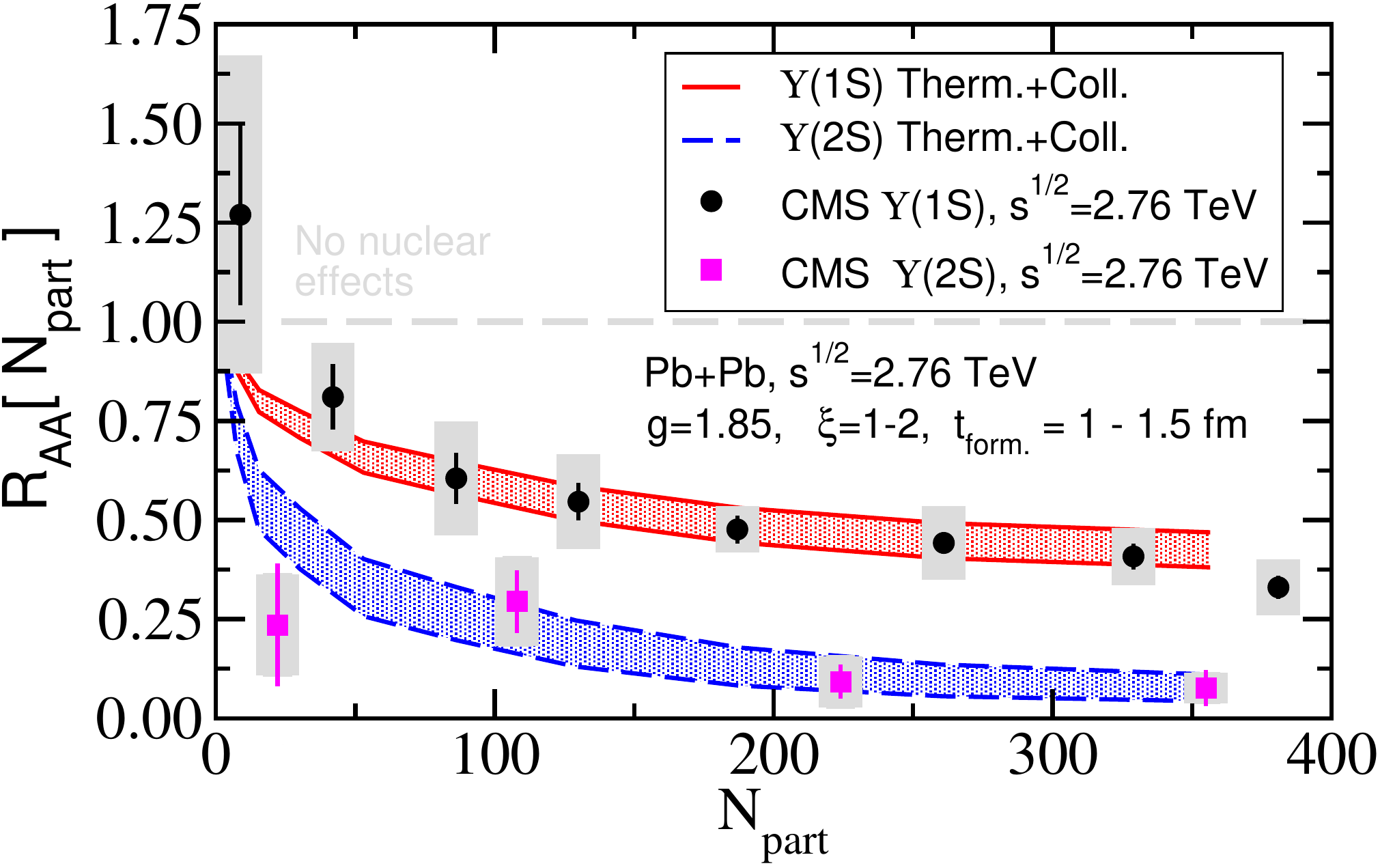}
\caption{$R_{AA}$ versus $p_T$ (left panel) and $N_{\rm{part}}$ (right panel)
for PbPb collisions at $2.76$TeV at the LHC for $\Upsilon$ states.}
\label{fig:Upsilon}
\end{figure}

Solving Eq.~\ref{rateq} for each of the mesonic states gives the $p_T$
differential yields for them. Finally, after taking into account the feed-down
from the excited states to the ground states we can find the final $p_T$ differential
yields in $AA$ collisions. 

Dividing the yields in $AA$ with the normalized yields in $pp$ gives $R_{AA}$.
In the next section, we show the results from our calculation for at the LHC.

\subsection{Results}
We now show a set of illustrative results of the calculation and comparisons
with the observations. 

\begin{figure}
\includegraphics[width=0.45\linewidth]{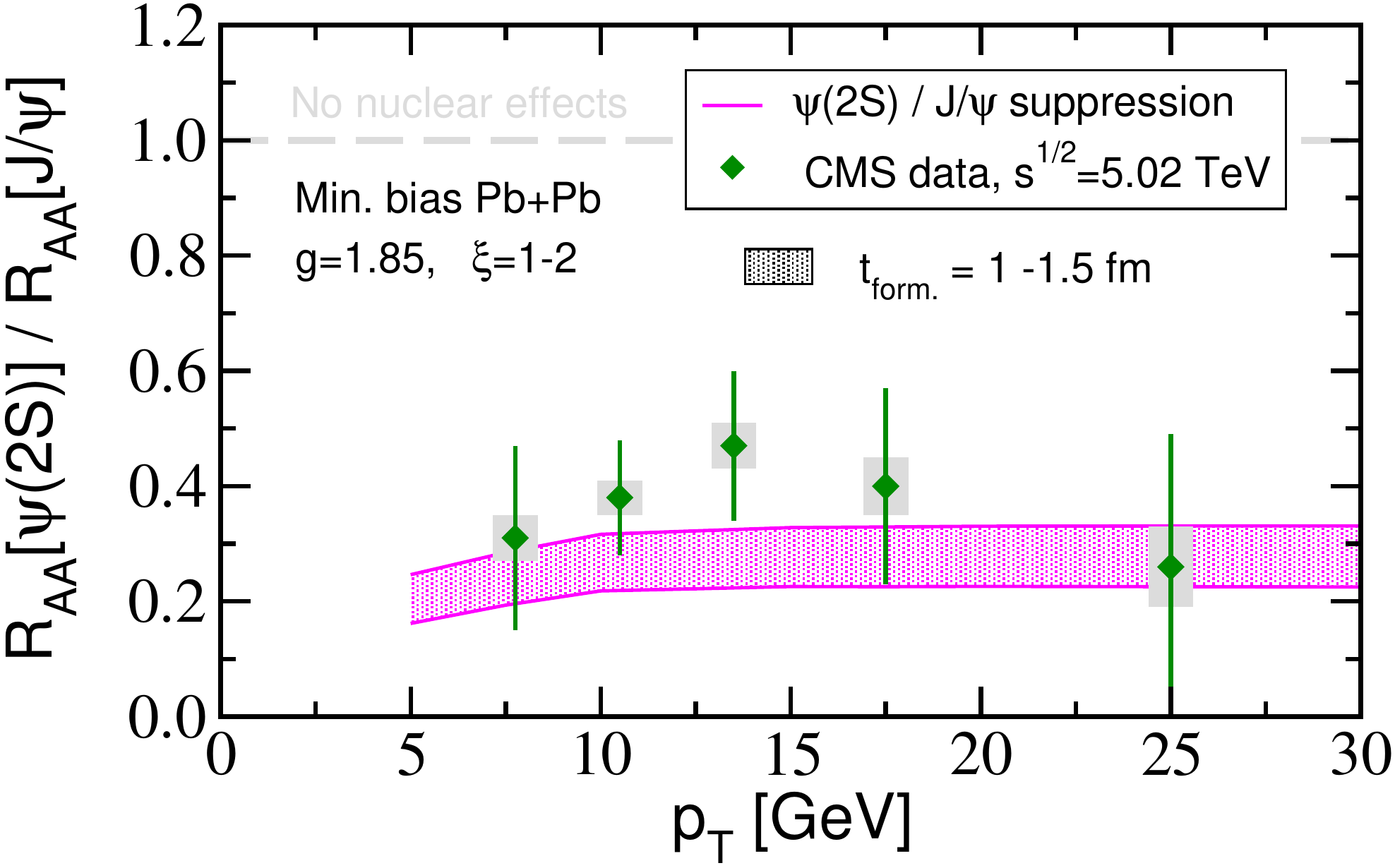}
\includegraphics[width=0.5\linewidth]{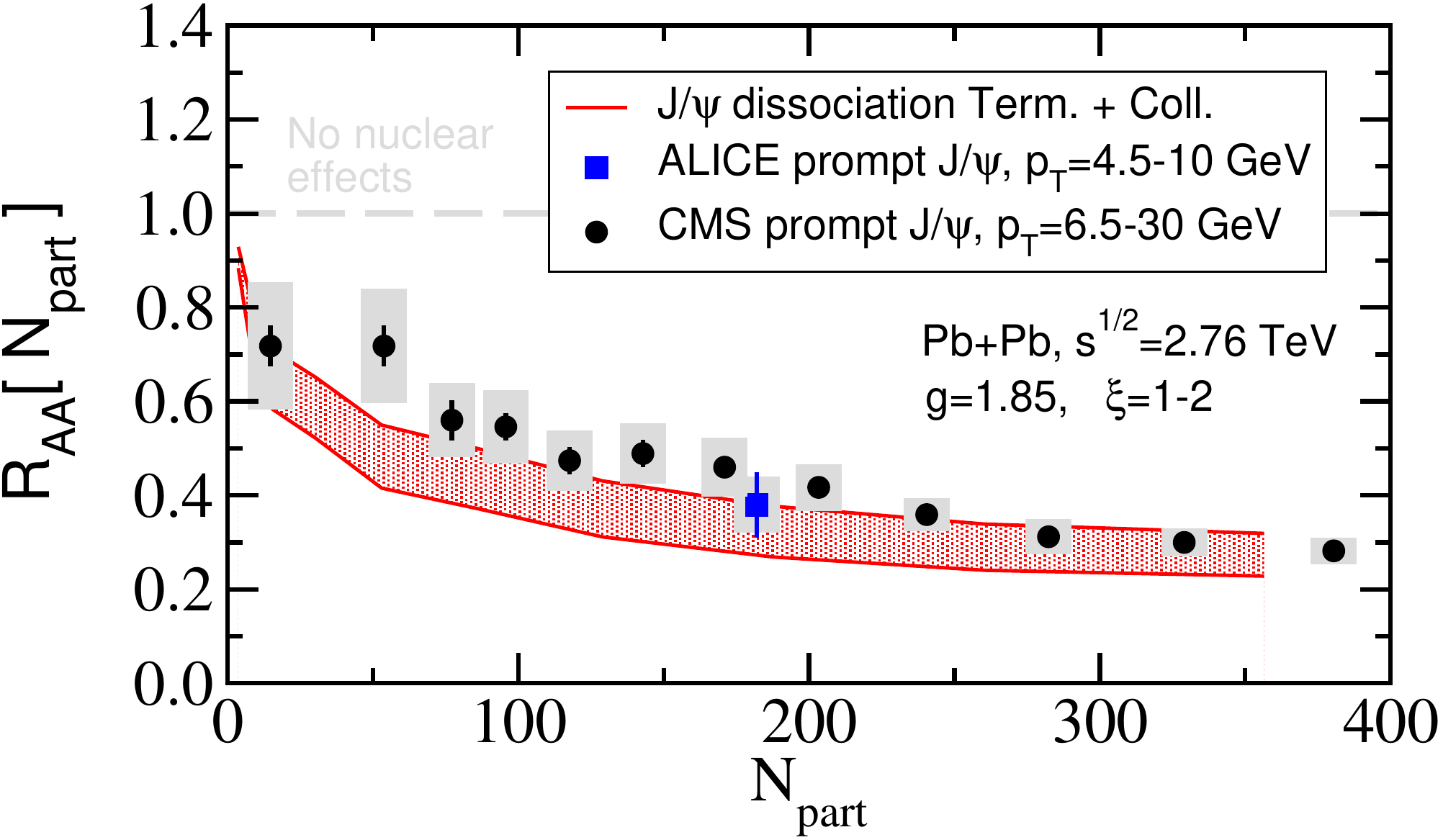}
\caption{The relative suppression of $J/\psi$ and $\psi(2s)$ versus $p_T$ 
in $5.02$TeV PbPb collisions (left panel) and $R_{AA}$ versus $N_{\rm{part}}$ 
(right panel) in $2.76$TeV PbPb collisions for $J/\psi$ at the LHC}
\label{fig:Jpsi}
\end{figure}

First considering the $\Upsilon$ states, we show results as a function of $p_T$
as well as $N_{\rm{part}}$ in Fig.~\ref{fig:Upsilon}.  We show similar results
for charmonia in Fig.~\ref{fig:Jpsi}. One nice feature is both $J/\psi$ and
$\Upsilon$ ground states and the first excited states are described by the
model. Both screening and dissociation play a role in getting the relative
suppression between the two.

Looking ahead, one can think of improvements to the calculation. First, 
the largest systematic uncertainty in the calculation comes from the ignorance
about the formation dynamics. These have been parameterized by a single
parameter in the model, $t_{\rm{form}}$. Reducing this uncertainty will require
modelling the conversion of the short distance color-singlet and
color-octet states to a bound color singlet state. Second, the treatment of the 
color degrees of freedom in the model above was too simplified. A recently
developed EFT~\cite{Makris:2019ttx,Makris:2019kap} for the propagation of high $p_T$ quarkonia in the QGP
provides a framework in which the color structure and the proper 
kinematics of the relevant modes can be systematically treated. Finally, the
dynamics described here are given by rate equations which miss the coherent
evolution of the $\qqb$ state. As we discuss below, in a simple example, 
these quantum effects might be important in a rapidly evolving QGP.

\section{Quarkonia in the QGP as open quantum systems~\label{sec:open}}
In the phenomenological applications discussed above, the dissociation rate 
for quarkonia are given by the form
\begin{equation}
\Gamma\sim \sum_f \overline{|\langle \Psi|\hat{O}|f\rangle|}^2
~\label{eq:Gamma}
\end{equation}
where $\hat{O}$ is a transition operator $|f\rangle$ refers to a dissociated
state. For example, $\hat{O}=\bfE\cdot\bfr$ and $|f\rangle$ correspond to 
color-octet states in the case of gluo-dissociation. The overline corresponds
to taking a thermal expectation value. 

In a time independent medium, this gives the leading order (in $\hat{O}$) decay rate for an initial
state $|\Psi\rangle$ which is typically chosen to be the quarkonium wavefunction
in vacuum.

However, in a time dependent medium, there are additional effects. As discussed
above, both the real and imaginary parts of the complex potential depend on
time. Even so, there are two cases where Eq.~\ref{eq:Gamma} gives the
instantaneous decay rate in the medium.

The first case is when the background potential changes very rapidly compared
to the energy difference between levels ($\Delta E$) of the system. Then an ``instant''
formalism can be used and $|\Psi\rangle$ can be taken as the vacuum state at any
given time. The other case is when the background potential changes very
slowly, and one can assume that $|\Psi\rangle$ represents the eigenstate of the 
instantaneous potential (adiabatic approximation)~\cite{Dutta:2012nw}.

For Bjorken evolution, the rate of change is $\sim \frac{1}{T}
\frac{dT}{dt}=-\frac{1}{3 \tau}$. In the absence of a clear hierarchy between
$1/\tau$ and $\Delta E$, one needs to follow the quantum state as the medium
evolves, while keeping track of the exchange of energy and momentum between the
state and the medium. The theory of open quantum systems commonly used in the
study of non-equilibrium condensed matter systems~\cite{Breuer:2002pc} provides
such a formalism.

In this formalism, the system is described by the density matrix ($\rho$) of
the $\qqb$ state obtained by tracing out the environmental degrees of freedom.
The evolution of $\rho$ is described by master equations
~\cite{Akamatsu:2011se,Akamatsu:2012vt,Akamatsu:2013,Akamatsu:2015kaa,Kajimoto:2017rel,Akamatsu:2018xim,Miura:2019ssi,Brambilla:2016wgg,Brambilla:2017zei,Brambilla:2019tpt,Brambilla:2019oaa,Blaizot:2018oev}
obtained after this tracing procedure. (See Ref.~\cite{Akamatsu:2020ypb}
for a recent review.)

There are two regimes explored widely
in the literature~\cite{Breuer:2002pc,Akamatsu:2013}.  One is when the system
relaxation time is longer than $1/\Delta E$. This is known as the quantum
optical limit.  The\ second case is when the system relaxation time is shorter
than $1/\Delta E$. This is known as the quantum Boltzmann limit.  It is
typically assumed that the relaxation time for the environment is short
compared to the system relaxation time, otherwise the evolution equations at a
given instant depend on the full history of the evolution: i.e. the equations
are not Markovian. [In the presence of long lived excitations in the
environment the Markovian approximation may be violated. (See
Sec.~\ref{sec:correlated}.)]

In the quantum optical regime, a convenient basis for $\rho$ is the eigenstates
of the system Hamiltonian. Interactions with the environment lead to
transitions between the states as well as transitions out of the Hilbert space
of bound states. This formalism was used to write master equations for the
system density matrix in terms of the transition rates between the states and
calculate the relative yields of $c\bar{c}$ and $b\bar{b}$ states in
Ref.~\cite{Borghini:2011yq,Borghini:2011ms}. This approach is of particular
interest for $b\bar{b}$ states in the QGP if they ``survive'' for temperatures
sufficiently above the crossover temperature.

In the quantum Boltzmann regime, it is more convenient to  write the density
matrix in the position basis. A model where a $\qqb$ interacting with an
environment with an ohmic spectral function was discussed in
Ref.~\cite{Young:2010jq}.  The authors wrote the formal expression for the
system density functional in terms of a path integral. They subsequently used
Monte Carlo techniques to integrate out the environment numerically and
calculated the Euclidean current-current correlation function
$\bar{\psi}\gamma^\mu\psi(\tau,\bfr)\bar{\psi}\gamma^\mu\psi(0,0)$. (This is
closely related to $\calC(q^\mu)$ defined in Eq.~\ref{eq:Ctr}.)

In general, the process of explicitly tracing out the environmental degrees is
complicated. However, evolution equations for the $\qqb$ density matrix
evolution using two different approaches. 

One approach was developed in a series of
papers~\cite{Akamatsu:2011se,Akamatsu:2012vt,Akamatsu:2013,Akamatsu:2015kaa,Kajimoto:2017rel,Akamatsu:2018xim,Miura:2019ssi},
in which master equations for the density matrices for isolated heavy quarks
and $\qqb$ pairs were derived and applied in weak coupling.  We summarize an
important result from Ref.~\cite{Akamatsu:2013} relevant for
Sec.~\ref{sec:correlated} and refer the reader to the original papers for more
details.

The key assumptions in deriving the master equations are
\begin{enumerate}
\item{The strong coupling $g$ is assumed to be small. Thus the real part of the
$\qqb$ potential has the screened Coulomb form. Furthermore, the gluonic
propagators are replaced by the weak coupling form (for eg. Eq.~\ref{eq:PiL}).}
\item{It was assumed that $E_b\ll m_D$ (consistent with the quantum Boltzmann
regime)}
\item{Furthermore, we will focus on the regime where quantum dissipation does not play
an important role. This corresponds to dropping ${\cal{O}}(v)$ and
${\cal{O}}(v^2)$ terms in the path integral used to derive the master equation
~\cite{Akamatsu:2015kaa} and can be formally justified if $v$ is small. However 
dissipation can have important quantitative effects~\cite{Miura:2019ssi}}
\end{enumerate}
 
Under these approximations, the master-equation for the $\qqb$ pair can
be written in Lindblad form~\cite{Akamatsu:2015kaa,Lindblad:1975ef}
\begin{eqnarray}
\label{eq:reduced_density}
\frac{\partial}{\partial t}\left(\begin{array}{c}{\rho_{1}} 
\\ {\rho_{8}}\end{array}\right)_{(t, \vec{r}, \overline{s})}
&=&\left(i \frac{\vec{\nabla}_{r}^{2}-\vec{\nabla}_{s}^{2}}{M}\right)
\left(\begin{array}{c}{\rho_{1}} \\ {\rho_{8}}\end{array}\right)_{(t, \vec{r})}
+i(V(\vec{r})-V(\vec{s}))\left[\begin{array}{cc}{C_{\mathrm{F}}} & {0} \\ 
{0} & {-1 / 2 N_{\mathrm{c}}}\end{array}\right]\left(\begin{array}{c}{\rho_{1}} \\ 
{\rho_{8}}\end{array}\right)_{(t, \vec{r}, \vec{s})
}\\
\nonumber
&&+{\mathcal{D}}(\vec{r}, \vec{s})\left(\begin{array}{c}{\rho_{1}} \\
{\rho_{8}}\end{array}\right)_{(t, \vec{r}, \vec{s})}.
\end{eqnarray}
Here $\vec{r}$ corresponds to the relative separation between the $\qqb$ in the
``ket'' space and $\vec{s}$ is the separation in the ``bra'' space.  
$\rho_{1,\;8}=\rho_{1,\;8}(t,\vec{r},\vec{s})$ are the singlet and octet
components of the $\qqb$ density matrix in position space. $V(\vec{r}),
V(\vec{s})$ correspond to the potential between $Q$ and $\bar{Q}$. We consider the $\qqb$ pair at rest in
the medium and hence the center-of-mass coordinates $\vec{R}$, $\vec{S}$ do not
play a role and we have suppressed the dependence on them.

${\mathcal{D}}(\vec{r},\vec{s})$ are terms related to  decoherence of the $\qqb$
state~\cite{Akamatsu:2015kaa},
\begin{equation}
\begin{split}
 {\mathcal{D}}(\vec{r}, \vec{s})=&
2 C_{\mathrm{F}} D(\overrightarrow{0})
-(D(\vec{r})+D(\vec{s}))
\left[\begin{array}{cc}{C_{\mathrm{F}}} & {0} \\
{0} & {-1 / 2 N_{\mathrm{c}}}\end{array}\right]
-2 D\left(\frac{\vec{r}-\vec{s}}{2}\right)
\left[\begin{array}{cc}{0} & {1 / 2 N_{\mathrm{c}}} \\ 
{C_{\mathrm{F}}} & {C_{\mathrm{F}}-1 / 2 N_{\mathrm{c}}}\end{array}\right]
\\
&+2 D\left(\frac{\vec{r}+\vec{s}}{2}\right)
\left[\begin{array}{cc}{0} & {1 / 2 N_{\mathrm{c}}} \\
{C_{\mathrm{F}}} & {-1 / N_{\mathrm{c}}}\end{array}\right].
\end{split}
\label{eq:aka_decoherence_terms}
\end{equation}
The function $D(\vec{r})$ is related to the imaginary part of gluonic
self-energy. In the HTL approximation,
\begin{equation}
\label{eq:akadrHTL}
     D(\vec{r}) = -g^2 T\int \frac{d^{3} k}{(2 \pi)^{3}} \frac{\pi
     m_{\mathrm{D}}^{2} e^{i \vec{k} \cdot
     \vec{r}}}{k\left(k^{2}+m_{\mathrm{D}}^{2}\right)^{2}}\;.
\end{equation} 
One can immediately identify $D(\vec{r})$ as the Fourier transform of the small
$q^0$ form of the longitudinal gluonic spectral function multiplied by $n_B(q^0)$
(Eqs.~\ref{eq:rho0HTL},~\ref{eq:smallq0}). 

Eq.~\ref{eq:reduced_density} can be solved by introducing noise fields
$\theta^a(t, \vec{r})$~\cite{Akamatsu:2015kaa} which are picked from an
ensemble which is specified by the following expectation values,
\begin{eqnarray}
\label{eq:noise_full_correlation}
\langle\langle \theta^a(t, \vec{r}) \rangle\rangle &=& 0 \nonumber\\
\langle\langle\theta^a(t, \vec{r})\theta^b(t', \vec{r}') \rangle\rangle &=&
\delta^{ab}D(\vec{r}-\vec{r}')\delta(t-t')\;, 
\end{eqnarray}
where $\langle\langle .. \rangle\rangle$ signifies taking the stochastic average over the noise fields.

For each member of the ensemble $\theta^a(t, \vec{r})$, $\psi$ is evolved using
the Schr\"{o}dinger equation,
\begin{eqnarray}
\label{eq:stochasticevolution}
\psi(t+dt) && = e^{-i H_{\theta}(t)dt} \psi(t)  \nonumber\\
H_{\theta}(\vec{r},t)&&=-\frac{\vec{\nabla}_{r}^{2}}{M} + V(r)(t^{a}\otimes t^{a,\ast}) 
+\theta^{a}(t,
\frac{\vec{r}}{2})\left(t^{a} \otimes 1\right)-\theta^{a}(t,-\frac{\vec{r}}{2})\left(1
\otimes t^{a *}\right).\;
\end{eqnarray}

The density matrix can be obtained by taking a stochastic average of the outer
product 
 \begin{equation}
 \label{eq:stochastic_density_matrix}
 \rho(t, \vec{r}, \vec{s}) = \langle\langle\;\; 
    \ket{\psi(t, \vec{r}) }\bra{\psi(t, \vec{s})} \;\;\rangle\rangle\;.
\end{equation}

A simplified version of Eq.~\ref{eq:stochasticevolution} was simulated in
\cite{Akamatsu:2018xim}, where the wavefunction was assumed to be
one-dimensional and the color-structure of $\qqb$ pair was neglected.
In~\cite{Sharma:2019xum} the calculation was extended to include the full color
structure and three dimensional wavefunction. In addition the evolution was
extended to noise correlated in time. We will describe this work in
Sec.~\ref{sec:correlated}.

The second approach was developed in a series of
papers~\cite{Brambilla:2016wgg,Brambilla:2017zei,Brambilla:2019tpt,Brambilla:2019oaa}.
The authors started from the pNRQCD lagrangian (Eq.~\ref{eq:LpNRQCD}). At the
leading order in the multipole expansion, the evolution of the singlet and the
octet components of density matrix are given by the singlet and the octet
hamiltonians respectively. The first order correction arises due to the dipole
term. The diagrams are similar to Fig.~\ref{fig:SOS} where the propagators now
refer to the propagation of the density matrix in time. (See
Refs.~\cite{Brambilla:2016wgg,Brambilla:2017zei} for the details of the
calculation.)

Refs.~\cite{Brambilla:2016wgg,Brambilla:2017zei} were the first papers to
include both gluo-dissociation and Landau damping in a consistent framework
taking into account the full non-Abelian and quantum nature of the problem 
of a $\qqb$ propagating in a thermal medium. This approach
conserves the number of $Q$ and $\bar{Q}$ during the evolution. At the same
time, since the evolution maintains the coherence, both ``dissociation'' and
``recombination'' occur during evolution. Finally, it has the advantage that
various hierarchies between $T$, $m_D$, and $E_b$ can be considered.

In
Refs.~\cite{Brambilla:2016wgg,Brambilla:2017zei,Brambilla:2019tpt,Brambilla:2019oaa},
it was shown that in the hierarchy $1/r\gg \pi T \sim m_D\gg
E_b,\;\Lambda_{QCD}$, the master equations at a given $T$ are specified by only 
two constants, $\kappa$ and $\gamma$, the $\Re e$ and the $\Im m$ parts of the
chromo-electric field correlator (Eq.~\ref{eq:GEEStaticLocal}).

In Refs.~\cite{Brambilla:2016wgg,Brambilla:2017zei} this technique was applied
to the phenomenology of bottomonia at LHC by calculating $R_{AA}$ for
$\Upsilon(1S)$ and $\Upsilon(2S)$ states (see Fig.~$7$ in
Ref.~\cite{Brambilla:2017zei}). 

\subsection{Correlated and uncorrelated noise~\label{sec:correlated}}
We first simplify the stochastic evolution equation
(Eq.~\ref{eq:stochasticevolution}) (and therefore its corresponding master
equation) by expanding the decoherence terms in small $\vec{r}, \vec{s}$. This
approximation is motivated by the hierarchy between the inverse size of the
states and the temperature $1/r\gg T$. 

This allows us to extend the calculation to a three dimensional system while
keeping all the color structure of $\qqb$ pair intact without a high
computational cost. The calculation allows for transitions between different
angular-momentum states ($l = 0,1$).  Transitions which change the
angular-momentum by two units or more are
suppressed~\cite{Brambilla:2016wgg,Brambilla:2017zei} by
${\mathcal{O}}(r^2T^2)$. 

The stochastic evolution operator up to ${\mathcal{O}}(\vec{r}^2)$ for a $l=0$
initial state is,
\begin{eqnarray}
\label{eq:paper_decoherence_1}
  && \psi(t+dt)  = e^{-i H_{\theta}dt}\psi(t) \nonumber \\  
  &&  H_{\theta} =  (\frac{-\nabla^2}{ M} (1_Q \otimes 1_{\bar{Q}}) + V(r)(t^a \otimes t^{\ast,a})
   +D^{a} \ \frac{\vec{r}}{2}\cdot \vec{\theta}^{a}(t)+ F^{a} \  \theta^a(t)+\mathcal{O}(\vec{r}^2) ) \nonumber \\
&&\theta^{a}(t) = \theta^{a}(\vec{r},t)|_{\vec{r}=0},\quad \vec{\theta}_{i}(t)
= \vec{\nabla}_{i}\theta(\vec{r},t)|_{\vec{r}=0}\;,
\end{eqnarray}
where the noise field $\theta(\vec{r},t)$ was defined in
Eq.~\ref{eq:noise_full_correlation}. ($F^{a}= (t^a_Q\otimes 1_{\bar{Q}} -
1_{Q}\otimes t^{\ast\;a}_{\bar{Q}})$ and $D^{a} = (t^a_Q\otimes 1_{\bar{Q}} +
1_{Q}\otimes t^{\ast\;a}_{\bar{Q}})$ are operators in the color-space of $\qqb$
pair. The subscript $i$ refers to the spatial index, and we refer
$\vec{\nabla}_i\theta({\bf{r}}, t)|_{{\bf{r}}=0}$ as the three tuple $\theta_i(t)$ for
notational convenience.)

The noises appearing in Eq.~\ref{eq:paper_decoherence_1} can be
generated as random-fluctuations uncorrelated at unequal times,
\begin{eqnarray}
\label{eq:paper_noise_correlation}
 %\langle\langle {\theta}^a(t) \rangle\rangle &=& 0 \nonumber\\
 \langle\langle {\theta}^a_i(t) \rangle\rangle &=& 0 \nonumber\\
 %\langle\langle \theta^a(t)\theta^b(t')\rangle\rangle &=& \delta^{ab}\delta(t-t')D(\vec{0}),\nonumber \\
 \langle\langle \vec{\theta_i}^a(t) \vec{\theta_j}^b(t')\rangle\rangle &=& \delta^{ab}\delta(t-t')\delta_{ij}\frac{-\nabla^2}{3}D(\vec{0}).
\end{eqnarray}
The noise $\vec{\theta_i}^a(t)$ can be interpreted as the $gE^{i\; a}(t)$ at
the center of mass of the $Q$ and $\bar{Q}$, and $-\frac{1}{3}\nabla^2 D$ is simply
the chromo-electric correlator. The noise correlated only at equal time and this
is reminiscent of the static limit discussed in Sec.~\ref{sec:screening}. This
is a direct consequence of the assumption that $E_b\ll m_D$: the system
evolution is slow compared to the relaxation and hence the noise is
uncorrelated at unequal times.

The above Hamiltonian evolution is written for a three dimensional system.
Since, $V(\vec{r})$ is rotationally invariant, we can separate the radial part
of the three dimensional wavefunction from its angular part. The wavefunction
in position space can be written as
\begin{equation}
    \label{eq:3d_wavefunctions}
    \Psi(\vec{r},t) = \frac{\psi(r)}{r}\Theta(\beta,\phi),
\end{equation}
where $\psi(r)$ is the radial wavefunction and $\Theta$ is the wavefunction in angular momentum space, with $\beta$ being the polar angle and $\phi$ azimuthal angle. We also define the normalized color states for $\qqb$ octet and singlet wavefunction as,
\begin{equation}
\label{eq:color_wavefunctions}
\ket{S} = \frac{1}{\sqrt{N_c}}\sum_{lk}\ket{lk} \quad \ket{O^a} =  \frac{1}{\sqrt{T_F}}\sum_{lk} (t^a)_{lk}\ket{lk}.
\end{equation}
The indices $l,k$ denotes the color states of a single quark or antiquark.  

Finally, we project the evolution operator in the
Eq.~\ref{eq:paper_decoherence_1} into the color and angular momentum space of
$\qqb$ pair,
\begin{equation}
\label{eq:paper_matrix_eqn1}
H_{\theta}(r,t) =  \left(\begin{array}{cccc}
H^{S}_{0}(r,t) & 0 & 0 & \frac{1}{\sqrt{2N_c}} r|\vec{\theta}^{c}(t)|\delta_{ac} \\
0 & H^{S}_{1}(r,t) & \frac{1}{\sqrt{2N_c}} r |\vec{\theta}^{c}(t)|\delta_{ac} & 0 \\
0 & \frac{1}{\sqrt{2N_c}} r |\vec{\theta}^{c}(t)|\delta_{ac} & H^{O}_{0}(r,t)+f^{abc}\theta^{c}(t)& \frac{d^{abc}}{2} r |\vec{\theta}(t)| \\
\frac{1}{\sqrt{2N_c}} r|\vec{\theta}^{a}(t)|\delta_{ac} & 0  &\frac{d^{abc}}{2}r|\vec{\theta}(t)|&  H^{O}_{1}(r,t)+f^{abc}\theta^{c}(t)  
\end{array}{} \right). 
\end{equation}

This Hamiltonian acts on the wavefunction given in the form 
 \begin{equation}
 \label{eq:wavefunctions_section1}
 \psi(r,t)  = \left(\psi^{S}_{l=0}(r,t), \psi^{S}_{l=1}(r,t),\psi^{O^{a}}_{l=0}(r,t),
      \psi^{O^{a}}_{l=1}(r,t) \right).
 \end{equation}
 
Here, $\psi^S(r,t)$ and $\psi^{O^a}(r,t)$ denote radial wavefunctions
for $\qqb$ pair in singlet and octet states respectively and the index $a$ runs
from $1$ to $(N_c^2-1)$ for different color-octet states. $l$ denotes the angular
momentum states, which take the values $l = 0,1$. The Hamiltonians for the
singlet and octet states are 
\begin{eqnarray}
\label{eq:section1_hamiltonians}
H^S_l && = -\frac{1}{M}\frac{\partial^2}{\partial r^2} - \frac{C_{\mathrm{F} }\alpha}{r}e^{-m_D r} + \frac{l(l+1)}{ M r^2} \nonumber, \\
H^O_l && = -\frac{1}{M}\frac{\partial^2}{\partial r^2} + \frac{\alpha}{2 N_{\mathrm{c}}r}e^{-m_D r} + \frac{l(l+1)}{M r^2}. 
\end{eqnarray}

Under the approximations considered, the correlation functions
(Eqs.~\ref{eq:paper_noise_correlation}) are the most important quantities which
control the suppression pattern. As discussed above, when $E_b\ll m_D$ the noise
correlator is local in time
(Eqs.~\ref{eq:stochasticevolution},~\ref{eq:paper_noise_correlation}). When
this hierarchy is not satisfied, the noise terms need to carry information
about the correlations of the chromo-electric fields in time. 

This suggests a simple modification to include gluo-dissociation where on-shell
gluons are absorbed. To leading order,
\begin{eqnarray}
\label{eqn:EE_finite_frequency}
  &&\langle\langle {\vec{\theta}}^a_i(t){\vec{\theta}}^b_j(t')\rangle\rangle \nonumber\\
  =&&{\rm{Tr}} \langle e^{-\mathcal{H}/T}[g\vec{E}^a_i(t)][g\vec{E}^b_j(t')]\rangle \nonumber \\
  =&& 
  \delta_{ab}\delta_{ij}\frac{g^2 T^4}{6 N_{\rm{c}}\pi} 
  \int_{0}^{\infty} d\xi \ x^3 \cos(\xi\;T\;(t-t')))\frac{1}{e^{\xi}-1} 
  . \nonumber \\
 &&
\end{eqnarray}

We will explore the two cases (Eq.~\ref{eq:paper_noise_correlation} labelled
decoherence and Eq.~\ref{eqn:EE_finite_frequency} labelled as
gluo-dissociation) below.

\begin{figure}[h]
\includegraphics[width = 0.5\textwidth]{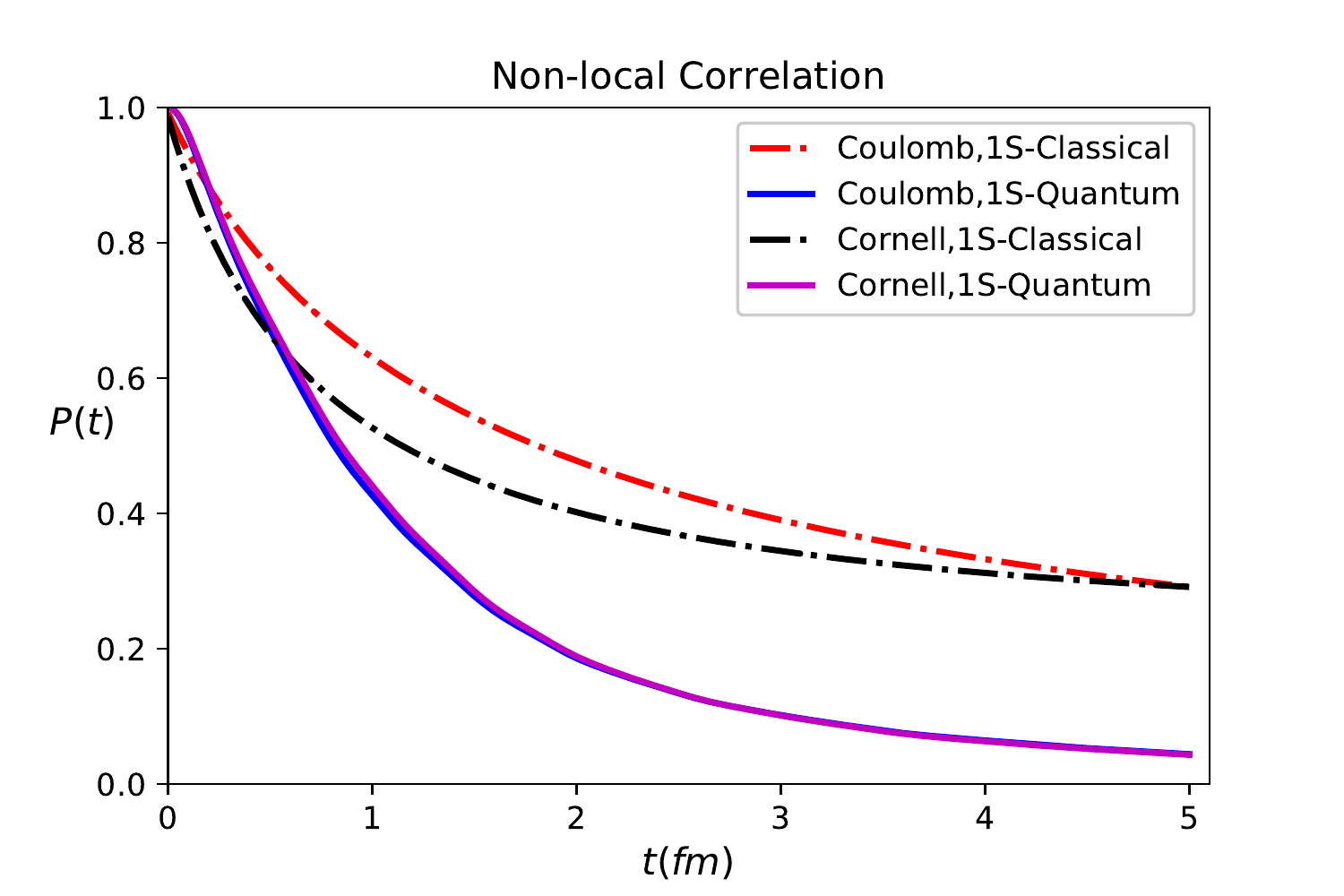}
\includegraphics[width = 0.5\textwidth]{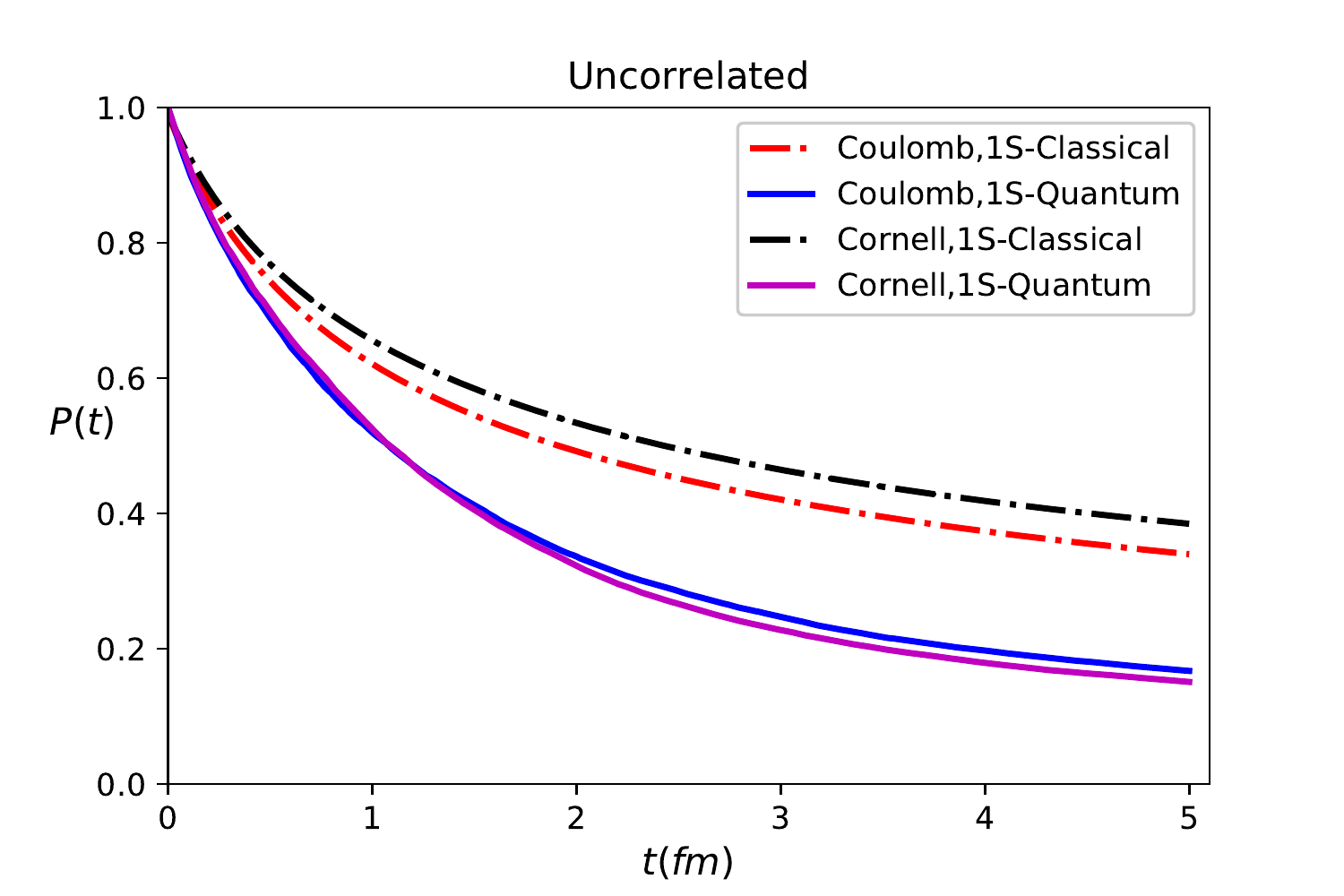}
\caption{\label{fig:final_results_qvsc_gluodissociation}
(color online) Comparison of $P(t)$ between
the classical (black dot dashed for Cornell and red dot dashed for Coulomb) and
quantum (pink solid for Cornell and blue solid for Coulomb) approach for the
case of gluo-dissociation (left panel) and decoherence (right panel). The classical results at early
time differ substantially for the two initial wavefunctions we used, whereas
the quantum results are closer.}
\end{figure}

\subsection{Results}
The survival probability at any given time $t$ is defined as 
\begin{equation}
    \label{eq:survival_probability}
    P(t) = \langle\langle \  |\braket{\psi_{0}|\psi_{\theta}(t)}|^2  \ \rangle\rangle,
\end{equation}
$\psi_0$ is the wavefunction in vacuum. The quantity $P(t)$ at freezeout is
related to the observed suppression $R_{AA}$ of quarkonium states.

We modelled the background system evolution as a Bjorken expanding medium. The
calculation was performed for $\Upsilon(1S)$ states. For the initial state, we
considered two commonly studied forms. We chose the initial state as the
eigenstate for a phenomenological vacuum Cornell potential and for comparison
also considered the eigenstate of the Coulomb potential. (See
Ref.~\cite{Sharma:2019xum} for details of the chosen parameters and comparisons
of the wavefunction forms.)

We compare the results obtained using the stochastic Schor\"{o}dinger equation
(``quantum'' evolution) to the results  obtained by solving the rate equation
(Eq.~\ref{eq:classical_RAA}) which are labelled as ``classical'' evolution as
it does not carry information about the coherent evolution of the wavefunction.

We have presented our comparison of survival probability $P(t)$ between the
classical and quantum approach in
Fig.~\ref{fig:final_results_qvsc_gluodissociation} for decoherence (left panel)
and gluo-dissociation (right panel). 

The main conclusions are the following. First, we note that the rate equations
give a larger survival probability than the quantum calculation even though the
correlation of the noise term in the quantum calculation, and $\Gamma$ in the
classical calculation are both given by chromo-electric field correlator. As mentioned
above, this difference is because the quantum evolution tracks how the
$Q\bar{Q}$ wavefunction evolves with time while in the classical case we use
the initial wavefunction. Second, we note that 
experimental results for $\Upsilon(1S)$
(Refs.~\cite{Chatrchyan:2011pe,Khachatryan:2016xxp,Sirunyan:2017lzi}
suggest $R_{AA}\sim 0.5$ for $1S$ states) which is substantially larger than
the results for both correlated (gluo-dissociation) and the uncorrelated noise
(decoherence) although the results for decoherence are closer to the
experimental value (Fig.~\ref{fig:final_results_qvsc_gluodissociation}). 
However, there are additional effects one needs to consider before
a quantitative comparison with the phenomenology can be made. For example, it 
was demonstrated in Ref.~\cite{Miura:2019ssi} that in the Abelian theory 
that inclusion of quantum dissipation can lead to an increase in $R_{AA}$
compared to the result for decoherence. Additionally, one needs to include
feed-down contributions from excited states and take a realistic 
medium evolution model. Third, we note that the results do not depend strongly on the choice of the
initial state (Cornell and Coulomb eigenstates) since these are narrow states.
However, a similar calculation for $2S$ states shows a significant difference
between the two. This implies that more effort needs to be made to understand
the choice of the initial state. Finally, we note that for the quantum
calculation with the parameters used, gluo-dissociation is a stronger effect.
This shows that if the hierarchy between $m_D$ and $E_b$ is not very
strong, a formalism to handle both effects is important.

While improvements along the suggested lines above are ongoing, the calculation
makes it clear that quantum effects play an important role in the evolution of
$\qqb$ states.

\section{Conclusions}
The $\qqb$ system in a static thermal medium is characterized by various energy
scales, $M$, $1/r$, $E_b$, which need to be compared with the thermal scales
$m_D$ and $T$.  Since $M$ is much larger than the other scales, the system is
amenable to treatments using non-relativistic EFTs. 

If $1/r\gg E_b, T$ the EFT of the system is potential NRQCD. The potentials in
this theory are complex, reflecting the fact that the system can exchange
energy and momentum with the environment. pNRQCD (Sec.~\ref{sec:pNRQCD}) gives a useful framework to
connect the coefficients to the decay rates.

One can calculate decay rates and relate them to the observed $R_{AA}$ in heavy
ion collisions. These decay rates have been calculated to leading order using perturbation
theory and using EFTs and applied to quarkonium phenomenology
(Sec.~\ref{sec:lowpT}).

Given that the coupling $g$ is strong at these scales, non-perturbative
corrections are likely to be important. Over the last several years, significant
progress has occurred in the calculation of these coefficients using lattice
QCD.   

In a medium evolving in time, there is an additional scale associated with the
time scale of change of the background medium. If this time scale is comparable
to the system  scale ($1/E_b$),  it becomes necessary to follow the coherent
evolution of the system while taking into account decoherence due to the
interactions with the environment. A natural framework for this is the theory
of open quantum systems. The state of the system is described by a density
matrix whose evolution equation can be obtained by integrating out the
environment. 

Over the last few years, master equations for the density matrices have been
derived for quarkonia in the weak coupling limit. Similar equations can also be
obtained under various hierarchies of the energy scales
directly from pNRQCD. (See Sec.~\ref{sec:open})

Combined with the progress in the non-perturbative calculation of the complex
potentials and the chromo-electric correlator, these developments have brought
the goal of the calculation of the observed yields of quarkonia in $AA$
collisions from first principles closer to fruition. 

\section{Acknowledgements}
We acknowledge collaborators on various projects related to heavy quark and
quarkonium physics, Samuel Aronson, Evan Borras, Sourendu Gupta, Brian Odegard,
Anurag Tiwari, Ivan Vitev, and Ben-Wei Zhang. We also thank Dibyendu Bala,
Saumen Datta for several illuminating discussions. We also acknowledge
discussions and exchanges with Yukiano Akamatsu, Jean-Paul Blaizot, Nora
Brambilla, Alexander Rothkopf, and Peter Petreczky. 

\section{Author contribution statement}
This single author of this review is Rishi Sharma.

\appendix


\begin{thebibliography}{}

%-- Hydro
%%\cite{Habich:2014jna}
%\bibitem{Habich:2014jna} 
%  M.~Habich, J.~L.~Nagle and P.~Romatschke,
%  %``Particle spectra and HBT radii for simulated central nuclear collisions of C + C, Al + Al, Cu + Cu, Au + Au, and Pb + Pb from $\sqrt{s}=62.4$ - $2760$ GeV,''
%  Eur.\ Phys.\ J.\ C {\bf 75}, no. 1, 15 (2015)
%  doi:10.1140/epjc/s10052-014-3206-7
%  [arXiv:1409.0040 [nucl-th]].
%  %%CITATION = doi:10.1140/epjc/s10052-014-3206-7;%%
%  %21 citations counted in INSPIRE as of 15 Aug 2017

%\cite{Romatschke:2009im}
\bibitem{Romatschke:2009im}
P.~Romatschke,
%``New Developments in Relativistic Viscous Hydrodynamics,''
Int. J. Mod. Phys. E \textbf{19} (2010), 1-53
doi:10.1142/S0218301310014613
[arXiv:0902.3663 [hep-ph]].
%408 citations counted in INSPIRE as of 01 Oct 2020

%\cite{Teaney:2009qa}
\bibitem{Teaney:2009qa}
D.~A.~Teaney,
%``Viscous Hydrodynamics and the Quark Gluon Plasma,''
doi:10.1142/9789814293297\_0004
[arXiv:0905.2433 [nucl-th]].
%227 citations counted in INSPIRE as of 01 Oct 2020

%\cite{Hirano:2012qz}
\bibitem{Hirano:2012qz}
T.~Hirano,
%``Dynamics of relativistic heavy ion collisions and the quark gluon plasma,''
Prog. Theor. Phys. Suppl. \textbf{195} (2012), 1-18
doi:10.1143/PTPS.195.1
%2 citations counted in INSPIRE as of 01 Oct 2020

%%\cite{Song:2012tv}
%\bibitem{Song:2012tv}
%H.~Song,
%%``Hydrodynamic Modeling and the QGP Shear Viscosity,''
%Eur. Phys. J. A \textbf{48}, 163 (2012)
%doi:10.1140/epja/i2012-12163-9
%[arXiv:1207.2396 [nucl-th]].
%%19 citations counted in INSPIRE as of 01 Oct 2020

%\cite{Song:2013gia}
\bibitem{Song:2013gia}
H.~Song,
%``Hydrodynamic modelling for relativistic heavy-ion collisions at RHIC and LHC,''
Pramana \textbf{84}, 703-715 (2015)
doi:10.1007/s12043-015-0971-2
[arXiv:1401.0079 [nucl-th]].
%34 citations counted in INSPIRE as of 01 Oct 2020

%\cite{Gale:2013da}
\bibitem{Gale:2013da}
C.~Gale, S.~Jeon and B.~Schenke,
%``Hydrodynamic Modeling of Heavy-Ion Collisions,''
Int. J. Mod. Phys. A \textbf{28}, 1340011 (2013)
doi:10.1142/S0217751X13400113
[arXiv:1301.5893 [nucl-th]].
%492 citations counted in INSPIRE as of 01 Oct 2020

%\cite{Heinz:2013th}
\bibitem{Heinz:2013th}
U.~Heinz and R.~Snellings,
%``Collective flow and viscosity in relativistic heavy-ion collisions,''
Ann. Rev. Nucl. Part. Sci. \textbf{63}, 123-151 (2013)
doi:10.1146/annurev-nucl-102212-170540
[arXiv:1301.2826 [nucl-th]].
%755 citations counted in INSPIRE as of 01 Oct 2020

%\cite{Shen:2014vra}
\bibitem{Shen:2014vra} 
  C.~Shen, Z.~Qiu, H.~Song, J.~Bernhard, S.~Bass and U.~Heinz,
  %``The iEBE-VISHNU code package for relativistic heavy-ion collisions,''
  Comput.\ Phys.\ Commun.\  {\bf 199}, 61 (2016)
  doi:10.1016/j.cpc.2015.08.039
  [arXiv:1409.8164 [nucl-th]].
  %%CITATION = doi:10.1016/j.cpc.2015.08.039;%%
  %85 citations counted in INSPIRE as of 15 Aug 2017

%\cite{Jeon:2015dfa}
\bibitem{Jeon:2015dfa}
S.~Jeon and U.~Heinz,
%``Introduction to Hydrodynamics,''
Int. J. Mod. Phys. E \textbf{24}, no.10, 1530010 (2015)
doi:10.1142/S0218301315300106
[arXiv:1503.03931 [hep-ph]].
%105 citations counted in INSPIRE as of 01 Oct 2020

%\cite{Jaiswal:2016hex}
\bibitem{Jaiswal:2016hex}
A.~Jaiswal and V.~Roy,
%``Relativistic hydrodynamics in heavy-ion collisions: general aspects and recent developments,''
Adv. High Energy Phys. \textbf{2016}, 9623034 (2016)
doi:10.1155/2016/9623034
[arXiv:1605.08694 [nucl-th]].
%57 citations counted in INSPIRE as of 01 Oct 2020

%Hydro examples
%\cite{Romatschke:2007mq}
\bibitem{Romatschke:2007mq}
P.~Romatschke and U.~Romatschke,
%``Viscosity Information from Relativistic Nuclear Collisions: How Perfect is the Fluid Observed at RHIC?,''
Phys. Rev. Lett. \textbf{99}, 172301 (2007)
doi:10.1103/PhysRevLett.99.172301
[arXiv:0706.1522 [nucl-th]].
%935 citations counted in INSPIRE as of 01 Oct 2020

%\cite{Heinz:2009cv}
\bibitem{Heinz:2009cv}
U.~W.~Heinz, J.~S.~Moreland and H.~Song,
%``Viscosity from elliptic flow: The Path to precision,''
Phys. Rev. C \textbf{80}, 061901 (2009)
doi:10.1103/PhysRevC.80.061901
[arXiv:0908.2617 [nucl-th]].
%30 citations counted in INSPIRE as of 01 Oct 2020

%\cite{Schenke:2010nt}
\bibitem{Schenke:2010nt} 
  B.~Schenke, S.~Jeon and C.~Gale,
  %``(3+1)D hydrodynamic simulation of relativistic heavy-ion collisions,''
  Phys.\ Rev.\ C {\bf 82}, 014903 (2010)
  doi:10.1103/PhysRevC.82.014903
  [arXiv:1004.1408 [hep-ph]].
  %%CITATION = doi:10.1103/PhysRevC.82.014903;%%
  %160 citations counted in INSPIRE as of 15 Aug 2017

%% BDMPS
%%\cite{Baier:1996vi}
%\bibitem{Baier:1996vi}
%R.~Baier, Y.~L.~Dokshitzer, A.~H.~Mueller, S.~Peigne and D.~Schiff,
%%``The Landau-Pomeranchuk-Migdal effect in QED,''
%Nucl. Phys. B \textbf{478}, 577-597 (1996)
%doi:10.1016/0550-3213(96)00426-9
%[arXiv:hep-ph/9604327 [hep-ph]].
%%127 citations counted in INSPIRE as of 02 Oct 2020
%
%%\cite{Baier:1996kr}
%\bibitem{Baier:1996kr}
%R.~Baier, Y.~L.~Dokshitzer, A.~H.~Mueller, S.~Peigne and D.~Schiff,
%%``Radiative energy loss of high-energy quarks and gluons in a finite volume quark - gluon plasma,''
%Nucl. Phys. B \textbf{483}, 291-320 (1997)
%doi:10.1016/S0550-3213(96)00553-6
%[arXiv:hep-ph/9607355 [hep-ph]].
%%948 citations counted in INSPIRE as of 02 Oct 2020
%
%%\cite{Baier:1996sk}
%\bibitem{Baier:1996sk}
%R.~Baier, Y.~L.~Dokshitzer, A.~H.~Mueller, S.~Peigne and D.~Schiff,
%%``Radiative energy loss and p(T) broadening of high-energy partons in nuclei,''
%Nucl. Phys. B \textbf{484}, 265-282 (1997)
%doi:10.1016/S0550-3213(96)00581-0
%[arXiv:hep-ph/9608322 [hep-ph]].
%%1147 citations counted in INSPIRE as of 02 Oct 2020
%
%%\cite{Zakharov:1997uu}
%\bibitem{Zakharov:1997uu}
%B.~G.~Zakharov,
%%``Radiative energy loss of high-energy quarks in finite size nuclear matter and quark - gluon plasma,''
%JETP Lett. \textbf{65}, 615-620 (1997)
%doi:10.1134/1.567389
%[arXiv:hep-ph/9704255 [hep-ph]].
%%585 citations counted in INSPIRE as of 02 Oct 2020
%
%%\cite{Baier:1998kq}
%\bibitem{Baier:1998kq}
%R.~Baier, Y.~L.~Dokshitzer, A.~H.~Mueller and D.~Schiff,
%%``Medium induced radiative energy loss: Equivalence between the BDMPS and Zakharov formalisms,''
%Nucl. Phys. B \textbf{531}, 403-425 (1998)
%doi:10.1016/S0550-3213(98)00546-X
%[arXiv:hep-ph/9804212 [hep-ph]].
%%285 citations counted in INSPIRE as of 02 Oct 2020

%qhat definition
%\cite{Wiedemann:2000ez}
\bibitem{Wiedemann:2000ez}
U.~A.~Wiedemann,
%``Transverse dynamics of hard partons in nuclear media and the QCD dipole,''
Nucl. Phys. B \textbf{582}, 409-450 (2000)
doi:10.1016/S0550-3213(00)00286-8
[arXiv:hep-ph/0003021 [hep-ph]].
%88 citations counted in INSPIRE as of 02 Oct 2020

%%\cite{Kovner:2001vi}
%\bibitem{Kovner:2001vi}
%A.~Kovner and U.~A.~Wiedemann,
%%``Eikonal evolution and gluon radiation,''
%Phys. Rev. D \textbf{64}, 114002 (2001)
%doi:10.1103/PhysRevD.64.114002
%[arXiv:hep-ph/0106240 [hep-ph]].
%%139 citations counted in INSPIRE as of 02 Oct 2020
%
%%\cite{Arnold:2002ja}
%\bibitem{Arnold:2002ja}
%P.~B.~Arnold, G.~D.~Moore and L.~G.~Yaffe,
%%``Photon and gluon emission in relativistic plasmas,''
%JHEP \textbf{06}, 030 (2002)
%doi:10.1088/1126-6708/2002/06/030
%[arXiv:hep-ph/0204343 [hep-ph]].
%%437 citations counted in INSPIRE as of 02 Oct 2020

%%\cite{Aurenche:2002pd}
%\bibitem{Aurenche:2002pd}
%P.~Aurenche, F.~Gelis and H.~Zaraket,
%%``A Simple sum rule for the thermal gluon spectral function and applications,''
%JHEP \textbf{05}, 043 (2002)
%doi:10.1088/1126-6708/2002/05/043
%[arXiv:hep-ph/0204146 [hep-ph]].
%%116 citations counted in INSPIRE as of 02 Oct 2020
%
%%\cite{Idilbi:2008vm}
%\bibitem{Idilbi:2008vm}
%A.~Idilbi and A.~Majumder,
%%``Extending Soft-Collinear-Effective-Theory to describe hard jets in dense QCD media,''
%Phys. Rev. D \textbf{80}, 054022 (2009)
%doi:10.1103/PhysRevD.80.054022
%[arXiv:0808.1087 [hep-ph]].
%%104 citations counted in INSPIRE as of 02 Oct 2020

%\cite{CaronHuot:2008ni}
\bibitem{CaronHuot:2008ni}
S.~Caron-Huot,
%``O(g) plasma effects in jet quenching,''
Phys. Rev. D \textbf{79}, 065039 (2009)
doi:10.1103/PhysRevD.79.065039
[arXiv:0811.1603 [hep-ph]].
%114 citations counted in INSPIRE as of 02 Oct 2020

%\cite{Majumder:2012sh}
\bibitem{Majumder:2012sh}
A.~Majumder,
%``Calculating the jet quenching parameter q̂ in lattice gauge theory,''
Phys. Rev. C \textbf{87}, 034905 (2013)
doi:10.1103/PhysRevC.87.034905
[arXiv:1202.5295 [nucl-th]].
%54 citations counted in INSPIRE as of 02 Oct 2020

%NRQCD, potentials
%\cite{Eichten:2007qx}
\bibitem{Eichten:2007qx}
  E.~Eichten, S.~Godfrey, H.~Mahlke and J.~L.~Rosner,
  %``Quarkonia and their transitions,''
  Rev.\ Mod.\ Phys.\  {\bf 80} (2008) 1161
  doi:10.1103/RevModPhys.80.1161
  [hep-ph/0701208].
  %%CITATION = doi:10.1103/RevModPhys.80.1161;%%
  %280 citations counted in INSPIRE as of 02 Feb 2020

%\cite{Bodwin:1994jh}
\bibitem{Bodwin:1994jh}
  G.~T.~Bodwin, E.~Braaten and G.~P.~Lepage,
  %``Rigorous QCD analysis of inclusive annihilation and production of heavy
  %quarkonium,''
  Phys.\ Rev.\  D {\bf 51}, 1125 (1995)
  [Erratum-ibid.\  D {\bf 55}, 5853 (1997)].
  %%CITATION = PHRVA,D51,1125;%%

%\cite{Quigg:1979vr}
\bibitem{Quigg:1979vr}
  C.~Quigg and J.~L.~Rosner,
  %``Quantum Mechanics with Applications to Quarkonium,''
  Phys.\ Rept.\  {\bf 56} (1979) 167.
  doi:10.1016/0370-1573(79)90095-4
  %%CITATION = doi:10.1016/0370-1573(79)90095-4;%%
  %687 citations counted in INSPIRE as of 02 Feb 2020

%\cite{Eichten:1979ms}
\bibitem{Eichten:1979ms}
E.~Eichten, K.~Gottfried, T.~Kinoshita, K.~D.~Lane and T.~M.~Yan,
%``Charmonium: Comparison with Experiment,''
Phys. Rev. D \textbf{21}, 203 (1980)
doi:10.1103/PhysRevD.21.203
%1776 citations counted in INSPIRE as of 04 Oct 2020

%%\cite{Lucha:1991}
%\bibitem{Lucha:1991}
%Lucha, Wolfgang, Franz F. Schöberl, and Dieter Gromes. 
%%"Bound states of quarks." 
%Physics reports 200.4 (1991): 127-240.

%\cite{Otto:1984qr}
\bibitem{Otto:1984qr}
  S.~W.~Otto and J.~D.~Stack,
  %``The SU(3) Heavy Quark Potential with High Statistics,''
  Phys.\ Rev.\ Lett.\  {\bf 52} (1984) 2328.
  doi:10.1103/PhysRevLett.52.2328
  %%CITATION = doi:10.1103/PhysRevLett.52.2328;%%
  %171 citations counted in INSPIRE as of 02 Feb 2020

% Additional reviws/papers on quarkonium spectro..
%%\cite{Eichten:1974af}
%\bibitem{Eichten:1974af}
%E.~Eichten, K.~Gottfried, T.~Kinoshita, J.~B.~Kogut, K.~D.~Lane and T.~M.~Yan,
%%``The Spectrum of Charmonium,''
%Phys. Rev. Lett. \textbf{34}, 369-372 (1975)
%[erratum: Phys. Rev. Lett. \textbf{36}, 1276 (1976)]
%doi:10.1103/PhysRevLett.34.369
%%1151 citations counted in INSPIRE as of 04 Oct 2020
%
%%\cite{Eichten:1978tg}
%\bibitem{Eichten:1978tg}
%E.~Eichten, K.~Gottfried, T.~Kinoshita, K.~D.~Lane and T.~M.~Yan,
%%``Charmonium: The Model,''
%Phys. Rev. D \textbf{17}, 3090 (1978)
%[erratum: Phys. Rev. D \textbf{21}, 313 (1980)]
%doi:10.1103/PhysRevD.17.3090
%%1512 citations counted in INSPIRE as of 04 Oct 2020

%\cite{Bodwin:1994jh}

%-- Review of T=0 and vacuum quarkonia
%\cite{Voloshin:2007dx}
\bibitem{Voloshin:2007dx}
M.~B.~Voloshin,
%``Charmonium,''
Prog. Part. Nucl. Phys. \textbf{61} (2008), 455-511
doi:10.1016/j.ppnp.2008.02.001
[arXiv:0711.4556 [hep-ph]].
%438 citations counted in INSPIRE as of 01 Jan 2021

%\cite{Brambilla:2010cs}
\bibitem{Brambilla:2010cs}
  N.~Brambilla {\it et al.},
  %``Heavy Quarkonium: Progress, Puzzles, and Opportunities,''
  Eur.\ Phys.\ J.\ C {\bf 71} (2011) 1534
  doi:10.1140/epjc/s10052-010-1534-9
  [arXiv:1010.5827 [hep-ph]].
  %%CITATION = doi:10.1140/epjc/s10052-010-1534-9;%%
  %1433 citations counted in INSPIRE as of 28 Dec 2019

%-- pNRQCD vacuum first and review
%\cite{Brambilla:1999xf}
\bibitem{Brambilla:1999xf}
  N.~Brambilla, A.~Pineda, J.~Soto and A.~Vairo,
  %``Potential NRQCD: An Effective theory for heavy quarkonium,''
  Nucl.\ Phys.\ B {\bf 566} (2000) 275
  doi:10.1016/S0550-3213(99)00693-8
  [hep-ph/9907240].
  %%CITATION = doi:10.1016/S0550-3213(99)00693-8;%%
  %606 citations counted in INSPIRE as of 07 Feb 2020

%\cite{Brambilla:2004jw}
\bibitem{Brambilla:2004jw}
  N.~Brambilla, A.~Pineda, J.~Soto and A.~Vairo,
  %``Effective Field Theories for Heavy Quarkonium,''
  Rev.\ Mod.\ Phys.\  {\bf 77} (2005) 1423
  doi:10.1103/RevModPhys.77.1423
  [hep-ph/0410047].
  %%CITATION = doi:10.1103/RevModPhys.77.1423;%%
  %518 citations counted in INSPIRE as of 28 Dec 2019

% Potentials
%\cite{Brambilla:2009bi}
\bibitem{Brambilla:2009bi}
N.~Brambilla, A.~Vairo, X.~Garcia Tormo, i and J.~Soto,
%``The QCD static energy at NNNLL,''
Phys. Rev. D \textbf{80} (2009), 034016
doi:10.1103/PhysRevD.80.034016
[arXiv:0906.1390 [hep-ph]].
%70 citations counted in INSPIRE as of 01 Jan 2021

%\cite{Brambilla:1999xj}
\bibitem{Brambilla:1999xj}
N.~Brambilla, A.~Pineda, J.~Soto and A.~Vairo,
%``The Heavy quarkonium spectrum at order m alpha**5(s) l n alpha(s),''
Phys. Lett. B \textbf{470} (1999), 215
doi:10.1016/S0370-2693(99)01301-5
[arXiv:hep-ph/9910238 [hep-ph]].
%141 citations counted in INSPIRE as of 31 Dec 2020

%\cite{Kniehl:2002br}
\bibitem{Kniehl:2002br}
B.~A.~Kniehl, A.~A.~Penin, V.~A.~Smirnov and M.~Steinhauser,
%``Potential NRQCD and heavy quarkonium spectrum at next-to-next-to-next-to-leading order,''
Nucl. Phys. B \textbf{635} (2002), 357-383
doi:10.1016/S0550-3213(02)00403-0
[arXiv:hep-ph/0203166 [hep-ph]].
%159 citations counted in INSPIRE as of 31 Dec 2020

%\cite{Radford:2007vd}
\bibitem{Radford:2007vd}
S.~F.~Radford and W.~W.~Repko,
%``Potential model calculations and predictions for heavy quarkonium,''
Phys. Rev. D \textbf{75}, 074031 (2007)
doi:10.1103/PhysRevD.75.074031
[arXiv:hep-ph/0701117 [hep-ph]].
%103 citations counted in INSPIRE as of 04 Oct 2020
%
%%\cite{Radford:2009qi}
%\bibitem{Radford:2009qi}
%S.~F.~Radford and W.~W.~Repko,
%%``Hyperfine splittings in the b anti-b system,''
%Nucl. Phys. A \textbf{865} (2011), 69-75
%doi:10.1016/j.nuclphysa.2011.06.032
%[arXiv:0912.2259 [hep-ph]].
%%23 citations counted in INSPIRE as of 01 Jan 2021

%\cite{Repko:2012rk}
\bibitem{Repko:2012rk}
W.~W.~Repko, M.~D.~Santia and S.~F.~Radford,
%``Three-loop static QCD potential in heavy quarkonia,''
Nucl. Phys. A \textbf{924} (2014), 65-73
doi:10.1016/j.nuclphysa.2014.01.005
[arXiv:1211.6373 [hep-ph]].
%11 citations counted in INSPIRE as of 31 Dec 2020

%%\cite{Wei-Zhao:2013sta}
%\bibitem{Wei-Zhao:2013sta}
%T.~Wei-Zhao, C.~Lu, Y.~You-Chang and C.~Hong,
%%``Bottomonium states versus recent experimental observations in the QCD-inspired potential model,''
%Chin. Phys. C \textbf{37} (2013), 083101
%doi:10.1088/1674-1137/37/8/083101
%[arXiv:1308.0960 [hep-ph]].
%%13 citations counted in INSPIRE as of 01 Jan 2021
%
%\cite{Brambilla:2020xod}
\bibitem{Brambilla:2020xod}
N.~Brambilla, H.~S.~Chung, D.~M\"uller and A.~Vairo,
%``Decay and electromagnetic production of strongly coupled quarkonia in pNRQCD,''
JHEP \textbf{04} (2020), 095
doi:10.1007/JHEP04(2020)095
[arXiv:2002.07462 [hep-ph]].
%4 citations counted in INSPIRE as of 01 Jan 2021

%\cite{Patrignani:2012an}
\bibitem{Patrignani:2012an}
C.~Patrignani, T.~K.~Pedlar and J.~L.~Rosner,
%``Recent Results in Bottomonium,''
Ann. Rev. Nucl. Part. Sci. \textbf{63}, 21-44 (2013)
doi:10.1146/annurev-nucl-102212-170609
[arXiv:1212.6552 [hep-ex]].
%30 citations counted in INSPIRE as of 04 Oct 2020
%
%%\cite{Eiglsperger:2007ay}
%\bibitem{Eiglsperger:2007ay}
%J.~Eiglsperger,
%%``Quarkonium Spectroscopy: Beyond One-Gluon Exchange,''
%[arXiv:0707.1269 [hep-ph]].
%%12 citations counted in INSPIRE as of 01 Jan 2021

%\cite{Burnier:2015tda}
\bibitem{Burnier:2015tda}
Y.~Burnier, O.~Kaczmarek and A.~Rothkopf,
%``Quarkonium at finite temperature: Towards realistic phenomenology from first principles,''
JHEP \textbf{12} (2015), 101
doi:10.1007/JHEP12(2015)101
[arXiv:1509.07366 [hep-ph]].
%63 citations counted in INSPIRE as of 01 Jan 2021

%-- Lattice Kaczmarek
%\cite{Kaczmarek:2002mc}
\bibitem{Kaczmarek:2002mc}
  O.~Kaczmarek, F.~Karsch, P.~Petreczky and F.~Zantow,
  %``Heavy quark anti-quark free energy and the renormalized Polyakov loop,''
  Phys.\ Lett.\ B {\bf 543} (2002) 41
  doi:10.1016/S0370-2693(02)02415-2
  [hep-lat/0207002].
  %%CITATION = doi:10.1016/S0370-2693(02)02415-2;%%
  %400 citations counted in INSPIRE as of 26 Jan 2020

%\cite{Digal:2003jc}
\bibitem{Digal:2003jc}
  S.~Digal, S.~Fortunato and P.~Petreczky,
  %``Heavy quark free energies and screening in SU(2) gauge theory,''
  Phys.\ Rev.\ D {\bf 68} (2003) 034008
  doi:10.1103/PhysRevD.68.034008
  [hep-lat/0304017].
  %%CITATION = doi:10.1103/PhysRevD.68.034008;%%
  %70 citations counted in INSPIRE as of 08 Feb 2020

%\cite{Kaczmarek:2003ph}
\bibitem{Kaczmarek:2003ph}
  O.~Kaczmarek, S.~Ejiri, F.~Karsch, E.~Laermann and F.~Zantow,
  %``Heavy quark free energies and the renormalized Polyakov loop in full QCD,''
  Prog.\ Theor.\ Phys.\ Suppl.\  {\bf 153} (2004) 287
  doi:10.1143/PTPS.153.287
  [hep-lat/0312015].
  %%CITATION = doi:10.1143/PTPS.153.287;%%
  %92 citations counted in INSPIRE as of 26 Jan 2020

%\cite{Kaczmarek:2005ui}
\bibitem{Kaczmarek:2005ui}
  O.~Kaczmarek and F.~Zantow,
  %``Static quark anti-quark interactions in zero and finite temperature QCD. I. Heavy quark free energies, running coupling and quarkonium binding,''
  Phys.\ Rev.\ D {\bf 71} (2005) 114510
  doi:10.1103/PhysRevD.71.114510
  [hep-lat/0503017].
  %%CITATION = doi:10.1103/PhysRevD.71.114510;%%
  %428 citations counted in INSPIRE as of 02 Feb 2020

%\cite{Doring:2007uh}
\bibitem{Doring:2007uh}
  M.~Doring, K.~Huebner, O.~Kaczmarek and F.~Karsch,
  %``Color Screening and Quark-Quark Interactions in Finite Temperature QCD,''
  Phys.\ Rev.\ D {\bf 75} (2007) 054504
  doi:10.1103/PhysRevD.75.054504
  [hep-lat/0702009].
  %%CITATION = doi:10.1103/PhysRevD.75.054504;%%
  %60 citations counted in INSPIRE as of 08 Feb 2020

%\cite{Gupta:2007ax}
\bibitem{Gupta:2007ax}
  S.~Gupta, K.~Huebner and O.~Kaczmarek,
  %``Renormalized Polyakov loops in many representations,''
  Phys.\ Rev.\ D {\bf 77} (2008) 034503
  doi:10.1103/PhysRevD.77.034503
  [arXiv:0711.2251 [hep-lat]].
  %%CITATION = doi:10.1103/PhysRevD.77.034503;%%
  %120 citations counted in INSPIRE as of 08 Feb 2020

%-- Bhanot and Peskin
%\cite{Peskin:1979}
\bibitem{Peskin:1979}
  M. E. Peskin,
  %Short Distance Analysis for Heavy Quark Systems. 1. Diagrammatics.
  Nucl.\ Phys.\ B {\bf{156}}, 365 (1979). 
  %Mar 1979
  %HUTP-79/A008 

%\cite{Bhanot:1979vb}
\bibitem{Bhanot:1979vb}
  G.~Bhanot and M.~E.~Peskin,
  %``Short Distance Analysis for Heavy Quark Systems. 2. Applications,''
  Nucl.\ Phys.\ B {\bf 156} (1979) 391.
  doi:10.1016/0550-3213(79)90200-1
  %%CITATION = doi:10.1016/0550-3213(79)90200-1;%%
  %371 citations counted in INSPIRE as of 28 Dec 2019

%-- classic quarkonia
\bibitem{Matsui:1986}{
    {T.~{Matsui} and H.~{Satz},},
    %"{J/{$\psi$} suppression by quark-gluon plasma formation}",
    {Physics Letters B},
    {\bf{178}}
    (1986),
    {416-422}.
}

%%--Sequential melting
%\cite{Karsch:1987pv}
\bibitem{Karsch:1987pv}
F.~Karsch, M.~T.~Mehr and H.~Satz,
%``Color Screening and Deconfinement for Bound States of Heavy Quarks,''
Z. Phys. C \textbf{37}, 617 (1988)
doi:10.1007/BF01549722
%401 citations counted in INSPIRE as of 09 Oct 2020

%\cite{Karsch:1991}
\bibitem{Karsch:1991}
F.~Karsch and H.~Satz,
%``The spectral analysis of strongly interacting matter'',
Zeitschrift für Physik C Particles and Fields,
{\bf{51}} 2 (1991), 209.

%%\cite{Digal:2001iu}
%\bibitem{Digal:2001iu}
%S.~Digal, P.~Petreczky and H.~Satz,
%%``String breaking and quarkonium dissociation at finite temperatures,''
%Phys. Lett. B \textbf{514}, 57-62 (2001)
%doi:10.1016/S0370-2693(01)00803-6
%[arXiv:hep-ph/0105234 [hep-ph]].
%%168 citations counted in INSPIRE as of 09 Oct 2020

%\cite{Digal:2001ue}
\bibitem{Digal:2001ue}
S.~Digal, P.~Petreczky and H.~Satz,
%``Quarkonium feed down and sequential suppression,''
Phys. Rev. D \textbf{64}, 094015 (2001)
doi:10.1103/PhysRevD.64.094015
[arXiv:hep-ph/0106017 [hep-ph]].
%291 citations counted in INSPIRE as of 09 Oct 2020

%\bibitem{Karsch:2006}
%{
%F. Karsch, D. Kharzeev, H. Satz, 
%%Sequential charmonium dissociation, 
%Physics Letters B, 
%{\bf{637}} (2006), 75-80. 
%%ISSN 0370-2693, 10.1016/j.physletb.2006.03.078.
%%(http://www.sciencedirect.com/science/article/pii/S037026930600445X)
%}

%-- Regeneration reviews
%\cite{Rapp:2008tf}
\bibitem{Rapp:2008tf}
  R.~Rapp, D.~Blaschke and P.~Crochet,
  %``Charmonium and bottomonium production in heavy-ion collisions,''
  Prog.\ Part.\ Nucl.\ Phys.\  {\bf 65} (2010) 209
  doi:10.1016/j.ppnp.2010.07.002
  [arXiv:0807.2470 [hep-ph]].
  %%CITATION = doi:10.1016/j.ppnp.2010.07.002;%%
  %181 citations counted in INSPIRE as of 28 Jan 2020

%\cite{Rapp:2009my}
\bibitem{Rapp:2009my}
  R.~Rapp and H.~van Hees,
  %``Heavy Quarks in the Quark-Gluon Plasma,''
  [arXiv:0903.1096 [hep-ph]].
  %%CITATION = ARXIV:0903.1096;%%

%-- p+A. CNM. Cold nuclear Matter 
%\bibitem{Lourenco:2008sk}
%  C.~Lourenco, R.~Vogt and H.~K.~Woehri,
%  %``Energy dependence of J/psi absorption in proton-nucleus collisions,''
%  JHEP {\bf 0902} (2009) 014
%  doi:10.1088/1126-6708/2009/02/014
%  [arXiv:0901.3054 [hep-ph]].
%  %%CITATION = doi:10.1088/1126-6708/2009/02/014;%%
%  %105 citations counted in INSPIRE as of 05 Feb 2020
%
%\cite{ConesadelValle:2011fw}
\bibitem{ConesadelValle:2011fw}
  Z.~Conesa del Valle {\it et al.},
  %``Quarkonium production in high energy proton-proton and proton-nucleus
  %collisions,''
  Nucl.\ Phys.\ Proc.\ Suppl.\  {\bf 214}, 3 (2011).
  %%CITATION = NUPHZ,214,3;%% 

%-- Laine
%\cite{Laine:2006ns}
\bibitem{Laine:2006ns}
  M.~Laine, O.~Philipsen, P.~Romatschke and M.~Tassler,
  %``Real-time static potential in hot QCD,''
  JHEP {\bf 0703} (2007) 054.
  %%CITATION = JHEPA,0703,054;%%

%%\cite{Xu:1995eb}
%\bibitem{Xu:1995eb} 
%  X.~M.~Xu, D.~Kharzeev, H.~Satz and X.~N.~Wang,
%  %``J / psi suppression in an equilibrating parton plasma,''
%  Phys.\ Rev.\ C {\bf 53}, 3051 (1996)
%  doi:10.1103/PhysRevC.53.3051
%  [hep-ph/9511331].
%  %%CITATION = doi:10.1103/PhysRevC.53.3051;%%
%  %105 citations counted in INSPIRE as of 25 Aug 2017
%
%%\cite{Digal:2001iu}
%\bibitem{Digal:2001iu}
%  S.~Digal, P.~Petreczky and H.~Satz,
%  %``String breaking and quarkonium dissociation at finite temperatures,''
%  Phys.\ Lett.\ B {\bf 514} (2001) 57
%  doi:10.1016/S0370-2693(01)00803-6
%  [hep-ph/0105234].
%  %%CITATION = doi:10.1016/S0370-2693(01)00803-6;%%
%  %164 citations counted in INSPIRE as of 02 Feb 2020

%%\cite{Digal:2001ue}
%\bibitem{Digal:2001ue}
%  S.~Digal, P.~Petreczky and H.~Satz,
%  %``Quarkonium feed down and sequential suppression,''
%  Phys.\ Rev.\ D {\bf 64} (2001) 094015
%  doi:10.1103/PhysRevD.64.094015
%  [hep-ph/0106017].
%  %%CITATION = doi:10.1103/PhysRevD.64.094015;%%
%  %278 citations counted in INSPIRE as of 02 Feb 2020

%-- Lattice light current correlations and di-lepton rates
%%\cite{Nakahara:1999vy}
%\bibitem{Nakahara:1999vy}
%Y.~Nakahara, M.~Asakawa and T.~Hatsuda,
%%``Hadronic spectral functions in lattice QCD,''
%Phys. Rev. D \textbf{60} (1999), 091503
%doi:10.1103/PhysRevD.60.091503
%[arXiv:hep-lat/9905034 [hep-lat]].
%%166 citations counted in INSPIRE as of 31 Oct 2020
%
%%\cite{Asakawa:2000tr}
%\bibitem{Asakawa:2000tr}
%M.~Asakawa, T.~Hatsuda and Y.~Nakahara,
%%``Maximum entropy analysis of the spectral functions in lattice QCD,''
%Prog. Part. Nucl. Phys. \textbf{46} (2001), 459-508
%doi:10.1016/S0146-6410(01)00150-8
%[arXiv:hep-lat/0011040 [hep-lat]].
%%495 citations counted in INSPIRE as of 31 Oct 2020
%
%%\cite{deForcrand:2000akx}
%\bibitem{deForcrand:2000akx}
%P.~de Forcrand \textit{et al.} [QCD-TARO],
%%``Meson correlators in finite temperature lattice QCD,''
%Phys. Rev. D \textbf{63} (2001), 054501
%doi:10.1103/PhysRevD.63.054501
%[arXiv:hep-lat/0008005 [hep-lat]].
%%92 citations counted in INSPIRE as of 31 Oct 2020
%
%%\cite{Hashimoto:1993np}
%\bibitem{Hashimoto:1993np}
%T.~Hashimoto, A.~Nakamura and I.~O.~Stamatescu,
%%``QCD with dynamical quarks at finite temperature: Spectral structure in the mesonic channels,''
%Nucl. Phys. B \textbf{406} (1993), 325-339
%doi:10.1016/0550-3213(93)90170-T
%%23 citations counted in INSPIRE as of 31 Oct 2020
%
%%\cite{Karsch:2001uw}
%\bibitem{Karsch:2001uw}
%F.~Karsch, E.~Laermann, P.~Petreczky, S.~Stickan and I.~Wetzorke,
%%``A Lattice calculation of thermal dilepton rates,''
%Phys. Lett. B \textbf{530} (2002), 147-152
%doi:10.1016/S0370-2693(02)01326-6
%[arXiv:hep-lat/0110208 [hep-lat]].
%%158 citations counted in INSPIRE as of 31 Oct 2020
%
%%\cite{Karsch:2001uw}
%\bibitem{Karsch:2001uw}
%F.~Karsch, E.~Laermann, P.~Petreczky, S.~Stickan and I.~Wetzorke,
%%``A Lattice calculation of thermal dilepton rates,''
%Phys. Lett. B \textbf{530} (2002), 147-152
%doi:10.1016/S0370-2693(02)01326-6
%[arXiv:hep-lat/0110208 [hep-lat]].
%%158 citations counted in INSPIRE as of 31 Oct 2020

%-- Lattice current correlations early
%\cite{Datta:2003ww}
\bibitem{Datta:2003ww}
  S.~Datta, F.~Karsch, P.~Petreczky and I.~Wetzorke,
  %``Behavior of charmonium systems after deconfinement,''
  Phys.\ Rev.\ D {\bf 69} (2004) 094507
  doi:10.1103/PhysRevD.69.094507
  [hep-lat/0312037].
  %%CITATION = doi:10.1103/PhysRevD.69.094507;%%
  %439 citations counted in INSPIRE as of 02 Feb 2020

%\cite{Umeda:2002vr}
\bibitem{Umeda:2002vr}
T.~Umeda, K.~Nomura and H.~Matsufuru,
%``Charmonium at finite temperature in quenched lattice QCD,''
Eur. Phys. J. C \textbf{39S1} (2005), 9-26
doi:10.1140/epjcd/s2004-01-002-1
[arXiv:hep-lat/0211003 [hep-lat]].
%216 citations counted in INSPIRE as of 31 Oct 202

%A few early proceedings
%
%%\cite{Asakawa:2002xj}
%\bibitem{Asakawa:2002xj}
%M.~Asakawa, T.~Hatsuda and Y.~Nakahara,
%%``Hadronic spectral functions above the QCD phase transition,''
%Nucl. Phys. B Proc. Suppl. \textbf{119} (2003), 481-483
%doi:10.1016/S0375-9474(02)01526-9
%[arXiv:hep-lat/0208059 [hep-lat]].
%%152 citations counted in INSPIRE as of 31 Oct 20200
%
%%\cite{Petreczky:2003js}
%\bibitem{Petreczky:2003js}
%P.~Petreczky, S.~Datta, F.~Karsch and I.~Wetzorke,
%%``Charmonium at finite temperature,''
%Nucl. Phys. B Proc. Suppl. \textbf{129} (2004), 596-598
%doi:10.1016/S0920-5632(03)02653-7
%[arXiv:hep-lat/0309012 [hep-lat]].
%%25 citations counted in INSPIRE as of 31 Oct 2020
%
%%\cite{Petreczky:2003rq}
%\bibitem{Petreczky:2003rq}
%P.~Petreczky,
%%``Quarkonium at finite temperature,''
%[arXiv:hep-lat/0310059 [hep-lat]].
%%1 citations counted in INSPIRE as of 31 Oct 2020

%\cite{Asakawa:2003re}
\bibitem{Asakawa:2003re}
M.~Asakawa and T.~Hatsuda,
%``J / psi and eta(c) in the deconfined plasma from lattice QCD,''
Phys. Rev. Lett. \textbf{92} (2004), 012001
doi:10.1103/PhysRevLett.92.012001
[arXiv:hep-lat/0308034 [hep-lat]].
%512 citations counted in INSPIRE as of 31 Oct 2020

%\cite{Jakovac:2006sf}
\bibitem{Jakovac:2006sf}
  A.~Jakovac, P.~Petreczky, K.~Petrov and A.~Velytsky,
  %``Quarkonium correlators and spectral functions at zero and finite temperature,''
  Phys.\ Rev.\ D {\bf 75} (2007) 014506
  doi:10.1103/PhysRevD.75.014506
  [hep-lat/0611017].
  %%CITATION = doi:10.1103/PhysRevD.75.014506;%%
  %202 citations counted in INSPIRE as of 08 Feb 2020

%\cite{Mocsy:2007yj}
\bibitem{Mocsy:2007yj}
  A.~Mocsy and P.~Petreczky,
  %``Can quarkonia survive deconfinement ?,''
  Phys.\ Rev.\  D {\bf 77}, 014501 (2008).
  %%CITATION = PHRVA,D77,014501;%%

%\cite{Mocsy:2007jz}
\bibitem{Mocsy:2007jz}
A.~Mocsy and P.~Petreczky,
%``Color Screening Melts Quarkonium,''
Phys.\ Rev.\ Lett.\  {\bf 99}, 211602 (2007).
%%CITATION = PRLTA,99,211602;%%

%\cite{Petreczky:2010tk}
\bibitem{Petreczky:2010tk}
P.~Petreczky, C.~Miao and A.~Mocsy,
%``Quarkonium spectral functions with complex potential,''
Nucl. Phys. A \textbf{855} (2011), 125-132
doi:10.1016/j.nuclphysa.2011.02.028
[arXiv:1012.4433 [hep-ph]].
%85 citations counted in INSPIRE as of 01 Jan 2021

%\cite{Bazavov:2009us}
\bibitem{Bazavov:2009us}
  A.~Bazavov, P.~Petreczky and A.~Velytsky,
  %``Quarkonium at Finite Temperature,''
  doi:10.1142/9789814293297\_0002
  arXiv:0904.1748 [hep-ph].
  %%CITATION = doi:10.1142/9789814293297_0002;%%
  %41 citations counted in INSPIRE as of 08 Feb 2020

%\cite{Aarts:2011sm}
\bibitem{Aarts:2011sm}
  G.~Aarts, C.~Allton, S.~Kim, M.~P.~Lombardo, M.~B.~Oktay, S.~M.~Ryan, D.~K.~Sinclair and J.~I.~Skullerud,
  %``What happens to the Upsilon and eta_b in the quark-gluon plasma? Bottomonium spectral functions from lattice QCD,''
  JHEP {\bf 1111} (2011) 103
  doi:10.1007/JHEP11(2011)103
  [arXiv:1109.4496 [hep-lat]].
  %%CITATION = doi:10.1007/JHEP11(2011)103;%%
  %111 citations counted in INSPIRE as of 02 Feb 2020

%\cite{Karsch:2012na}
\bibitem{Karsch:2012na}
  F.~Karsch, E.~Laermann, S.~Mukherjee and P.~Petreczky,
  %``Signatures of charmonium modification in spatial correlation functions,''
  Phys.\ Rev.\ D {\bf 85} (2012) 114501
  doi:10.1103/PhysRevD.85.114501
  [arXiv:1203.3770 [hep-lat]].
  %%CITATION = doi:10.1103/PhysRevD.85.114501;%%
  %30 citations counted in INSPIRE as of 08 Feb 2020

%-- Lattice reviews 
%\cite{Petreczky:2012rq}
\bibitem{Petreczky:2012rq}
  P.~Petreczky,
  %``Lattice QCD at non-zero temperature,''
  J.\ Phys.\ G {\bf 39} (2012) 093002
  doi:10.1088/0954-3899/39/9/093002
  [arXiv:1203.5320 [hep-lat]].
  %%CITATION = doi:10.1088/0954-3899/39/9/093002;%%
  %153 citations counted in INSPIRE as of 02 Feb 2020

%\cite{Datta:2014wga}
\bibitem{Datta:2014wga} 
  S.~Datta,
  %``Quarkonia at finite temperature in relativistic heavy ion collisions,''
  Pramana {\bf 84}, no. 5, 881 (2015)
  doi:10.1007/s12043-015-0975-y
  [arXiv:1403.8151 [nucl-th]].
  %%CITATION = doi:10.1007/s12043-015-0975-y;%%

%\cite{Rothkopf:2019ipj}
\bibitem{Rothkopf:2019ipj}
  A.~Rothkopf,
  %``Heavy Quarkonium in Extreme Conditions,''
  arXiv:1912.02253 [hep-ph].
  %%CITATION = ARXIV:1912.02253;%%
  %4 citations counted in INSPIRE as of 02 Feb 2020

%-- Sharma reviewed
%\cite{Aronson:2017ymv}
\bibitem{Aronson:2017ymv}
  S.~Aronson, E.~Borras, B.~Odegard, R.~Sharma and I.~Vitev,
  %``Collisional and thermal dissociation of $J/\psi$ and $\Upsilon$ states at the LHC,''
  Phys.\ Lett.\ B {\bf 778} (2018) 384
  doi:10.1016/j.physletb.2018.01.038
  [arXiv:1709.02372 [hep-ph]].
  %%CITATION = doi:10.1016/j.physletb.2018.01.038;%%
  %20 citations counted in INSPIRE as of 20 Dec 2019

%--Quarkonia Open
%--Borghini
%\cite{Borghini:2011yq}
\bibitem{Borghini:2011yq} 
  N.~Borghini and C.~Gombeaud,
  %``Dynamical Evolution of Heavy Quarkonia in a Deconfined Medium,''
  arXiv:1103.2945 [hep-ph].
  %%CITATION = ARXIV:1103.2945;%%
  %10 citations counted in INSPIRE as of 29 Aug 2017

%\cite{Borghini:2011ms}
\bibitem{Borghini:2011ms}
  N.~Borghini and C.~Gombeaud,
  %``Heavy quarkonia in a medium as a quantum dissipative system: Master equation approach,''
  Eur.\ Phys.\ J.\ C {\bf 72} (2012) 2000
  doi:10.1140/epjc/s10052-012-2000-7
  [arXiv:1109.4271 [nucl-th]].
  %%CITATION = doi:10.1140/epjc/s10052-012-2000-7;%%
  %34 citations counted in INSPIRE as of 28 Dec 2019

%%\cite{Dutta:2012nw}
%\bibitem{Dutta:2012nw}
%  N.~Dutta and N.~Borghini,
%  %``Sequential suppression of quarkonia and high-energy nucleus–nucleus collisions,''
%  Mod.\ Phys.\ Lett.\ A {\bf 30} (2015) no.37,  1550205
%  doi:10.1142/S0217732315502053
%  [arXiv:1206.2149 [nucl-th]].
%  %%CITATION = doi:10.1142/S0217732315502053;%%
%  %15 citations counted in INSPIRE as of 28 Dec 2019

%-- Young Dusling
%\cite{Young:2010jq}
\bibitem{Young:2010jq}
  C.~Young and K.~Dusling,
  %``Quarkonium above deconfinement as an open quantum system,''
  Phys.\ Rev.\ C {\bf 87} (2013) no.6,  065206
  doi:10.1103/PhysRevC.87.065206
  [arXiv:1001.0935 [nucl-th]].
  %%CITATION = doi:10.1103/PhysRevC.87.065206;%%
  %31 citations counted in INSPIRE as of 28 Dec 2019

%--Akamatsu
%\cite{Akamatsu:2011se}
\bibitem{Akamatsu:2011se} 
  Y.~Akamatsu and A.~Rothkopf,
  %``Stochastic potential and quantum decoherence of heavy quarkonium in the quark-gluon plasma,''
  Phys.\ Rev.\ D {\bf 85}, 105011 (2012)
  doi:10.1103/PhysRevD.85.105011
  [arXiv:1110.1203 [hep-ph]].
  %%CITATION = doi:10.1103/PhysRevD.85.105011;%%
  %39 citations counted in INSPIRE as of 29 Aug 2017

%\cite{Akamatsu:2012vt}
\bibitem{Akamatsu:2012vt}
  Y.~Akamatsu,
  %``Real-time quantum dynamics of heavy quark systems at high temperature,''
  Phys.\ Rev.\ D {\bf 87} (2013) no.4,  045016
  doi:10.1103/PhysRevD.87.045016
  [arXiv:1209.5068 [hep-ph]].
  %%CITATION = doi:10.1103/PhysRevD.87.045016;%%
  %25 citations counted in INSPIRE as of 28 Dec 2019

%\cite{Akamatsu:2013}
\bibitem{Akamatsu:2013}
  Y.~Akamatsu,
  %``Heavy quark master equations in the Lindblad form at high temperatures,''
  Phys.\ Rev.\ D {\bf 91} (2015) no.5, 56002 
  doi:10.1103/PhysRevD.91.056002
  [arXiv: [hep-ph/1403.5783]].
  %%CITATION = doi:10.1103/PhysRevD.91.056002;%%

%\cite{Akamatsu:2015kaa}
\bibitem{Akamatsu:2015kaa}
  Y.~Akamatsu,
  %``Langevin dynamics and decoherence of heavy quarks at high temperatures,''
  Phys.\ Rev.\ C {\bf 92} (2015) no.4,  044911
  doi:10.1103/PhysRevC.92.044911
  [arXiv:1503.08110 [nucl-th]].
  %%CITATION = doi:10.1103/PhysRevC.92.044911;%%
  %18 citations counted in INSPIRE as of 28 Dec 2019


%\cite{Kajimoto:2017rel}
\bibitem{Kajimoto:2017rel} 
  S.~Kajimoto, Y.~Akamatsu, M.~Asakawa and A.~Rothkopf,
  %``Dynamical dissociation of quarkonia by wave function decoherence,''
  arXiv:1705.03365 [nucl-th].
  %%CITATION = ARXIV:1705.03365;%%

%\cite{Akamatsu:2018xim}
\bibitem{Akamatsu:2018xim}
  Y.~Akamatsu, M.~Asakawa, S.~Kajimoto and A.~Rothkopf,
  %``Quantum dissipation of a heavy quark from a nonlinear stochastic Schr\"odinger equation,''
  JHEP {\bf 1807} (2018) 029
  doi:10.1007/JHEP07(2018)029
  [arXiv:1805.00167 [nucl-th]].
  %%CITATION = doi:10.1007/JHEP07(2018)029;%%
  %9 citations counted in INSPIRE as of 28 Dec 2019

%\cite{Miura:2019ssi}
\bibitem{Miura:2019ssi}
  T.~Miura, Y.~Akamatsu, M.~Asakawa and A.~Rothkopf,
  %``Quantum Brownian motion of a heavy quark pair in the quark-gluon plasma,''
  arXiv:1908.06293 [nucl-th].
  %%CITATION = ARXIV:1908.06293;%%
  %2 citations counted in INSPIRE as of 28 Dec 2019

%\cite{Akamatsu:2020ypb}
\bibitem{Akamatsu:2020ypb}
Y.~Akamatsu,
%``Quarkonium in Quark-Gluon Plasma: Open Quantum System Approaches Re-examined,''
[arXiv:2009.10559 [nucl-th]].
%1 citations counted in INSPIRE as of 01 Nov 2020

%--Brambilla open
%\cite{Brambilla:2016wgg}
\bibitem{Brambilla:2016wgg} 
  N.~Brambilla, M.~A.~Escobedo, J.~Soto and A.~Vairo,
  %``Quarkonium suppression in heavy-ion collisions: an open quantum system approach,''
  Phys.\ Rev.\ D {\bf 96}, no. 3, 034021 (2017)
  doi:10.1103/PhysRevD.96.034021
  [arXiv:1612.07248 [hep-ph]].
  %%CITATION = doi:10.1103/PhysRevD.96.034021;%%
  %7 citations counted in INSPIRE as of 29 Aug 2017


%\cite{Brambilla:2017zei}
\bibitem{Brambilla:2017zei}
  N.~Brambilla, M.~A.~Escobedo, J.~Soto and A.~Vairo,
  %``Heavy quarkonium suppression in a fireball,''
  Phys.\ Rev.\ D {\bf 97} (2018) no.7,  074009
  doi:10.1103/PhysRevD.97.074009
  [arXiv:1711.04515 [hep-ph]].
  %%CITATION = doi:10.1103/PhysRevD.97.074009;%%
  %32 citations counted in INSPIRE as of 28 Dec 2019

%\cite{Brambilla:2019tpt}
\bibitem{Brambilla:2019tpt}
  N.~Brambilla, M.~A.~Escobedo, A.~Vairo and P.~Vander Griend,
  %``Transport coefficients from in medium quarkonium dynamics,''
  Phys.\ Rev.\ D {\bf 100} (2019) no.5,  054025
  doi:10.1103/PhysRevD.100.054025
  [arXiv:1903.08063 [hep-ph]].
  %%CITATION = doi:10.1103/PhysRevD.100.054025;%%
  %6 citations counted in INSPIRE as of 28 Dec 2019

%\cite{Brambilla:2019oaa}
\bibitem{Brambilla:2019oaa}
  N.~Brambilla {\it et al.} [TUMQCD Collaboration],
  %``Heavy quark momentum diffusion coefficient from the lattice,''
  arXiv:1912.00689 [hep-lat].
  %%CITATION = ARXIV:1912.00689;%%

%Blaizot open
%\cite{Blaizot:2015hya}
\bibitem{Blaizot:2015hya}
  J.~P.~Blaizot, D.~De Boni, P.~Faccioli and G.~Garberoglio,
  %``Heavy quark bound states in a quark–gluon plasma: Dissociation and recombination,''
  Nucl.\ Phys.\ A {\bf 946} (2016) 49
  doi:10.1016/j.nuclphysa.2015.10.011
  [arXiv:1503.03857 [nucl-th]].
  %%CITATION = doi:10.1016/j.nuclphysa.2015.10.011;%%
  %40 citations counted in INSPIRE as of 26 Jan 2020

%\cite{Blaizot:2017ypk}
\bibitem{Blaizot:2017ypk}
  J.~P.~Blaizot and M.~A.~Escobedo,
  %``Quantum and classical dynamics of heavy quarks in a quark-gluon plasma,''
  JHEP {\bf 1806} (2018) 034
  doi:10.1007/JHEP06(2018)034
  [arXiv:1711.10812 [hep-ph]].
  %%CITATION = doi:10.1007/JHEP06(2018)034;%%
  %23 citations counted in INSPIRE as of 27 Jan 2020

%\cite{Blaizot:2018oev}
\bibitem{Blaizot:2018oev}
  J.~P.~Blaizot and M.~A.~Escobedo,
  %``Approach to equilibrium of a quarkonium in a quark-gluon plasma,''
  Phys.\ Rev.\ D {\bf 98} (2018) no.7,  074007
  doi:10.1103/PhysRevD.98.074007
  [arXiv:1803.07996 [hep-ph]].
  %%CITATION = doi:10.1103/PhysRevD.98.074007;%%
  %17 citations counted in INSPIRE as of 27 Jan 2020

%%\cite{Escobedo:2019gzn}
%\bibitem{Escobedo:2019gzn}
%  M.~Á.~Escobedo and J.~P.~Blaizot,
%  %``Quantum and Classical Dynamics of Heavy Quarks in a Quark-Gluon Plasma,''
%  Nucl.\ Phys.\ A {\bf 982} (2019) 707.
%  doi:10.1016/j.nuclphysa.2018.10.025
%  %%CITATION = doi:10.1016/j.nuclphysa.2018.10.025;%%

%-- Mehen, Yao
%
%%\cite{Yao:2019jir}
%\bibitem{Yao:2019jir}
%  X.~Yao, W.~Ke, Y.~Xu, S.~Bass, T.~Mehen and B.~Müller,
%  %``Fate of Heavy Quark Bound States inside Quark-Gluon Plasma,''
%  arXiv:1912.01633 [hep-ph].
%  %%CITATION = ARXIV:1912.01633;%%

%%\cite{Yao:2018dap}
%\bibitem{Yao:2018dap}
%  X.~Yao, W.~Ke, Y.~Xu, S.~Bass and B.~Müller,
%  %``Quarkonium production in heavy ion collisions: coupled Boltzmann transport equations,''
%  PoS HardProbes {\bf 2018} (2018) 157
%  doi:10.22323/1.345.0157
%  [arXiv:1812.02238 [hep-ph]].
%  %%CITATION = doi:10.22323/1.345.0157;%%
%  %5 citations counted in INSPIRE as of 27 Jan 2020

%\cite{Yao:2017fuc}
\bibitem{Yao:2017fuc}
  X.~Yao and B.~Müller,
  %``Approach to equilibrium of quarkonium in quark-gluon plasma,''
  Phys.\ Rev.\ C {\bf 97} (2018) no.1,  014908
   Erratum: [Phys.\ Rev.\ C {\bf 97} (2018) no.4,  049903]
  doi:10.1103/PhysRevC.97.049903, 10.1103/PhysRevC.97.014908
  [arXiv:1709.03529 [hep-ph]].
  %%CITATION = doi:10.1103/PhysRevC.97.049903, 10.1103/PhysRevC.97.014908;%%
  %25 citations counted in INSPIRE as of 27 Jan 2020

%\cite{Yao:2018sgn}
\bibitem{Yao:2018sgn}
  X.~Yao and B.~Müller,
  %``Quarkonium inside the quark-gluon plasma: Diffusion, dissociation, recombination, and energy loss,''
  Phys.\ Rev.\ D {\bf 100} (2019) no.1,  014008
  doi:10.1103/PhysRevD.100.014008
  [arXiv:1811.09644 [hep-ph]].
  %%CITATION = doi:10.1103/PhysRevD.100.014008;%%
  %14 citations counted in INSPIRE as of 27 Jan 2020

%\cite{Yao:2018nmy}
\bibitem{Yao:2018nmy}
  X.~Yao and T.~Mehen,
  %``Quarkonium in-medium transport equation derived from first principles,''
  Phys.\ Rev.\ D {\bf 99} (2019) no.9,  096028
  doi:10.1103/PhysRevD.99.096028
  [arXiv:1811.07027 [hep-ph]].
  %%CITATION = doi:10.1103/PhysRevD.99.096028;%%
  %20 citations counted in INSPIRE as of 27 Jan 2020

%%\cite{Yao:2018zrg}
%\bibitem{Yao:2018zrg}
%  X.~Yao, W.~Ke, Y.~Xu, S.~Bass and B.~Müller,
%  %``Quarkonium production in heavy ion collisions: coupled Boltzmann transport equations,''
%  Nucl.\ Phys.\ A {\bf 982} (2019) 755
%  doi:10.1016/j.nuclphysa.2018.10.005
%  [arXiv:1807.06199 [nucl-th]].
%  %%CITATION = doi:10.1016/j.nuclphysa.2018.10.005;%%
%  %9 citations counted in INSPIRE as of 27 Jan 2020
 
%\cite{Yao:2018zze}
\bibitem{Yao:2018zze}
  X.~Yao and B.~Müller,
  %``Doubly charmed baryon production in heavy ion collisions,''
  Phys.\ Rev.\ D {\bf 97} (2018) no.7,  074003
  doi:10.1103/PhysRevD.97.074003
  [arXiv:1801.02652 [hep-ph]].
  %%CITATION = doi:10.1103/PhysRevD.97.074003;%%
  %21 citations counted in INSPIRE as of 27 Jan 2020

%\cite{Yao:2020eqy}
\bibitem{Yao:2020eqy}
X.~Yao and T.~Mehen,
%``Quarkonium Semiclassical Transport in Quark-Gluon Plasma: Factorization and Quantum Correction,''
[arXiv:2009.02408 [hep-ph]].
%3 citations counted in INSPIRE as of 10 Nov 2020


%--Strickland open
%\cite{Islam:2020bnp}
\bibitem{Islam:2020bnp}
A.~Islam and M.~Strickland,
%``Bottomonium suppression and elliptic flow using Heavy Quarkonium Quantum Dynamics,''
[arXiv:2010.05457 [hep-ph]].
%0 citations counted in INSPIRE as of 01 Nov 2020


%\cite{Sharma:2019xum}
\bibitem{Sharma:2019xum}
R.~Sharma and A.~Tiwari,
%``Quantum evolution of quarkonia with correlated and uncorrelated noise,''
Phys. Rev. D \textbf{101} (2020) no.7, 074004
doi:10.1103/PhysRevD.101.074004
[arXiv:1912.07036 [hep-ph]].
%8 citations counted in INSPIRE as of 01 Nov 2020

%-- AdS/CFT reviews
%\cite{CasalderreySolana:2011us}
\bibitem{CasalderreySolana:2011us}
  J.~Casalderrey-Solana, H.~Liu, D.~Mateos, K.~Rajagopal and U.~A.~Wiedemann,
  %``Gauge/String Duality, Hot QCD and Heavy Ion Collisions,''.
  [arXiv:1101:0618]
  %%CITATION = ARXIV:1101.0618;%%

%%\cite{Adare:2012qf}
%\bibitem{Adare:2012qf}
%  A.~Adare {\it et al.} [PHENIX Collaboration],
%  %``Transverse-Momentum Dependence of the $J/\psi$ Nuclear Modification in $d+$Au Collisions at $\sqrt{s_{NN}}=200$ GeV,''
%  Phys.\ Rev.\ C {\bf 87} (2013) no.3,  034904
%  doi:10.1103/PhysRevC.87.034904
%  [arXiv:1204.0777 [nucl-ex]].
%  %%CITATION = doi:10.1103/PhysRevC.87.034904;%%
%  %93 citations counted in INSPIRE as of 05 Feb 2020

%-- Recent pheno
%\cite{Mocsy:2013syh}
\bibitem{Mocsy:2013syh} 
  A.~Mocsy, P.~Petreczky and M.~Strickland,
  %``Quarkonia in the Quark Gluon Plasma,''
  Int.\ J.\ Mod.\ Phys.\ A {\bf 28}, 1340012 (2013)
  doi:10.1142/S0217751X13400125
  [arXiv:1302.2180 [hep-ph]].
  %%CITATION = doi:10.1142/S0217751X13400125;%%
  %64 citations counted in INSPIRE as of 25 Aug 2017

%\cite{Aarts:2016hap}
\bibitem{Aarts:2016hap}
  G.~Aarts {\it et al.},
  %``Heavy-flavor production and medium properties in high-energy nuclear collisions - What next?,''
  Eur.\ Phys.\ J.\ A {\bf 53} (2017) no.5,  93
  doi:10.1140/epja/i2017-12282-9
  [arXiv:1612.08032 [nucl-th]].
  %%CITATION = doi:10.1140/epja/i2017-12282-9;%%
  %44 citations counted in INSPIRE as of 28 Dec 2019

%\cite{Andronic:2015wma}
\bibitem{Andronic:2015wma} 
  A.~Andronic {\it et al.},
  %``Heavy-flavour and quarkonium production in the LHC era: from proton–proton to heavy-ion collisions,''
  Eur.\ Phys.\ J.\ C {\bf 76}, no. 3, 107 (2016)
  doi:10.1140/epjc/s10052-015-3819-5
  [arXiv:1506.03981 [nucl-ex]].
  %%CITATION = doi:10.1140/epjc/s10052-015-3819-5;%%
  %150 citations counted in INSPIRE as of 25 Aug 2017

%--Point split operator and potential
%\cite{Philipsen:2002az}
\bibitem{Philipsen:2002az}
O.~Philipsen,
%``Nonperturbative formulation of the static color octet potential,''
Phys. Lett. B \textbf{535} (2002), 138-144
doi:10.1016/S0370-2693(02)01777-X
[arXiv:hep-lat/0203018 [hep-lat]].
%65 citations counted in INSPIRE as of 31 Oct 2020

%Different channels
%\cite{Burnier:2007qm}
\bibitem{Burnier:2007qm}
Y.~Burnier, M.~Laine and M.~Vepsalainen,
%``Heavy quarkonium in any channel in resummed hot QCD,''
JHEP \textbf{01} (2008), 043
doi:10.1088/1126-6708/2008/01/043
[arXiv:0711.1743 [hep-ph]].
%129 citations counted in INSPIRE as of 31 Oct 2020

%-- Polyakov loop perturbative
%\cite{McLerran:1981pb}
\bibitem{McLerran:1981pb}
  L.~D.~McLerran and B.~Svetitsky,
  %``Quark Liberation at High Temperature: A Monte Carlo Study of SU(2) Gauge Theory,''
  Phys.\ Rev.\ D {\bf 24} (1981) 450.
  doi:10.1103/PhysRevD.24.450
  %%CITATION = doi:10.1103/PhysRevD.24.450;%%
  %620 citations counted in INSPIRE as of 08 Feb 2020


%-- Blaizot
%\cite{Beraudo:2007ky}
\bibitem{Beraudo:2007ky}
  A.~Beraudo, J.-P.~Blaizot and C.~Ratti,
  %``Real and imaginary-time Q anti-Q correlators in a thermal medium,''
  Nucl.\ Phys.\ A {\bf 806} (2008) 312
  doi:10.1016/j.nuclphysa.2008.03.001
  [arXiv:0712.4394 [nucl-th]].
  %%CITATION = doi:10.1016/j.nuclphysa.2008.03.001;%%
  %197 citations counted in INSPIRE as of 28 Dec 2019

%-- Kapusta book
%\cite{Kapusta:2006pm}
\bibitem{Kapusta:2006pm}
  J.~I.~Kapusta and C.~Gale,
  %``Finite-temperature field theory: Principles and applications,''
  doi:10.1017/CBO9780511535130
  %%CITATION = doi:10.1017/CBO9780511535130;%%
  %188 citations counted in INSPIRE as of 23 Jan 2020

%-- Brambilla
%\cite{Brambilla:2008cx}
\bibitem{Brambilla:2008cx} 
  N.~Brambilla, J.~Ghiglieri, A.~Vairo and P.~Petreczky,
  %``Static quark-antiquark pairs at finite temperature,''
  Phys.\ Rev.\ D {\bf 78}, 014017 (2008),
  %%CITATION = ARXIV:0804.0993;%%

%\cite{Escobedo:2008sy}
\bibitem{Escobedo:2008sy}
  M.~A.~Escobedo and J.~Soto,
  %``Non-relativistic bound states at finite temperature (I): The Hydrogen atom,''
  Phys.\ Rev.\ A {\bf 78} (2008) 032520
  doi:10.1103/PhysRevA.78.032520
  [arXiv:0804.0691 [hep-ph]].
  %%CITATION = doi:10.1103/PhysRevA.78.032520;%%
  %78 citations counted in INSPIRE as of 28 Dec 2019

%\cite{Brambilla:2010vq}
\bibitem{Brambilla:2010vq}
  N.~Brambilla, M.~A.~Escobedo, J.~Ghiglieri, J.~Soto and A.~Vairo,
  %``Heavy Quarkonium in a weakly-coupled quark-gluon plasma below the melting temperature,''
  JHEP {\bf 1009} (2010) 038
  doi:10.1007/JHEP09(2010)038
  [arXiv:1007.4156 [hep-ph]].
  %%CITATION = doi:10.1007/JHEP09(2010)038;%%
  %118 citations counted in INSPIRE as of 28 Dec 2019

%\cite{Escobedo:2011ie}  
\bibitem{Escobedo:2011ie}
  M.~A.~Escobedo, J.~Soto and M.~Mannarelli,
  %``Non-relativistic bound states in a moving thermal bath,''
  Phys.\ Rev.\  D {\bf 84}, 016008 (2011) and references therein.
  %%CITATION = PHRVA,D84,016008;%%

%\cite{Brambilla:2011sg}
\bibitem{Brambilla:2011sg}
  N.~Brambilla, M.~A.~Escobedo, J.~Ghiglieri and A.~Vairo,
  %``Thermal width and gluo-dissociation of quarkonium in pNRQCD,''
  JHEP {\bf 1112} (2011) 116
  doi:10.1007/JHEP12(2011)116
  [arXiv:1109.5826 [hep-ph]].
  %%CITATION = doi:10.1007/JHEP12(2011)116;%%
  %74 citations counted in INSPIRE as of 28 Dec 2019

%\cite{Brambilla:2011mk}
\bibitem{Brambilla:2011mk}
  N.~Brambilla, M.~A.~Escobedo, J.~Ghiglieri and A.~Vairo,
  %``The spin-orbit potential and Poincar\'e invariance in finite temperature pNRQCD,''
  JHEP {\bf 1107} (2011) 096
  doi:10.1007/JHEP07(2011)096
  [arXiv:1105.4807 [hep-ph]].
  %%CITATION = doi:10.1007/JHEP07(2011)096;%%
  %13 citations counted in INSPIRE as of 28 Dec 2019

%\cite{Brambilla:2013dpa}
\bibitem{Brambilla:2013dpa} 
  N.~Brambilla, M.~A.~Escobedo, J.~Ghiglieri and A.~Vairo,
  %``Thermal width and quarkonium dissociation by inelastic parton scattering,''
  JHEP {\bf 1305}, 130 (2013)
  doi:10.1007/JHEP05(2013)130
  [arXiv:1303.6097 [hep-ph]].
  %%CITATION = doi:10.1007/JHEP05(2013)130;%%
  %39 citations counted in INSPIRE as of 25 Aug 2017

%\cite{Biondini:2017qjh}
\bibitem{Biondini:2017qjh}
  S.~Biondini, N.~Brambilla, M.~A.~Escobedo and A.~Vairo,
  %``Momentum anisotropy effects for quarkonium in a weakly-coupled quark-gluon plasma below the melting temperature,''
  Phys.\ Rev.\ D {\bf 95} (2017) no.7,  074016
  doi:10.1103/PhysRevD.95.074016
  [arXiv:1701.06956 [hep-ph]].
  %%CITATION = doi:10.1103/PhysRevD.95.074016;%%
  %4 citations counted in INSPIRE as of 25 Jan 2020
 
%----Lattice kappa
%\cite{Banerjee:2011ra}
\bibitem{Banerjee:2011ra}
  D.~Banerjee, S.~Datta, R.~Gavai and P.~Majumdar,
  %``Heavy Quark Momentum Diffusion Coefficient from Lattice QCD,''
  Phys.\ Rev.\ D {\bf 85} (2012) 014510
  doi:10.1103/PhysRevD.85.014510
  [arXiv:1109.5738 [hep-lat]].
  %%CITATION = doi:10.1103/PhysRevD.85.014510;%%
  %117 citations counted in INSPIRE as of 26 Jan 2020

%\cite{Ding:2011hr}
\bibitem{Ding:2011hr}
  H.~T.~Ding, A.~Francis, O.~Kaczmarek, F.~Karsch, H.~Satz and W.~Soldner,
  %``Heavy Quark diffusion from lattice QCD spectral functions,''
  J.\ Phys.\ G {\bf 38} (2011) 124070
  doi:10.1088/0954-3899/38/12/124070
  [arXiv:1107.0311 [nucl-th]].
  %%CITATION = doi:10.1088/0954-3899/38/12/124070;%%
  %30 citations counted in INSPIRE as of 26 Jan 2020

%\cite{Francis:2015daa}
\bibitem{Francis:2015daa}
  A.~Francis, O.~Kaczmarek, M.~Laine, T.~Neuhaus and H.~Ohno,
  %``Nonperturbative estimate of the heavy quark momentum diffusion coefficient,''
  Phys.\ Rev.\ D {\bf 92} (2015) no.11,  116003
  doi:10.1103/PhysRevD.92.116003
  [arXiv:1508.04543 [hep-lat]].
  %%CITATION = doi:10.1103/PhysRevD.92.116003;%%
  %71 citations counted in INSPIRE as of 26 Jan 2020

%-- Lattice potential
%\cite{Rothkopf:2011db}
\bibitem{Rothkopf:2011db}
  A.~Rothkopf, T.~Hatsuda and S.~Sasaki,
  %``Complex Heavy-Quark Potential at Finite Temperature from Lattice QCD,''
  Phys.\ Rev.\ Lett.\  {\bf 108} (2012) 162001
  doi:10.1103/PhysRevLett.108.162001
  [arXiv:1108.1579 [hep-lat]].
  %%CITATION = doi:10.1103/PhysRevLett.108.162001;%%
  %137 citations counted in INSPIRE as of 26 Jan 2020

%\cite{Burnier:2013fca}
\bibitem{Burnier:2013fca}
  Y.~Burnier and A.~Rothkopf,
  %``A hard thermal loop benchmark for the extraction of the nonperturbative $Q\bar{Q}$ potential,''
  Phys.\ Rev.\ D {\bf 87} (2013) 114019
  doi:10.1103/PhysRevD.87.114019
  [arXiv:1304.4154 [hep-ph]].
  %%CITATION = doi:10.1103/PhysRevD.87.114019;%%
  %22 citations counted in INSPIRE as of 26 Jan 2020

%\cite{Burnier:2014ssa}
\bibitem{Burnier:2014ssa}
  Y.~Burnier, O.~Kaczmarek and A.~Rothkopf,
  %``Static quark-antiquark potential in the quark-gluon plasma from lattice QCD,''
  Phys.\ Rev.\ Lett.\  {\bf 114} (2015) no.8,  082001
  doi:10.1103/PhysRevLett.114.082001
  [arXiv:1410.2546 [hep-lat]].
  %%CITATION = doi:10.1103/PhysRevLett.114.082001;%%
  %76 citations counted in INSPIRE as of 26 Jan 2020

%Used above
%%\cite{Burnier:2015tda}
%\bibitem{Burnier:2015tda} 
%  Y.~Burnier, O.~Kaczmarek and A.~Rothkopf,
%  %``Quarkonium at finite temperature: Towards realistic phenomenology from first principles,''
%  JHEP {\bf 1512}, 101 (2015)
%  doi:10.1007/JHEP12(2015)101
%  [arXiv:1509.07366 [hep-ph]].
%  %%CITATION = doi:10.1007/JHEP12(2015)101;%%
%  %55 citations counted in INSPIRE as of 26 Jan 2020

%\cite{Burnier:2016mxc}
\bibitem{Burnier:2016mxc}
  Y.~Burnier and A.~Rothkopf,
  %``Complex heavy-quark potential and Debye mass in a gluonic medium from lattice QCD,''
  Phys.\ Rev.\ D {\bf 95} (2017) no.5,  054511
  doi:10.1103/PhysRevD.95.054511
  [arXiv:1607.04049 [hep-lat]].
  %%CITATION = doi:10.1103/PhysRevD.95.054511;%%
  %17 citations counted in INSPIRE as of 26 Jan 2020

%\cite{Bala:2019cqu}
\bibitem{Bala:2019cqu}
  D.~Bala and S.~Datta,
  %``Nonperturbative potential for study of quarkonia in QGP,''
  arXiv:1909.10548 [hep-lat].
  %%CITATION = ARXIV:1909.10548;%%
  %2 citations counted in INSPIRE as of 26 Jan 2020

%\cite{Bala:2019boe}
\bibitem{Bala:2019boe}
  D.~Bala and S.~Datta,
  %``Effective thermal potential between static $Q$ and $\bar Q$ in SU(3) gauge theory,''
  arXiv:1912.04826 [hep-lat].
  %%CITATION = ARXIV:1912.04826;%%

%-- Polyakov loop, free energy perturbative
%\cite{Brambilla:2010xn}
\bibitem{Brambilla:2010xn}
  N.~Brambilla, J.~Ghiglieri, P.~Petreczky and A.~Vairo,
  %``The Polyakov loop and correlator of Polyakov loops at next-to-next-to-leading order,''
  Phys.\ Rev.\ D {\bf 82} (2010) 074019
  doi:10.1103/PhysRevD.82.074019
  [arXiv:1007.5172 [hep-ph]].
  %%CITATION = doi:10.1103/PhysRevD.82.074019;%%
  %74 citations counted in INSPIRE as of 08 Feb 2020

%\cite{Berwein:2012mw}
\bibitem{Berwein:2012mw}
  M.~Berwein, N.~Brambilla, J.~Ghiglieri and A.~Vairo,
  %``Renormalization of the cyclic Wilson loop,''
  JHEP {\bf 1303} (2013) 069
  doi:10.1007/JHEP03(2013)069
  [arXiv:1212.4413 [hep-th]].
  %%CITATION = doi:10.1007/JHEP03(2013)069;%%
  %21 citations counted in INSPIRE as of 08 Feb 2020

%\cite{Berwein:2013xza}
\bibitem{Berwein:2013xza}
  M.~Berwein, N.~Brambilla and A.~Vairo,
  %``Renormalization of Loop Functions in QCD,''
  Phys.\ Part.\ Nucl.\  {\bf 45} (2014) no.4,  656
  doi:10.1134/S1063779614040029
  [arXiv:1312.6651 [hep-th]].
  %%CITATION = doi:10.1134/S1063779614040029;%%
  %10 citations counted in INSPIRE as of 08 Feb 2020

%\cite{Berwein:2015ayt}
\bibitem{Berwein:2015ayt}
  M.~Berwein, N.~Brambilla, P.~Petreczky and A.~Vairo,
  %``Polyakov loop at next-to-next-to-leading order,''
  Phys.\ Rev.\ D {\bf 93} (2016) no.3,  034010
  doi:10.1103/PhysRevD.93.034010
  [arXiv:1512.08443 [hep-ph]].
  %%CITATION = doi:10.1103/PhysRevD.93.034010;%%
  %17 citations counted in INSPIRE as of 08 Feb 2020

%\cite{Berwein:2017thy}
\bibitem{Berwein:2017thy}
  M.~Berwein, N.~Brambilla, P.~Petreczky and A.~Vairo,
  %``Polyakov loop correlator in perturbation theory,''
  Phys.\ Rev.\ D {\bf 96} (2017) no.1,  014025
  doi:10.1103/PhysRevD.96.014025
  [arXiv:1704.07266 [hep-ph]].
  %%CITATION = doi:10.1103/PhysRevD.96.014025;%%
  %7 citations counted in INSPIRE as of 08 Feb 2020

%-- Lattice free energy polyakov loop recent
%\cite{Bazavov:2016uvm}
\bibitem{Bazavov:2016uvm}
  A.~Bazavov, N.~Brambilla, H.-T.~Ding, P.~Petreczky, H.-P.~Schadler, A.~Vairo and J.~H.~Weber,
  %``Polyakov loop in 2+1 flavor QCD from low to high temperatures,''
  Phys.\ Rev.\ D {\bf 93} (2016) no.11,  114502
  doi:10.1103/PhysRevD.93.114502
  [arXiv:1603.06637 [hep-lat]].
  %%CITATION = doi:10.1103/PhysRevD.93.114502;%%
  %67 citations counted in INSPIRE as of 08 Feb 2020

%\cite{Bazavov:2018wmo}
\bibitem{Bazavov:2018wmo}
  A.~Bazavov {\it et al.} [TUMQCD Collaboration],
  %``Color screening in (2+1)-flavor QCD,''
  Phys.\ Rev.\ D {\bf 98} (2018) no.5,  054511
  doi:10.1103/PhysRevD.98.054511
  [arXiv:1804.10600 [hep-lat]].
  %%CITATION = doi:10.1103/PhysRevD.98.054511;%%
  %22 citations counted in INSPIRE as of 08 Feb 2020

%-- Experiment, Experimental Observations
%-- RHIC J/Psi
%\cite{Adler:2003rc}
\bibitem{Adler:2003rc}
  S.~S.~Adler {\it et al.} [PHENIX Collaboration],
  %``J / psi production in Au Au collisions at s(NN)**(1/2) = 200-GeV at the Relativistic Heavy Ion Collider,''
  Phys.\ Rev.\ C {\bf 69} (2004) 014901
  doi:10.1103/PhysRevC.69.014901
  [nucl-ex/0305030].
  %%CITATION = doi:10.1103/PhysRevC.69.014901;%%
  %103 citations counted in INSPIRE as of 01 Feb 2020

%\cite{Adare:2006ns}
\bibitem{Adare:2006ns}
  A.~Adare {\it et al.} [PHENIX Collaboration],
  %``$J/\psi$ Production vs Centrality, Transverse Momentum, and Rapidity in Au+Au Collisions at $\sqrt{s_{NN}} = 200$ GeV,''
  Phys.\ Rev.\ Lett.\  {\bf 98} (2007) 232301
  doi:10.1103/PhysRevLett.98.232301
  [nucl-ex/0611020].
  %%CITATION = doi:10.1103/PhysRevLett.98.232301;%%
  %669 citations counted in INSPIRE as of 01 Feb 2020

%\cite{Adare:2008sh}
\bibitem{Adare:2008sh} 
  A.~Adare {\it et al.} [PHENIX Collaboration],
  %``J/psi Production in s(NN)**(1/2) = 200-GeV Cu+Cu Collisions,''
  Phys.\ Rev.\ Lett.\  {\bf 101}, 122301 (2008)
  doi:10.1103/PhysRevLett.101.122301
  [arXiv:0801.0220 [nucl-ex]].
  %%CITATION = doi:10.1103/PhysRevLett.101.122301;%%
  %180 citations counted in INSPIRE as of 01 Feb 2020

%\cite{Abelev:2009qaa}
\bibitem{Abelev:2009qaa}
  B.~I.~Abelev {\it et al.} [STAR Collaboration],
  %``J/psi production at high transverse momentum in p+p and Cu+Cu collisions at s(NN)**1/2 = 200GeV,''
  Phys.\ Rev.\ C {\bf 80} (2009) 041902
  doi:10.1103/PhysRevC.80.041902
  [arXiv:0904.0439 [nucl-ex]].
  %%CITATION = doi:10.1103/PhysRevC.80.041902;%%
  %136 citations counted in INSPIRE as of 01 Feb 2020

%%\cite{Oda:2008kg}
%\bibitem{Oda:2008kg}
%  S.~X.~Oda [PHENIX Collaboration],
%  %``J/psi production at RHIC-PHENIX,''
%  J.\ Phys.\ G {\bf 35} (2008) 104134
%  doi:10.1088/0954-3899/35/10/104134
%  [arXiv:0804.4446 [nucl-ex]].
%  %%CITATION = doi:10.1088/0954-3899/35/10/104134;%%
%  %20 citations counted in INSPIRE as of 01 Feb 2020
%
%%\cite{Atomssa:2008dn}
%\bibitem{Atomssa:2008dn}
%  E.~T.~Atomssa [PHENIX Collaboration],
%  %``J/psi production measurements by the PHENIX experiment at RHIC,''
%  Eur.\ Phys.\ J.\ C {\bf 61} (2009) 683
%  doi:10.1140/epjc/s10052-009-0900-y
%  [arXiv:0805.4562 [nucl-ex]].
%  %%CITATION = doi:10.1140/epjc/s10052-009-0900-y;%%
%  %16 citations counted in INSPIRE as of 01 Feb 2020

%%\cite{Tang:2011kr}
%\bibitem{Tang:2011kr}
%  Z.~Tang  [STAR Collaboration],
%  %``J/psi production and correlation in p$+$p and Au$+$Au collisions at STAR,''
%  J.\ Phys.\ G {\bf 38}, 124107 (2011).
%  %%CITATION = JPHGB,G38,124107;%%

%\cite{Adare:2011yf}
\bibitem{Adare:2011yf}
  A.~Adare {\it et al.} [PHENIX Collaboration],
  %``$J/\psi$ suppression at forward rapidity in Au+Au collisions at $\sqrt{s_{NN}}=200$ GeV,''
  Phys.\ Rev.\ C {\bf 84} (2011) 054912
  doi:10.1103/PhysRevC.84.054912
  [arXiv:1103.6269 [nucl-ex]].
  %%CITATION = doi:10.1103/PhysRevC.84.054912;%%
  %233 citations counted in INSPIRE as of 01 Feb 2020

%\cite{Adamczyk:2012pw}
\bibitem{Adamczyk:2012pw}
  L.~Adamczyk {\it et al.} [STAR Collaboration],
  %``Measurement of $J/\psi$ Azimuthal Anisotropy in Au+Au Collisions at $\sqrt{s_{NN}}$ = 200 GeV,''
  Phys.\ Rev.\ Lett.\  {\bf 111} (2013) no.5,  052301
  doi:10.1103/PhysRevLett.111.052301
  [arXiv:1212.3304 [nucl-ex]].
  %%CITATION = doi:10.1103/PhysRevLett.111.052301;%%
  %51 citations counted in INSPIRE as of 01 Feb 2020

%\cite{Adamczyk:2012ey}
\bibitem{Adamczyk:2012ey}
  L.~Adamczyk {\it et al.} [STAR Collaboration],
  %``$J/\psi$ production at high transverse momenta in $p+p$ and Au+Au collisions at $\sqrt{s_{NN}} = 200$ GeV,''
  Phys.\ Lett.\ B {\bf 722} (2013) 55
  doi:10.1016/j.physletb.2013.04.010
  [arXiv:1208.2736 [nucl-ex]].
  %%CITATION = doi:10.1016/j.physletb.2013.04.010;%%
  %114 citations counted in INSPIRE as of 01 Feb 2020

%\cite{Adare:2012wf}
\bibitem{Adare:2012wf}
  A.~Adare {\it et al.} [PHENIX Collaboration],
  %``$J/\psi$ suppression at forward rapidity in Au+Au collisions at $\sqrt{s_{NN}}=39$ and 62.4 GeV,''
  Phys.\ Rev.\ C {\bf 86} (2012) 064901
  doi:10.1103/PhysRevC.86.064901
  [arXiv:1208.2251 [nucl-ex]].
  %%CITATION = doi:10.1103/PhysRevC.86.064901;%%
  %40 citations counted in INSPIRE as of 01 Feb 2020

%\cite{Adamczyk:2013tvk}
\bibitem{Adamczyk:2013tvk}
  L.~Adamczyk {\it et al.} [STAR Collaboration],
  %``$J/\psi$ production at low $p_T$ in Au + Au and Cu + Cu collisions at $\sqrt{s_{NN}}=200$ GeV with the STAR detector,''
  Phys.\ Rev.\ C {\bf 90} (2014) no.2,  024906
  doi:10.1103/PhysRevC.90.024906
  [arXiv:1310.3563 [nucl-ex]].
  %%CITATION = doi:10.1103/PhysRevC.90.024906;%%
  %76 citations counted in INSPIRE as of 01 Feb 2020

%\cite{Aidala:2014bqx}
\bibitem{Aidala:2014bqx}
  C.~Aidala {\it et al.} [PHENIX Collaboration],
  %``Nuclear matter effects on $J/\psi$ production in asymmetric Cu+Au collisions at $\sqrt{s_{_{NN}}}$ = 200 GeV,''
  Phys.\ Rev.\ C {\bf 90} (2014) no.6,  064908
  doi:10.1103/PhysRevC.90.064908
  [arXiv:1404.1873 [nucl-ex]].
  %%CITATION = doi:10.1103/PhysRevC.90.064908;%%
  %30 citations counted in INSPIRE as of 01 Feb 2020

%\cite{Adare:2015hva}
\bibitem{Adare:2015hva}
  A.~Adare {\it et al.} [PHENIX Collaboration],
  %``Forward $J/\psi$ production in U$+$U collisions at $\sqrt{s_{NN}}$=193 GeV,''
  Phys.\ Rev.\ C {\bf 93} (2016) no.3,  034903
  doi:10.1103/PhysRevC.93.034903
  [arXiv:1509.05380 [nucl-ex]].
  %%CITATION = doi:10.1103/PhysRevC.93.034903;%%
  %11 citations counted in INSPIRE as of 01 Feb 2020

%\cite{Adamczyk:2016srz}
\bibitem{Adamczyk:2016srz}
  L.~Adamczyk {\it et al.} [STAR Collaboration],
  %``Energy dependence of $J/\psi$ production in Au+Au collisions at $\sqrt{s_{NN}} =$ 39, 62.4 and 200 GeV,''
  Phys.\ Lett.\ B {\bf 771} (2017) 13
  doi:10.1016/j.physletb.2017.04.078
  [arXiv:1607.07517 [hep-ex]].
  %%CITATION = doi:10.1016/j.physletb.2017.04.078;%%
  %21 citations counted in INSPIRE as of 01 Feb 2020

%\cite{STAR:2019yox}
\bibitem{STAR:2019yox}
  J.~Adam {\it et al.} [STAR Collaboration],
  %``Observation of excess J/$\psi$ yield at very low transverse momenta in Au+Au collisions at $\sqrt{s_{\rm{NN}}} =$ 200 GeV and U+U collisions at $\sqrt{s_{\rm{NN}}} =$ 193 GeV,''
  Phys.\ Rev.\ Lett.\  {\bf 123} (2019) no.13,  132302
  doi:10.1103/PhysRevLett.123.132302
  [arXiv:1904.11658 [hep-ex]].
  %%CITATION = doi:10.1103/PhysRevLett.123.132302;%%
  %1 citations counted in INSPIRE as of 01 Feb 2020

%\cite{Adam:2019rbk}
\bibitem{Adam:2019rbk}
  J.~Adam {\it et al.} [STAR Collaboration],
  %``Measurement of inclusive $J/\psi$ suppression in Au+Au collisions at $\sqrt{s_{NN}}$ = 200 GeV through the dimuon channel at STAR,''
  Phys.\ Lett.\ B {\bf 797} (2019) 134917
  doi:10.1016/j.physletb.2019.134917
  [arXiv:1905.13669 [nucl-ex]].
  %%CITATION = doi:10.1016/j.physletb.2019.134917;%%
  %1 citations counted in INSPIRE as of 01 Feb 2020

% RHIC Upsilon
%%\cite{Reed:2011}
%\bibitem{Reed:2011} 
%  R.~Reed [STAR Collaboration],
%  %``Upsilon production in p + p, d + Au, Au + Au collisions at s(NN)**(1/2) = 200-GeV in STAR,''
%  J.\ Phys.\ Conf.\ Ser.\  {\bf 270}, 012026 (2011)
%  [Nucl.\ Phys.\ A {\bf 855}, 440 (2011)]; Quark Matter 2011.
%  %%CITATION = 00462,270,012026;%%
%
%%\cite{Whitaker:2012tt}
%\bibitem{Whitaker:2012tt}
%  S.~Whitaker,
%  %``$\Upsilon$ Measurements by the PHENIX Collaboration,''
%  Nucl.\ Phys.\ A {\bf 910-911} (2013) 462
%  doi:10.1016/j.nuclphysa.2012.12.108
%  [arXiv:1207.6649 [nucl-ex]].
%  %%CITATION = doi:10.1016/j.nuclphysa.2012.12.108;%%
%  %1 citations counted in INSPIRE as of 01 Feb 2020
%
%%\cite{Ye:2017fwv}
%\bibitem{Ye:2017fwv}
%  Z.~Ye [STAR Collaboration],
%  %``$\Upsilon$ measurements in p+p, p+Au and Au+Au collisions at $\sqrt {s_{NN}}$ = 200GeV with the STAR experiment,''
%  Nucl.\ Phys.\ A {\bf 967} (2017) 600.
%  doi:10.1016/j.nuclphysa.2017.06.040
%  %%CITATION = doi:10.1016/j.nuclphysa.2017.06.040;%%
%  %7 citations counted in INSPIRE as of 01 Feb 2020
 
%\cite{Adamczyk:2013poh}
\bibitem{Adamczyk:2013poh}
  L.~Adamczyk {\it et al.} [STAR Collaboration],
  %``Suppression of $\Upsilon$ production in d+Au and Au+Au collisions at $\sqrt{s_{NN}}$=200 GeV,''
  Phys.\ Lett.\ B {\bf 735} (2014) 127
   Erratum: [Phys.\ Lett.\ B {\bf 743} (2015) 537]
  doi:10.1016/j.physletb.2014.06.028, 10.1016/j.physletb.2015.01.046
  [arXiv:1312.3675 [nucl-ex]].
  %%CITATION = doi:10.1016/j.physletb.2014.06.028, 10.1016/j.physletb.2015.01.046;%%
  %98 citations counted in INSPIRE as of 01 Feb 2020

%\cite{Adamczyk:2016dzv}
\bibitem{Adamczyk:2016dzv}
  L.~Adamczyk {\it et al.} [STAR Collaboration],
  %``$\Upsilon$ production in U + U collisions at $\sqrt{{s}_{NN}}=$ 193 GeV measured with the STAR experiment,''
  Phys.\ Rev.\ C {\bf 94} (2016) no.6,  064904
  doi:10.1103/PhysRevC.94.064904
  [arXiv:1608.06487 [nucl-ex]].
  %%CITATION = doi:10.1103/PhysRevC.94.064904;%%
  %16 citations counted in INSPIRE as of 01 Feb 2020

%\cite{Adare:2014hje}
\bibitem{Adare:2014hje}
  A.~Adare {\it et al.} [PHENIX Collaboration],
  %``Measurement of $\Upsilon(1S+2S+3S)$ production in $p+p$ and Au$+$Au collisions at $\sqrt{s_{_{NN}}}=200$ GeV,''
  Phys.\ Rev.\ C {\bf 91} (2015) no.2,  024913
  doi:10.1103/PhysRevC.91.024913
  [arXiv:1404.2246 [nucl-ex]].
  %%CITATION = doi:10.1103/PhysRevC.91.024913;%%
  %54 citations counted in INSPIRE as of 01 Feb 2020

%-- LHC J/psi 
%\cite{Chatrchyan:2012np}
\bibitem{Chatrchyan:2012np} 
  S.~Chatrchyan {\it et al.} [CMS Collaboration],
  %``Suppression of non-prompt $J/\psi$, prompt $J/\psi$, and Y(1S) in PbPb collisions at $\sqrt{s_{NN}}=2.76$ TeV,''
  JHEP {\bf 1205}, 063 (2012)
  doi:10.1007/JHEP05(2012)063
  [arXiv:1201.5069 [nucl-ex]].
  %%CITATION = doi:10.1007/JHEP05(2012)063;%%
  %342 citations counted in INSPIRE as of 18 Aug 2017
  
%\cite{Khachatryan:2014bva}
\bibitem{Khachatryan:2014bva} 
  V.~Khachatryan {\it et al.} [CMS Collaboration],
  %``Measurement of Prompt $\psi(2S) \to J/\psi$ Yield Ratios in Pb-Pb and $p-p$ Collisions at $\sqrt {s_{NN}}=$ 2.76  TeV,''
  Phys.\ Rev.\ Lett.\  {\bf 113}, no. 26, 262301 (2014)
  doi:10.1103/PhysRevLett.113.262301
  [arXiv:1410.1804 [nucl-ex]].
  %%CITATION = doi:10.1103/PhysRevLett.113.262301;%%
  %44 citations counted in INSPIRE as of 29 Aug 2017

%\cite{Khachatryan:2016ypw}
\bibitem{Khachatryan:2016ypw} 
  V.~Khachatryan {\it et al.} [CMS Collaboration],
  %``Suppression and azimuthal anisotropy of prompt and nonprompt ${\mathrm{J}}/\psi $ production in PbPb collisions at $\sqrt{{s_{_{\text {NN}}}}} =2.76$ $\,\mathrm{TeV}$,''
  Eur.\ Phys.\ J.\ C {\bf 77}, no. 4, 252 (2017)
  doi:10.1140/epjc/s10052-017-4781-1
  [arXiv:1610.00613 [nucl-ex]].
  %%CITATION = doi:10.1140/epjc/s10052-017-4781-1;%%
  %14 citations counted in INSPIRE as of 29 Aug 2017

%\cite{Sirunyan:2016znt}
\bibitem{Sirunyan:2016znt} 
  A.~M.~Sirunyan {\it et al.} [CMS Collaboration],
  %``Relative Modification of Prompt ?(2S) and J/? Yields from pp to PbPb Collisions at $\sqrt{s_{NN}}=5.02$?TeV,''
  Phys.\ Rev.\ Lett.\  {\bf 118}, no. 16, 162301 (2017)
  doi:10.1103/PhysRevLett.118.162301
  [arXiv:1611.01438 [nucl-ex]].
  %%CITATION = doi:10.1103/PhysRevLett.118.162301;%%
  %10 citations counted in INSPIRE as of 18 Aug 2017

%\cite{Aad:2010px}
\bibitem{Aad:2010px}
  G.~Aad {\it et al.}  [ATLAS Collaboration],
  %``Measurement of the centrality dependence of J/{\psi} yields and observation
  %of Z production in lead-lead collisions with the ATLAS detector at the LHC,''
  Phys.\ Lett.\  B {\bf 697} (2011) 294.
  %%CITATION = PHLTA,B697,294;%%

%\cite{Aad:2010aa}
\bibitem{Aad:2010aa} 
  G.~Aad {\it et al.} [ATLAS Collaboration],
  %``Measurement of the centrality dependence of J/psi yields and observation of Z production in lead?lead collisions with the ATLAS detector at the LHC,''
  Phys.\ Lett.\ B {\bf 697}, 294 (2011)
  doi:10.1016/j.physletb.2011.02.006
  [arXiv:1012.5419 [hep-ex]].
  %%CITATION = doi:10.1016/j.physletb.2011.02.006;%%
  %175 citations counted in INSPIRE as of 18 Aug 2017

%\cite{Aaboud:2018quy}
\bibitem{Aaboud:2018quy}
  M.~Aaboud {\it et al.} [ATLAS Collaboration],
  %``Prompt and non-prompt $J/\psi $ and $\psi (2\mathrm {S})$ suppression at high transverse momentum in $5.02~\mathrm {TeV}$ Pb+Pb collisions with the ATLAS experiment,''
  Eur.\ Phys.\ J.\ C {\bf 78} (2018) no.9,  762
  doi:10.1140/epjc/s10052-018-6219-9
  [arXiv:1805.04077 [nucl-ex]].
  %%CITATION = doi:10.1140/epjc/s10052-018-6219-9;%%
  %44 citations counted in INSPIRE as of 01 Feb 2020

%\cite{Aaboud:2018ttm}
\bibitem{Aaboud:2018ttm}
  M.~Aaboud {\it et al.} [ATLAS Collaboration],
  %``Prompt and non-prompt $J/\psi $ elliptic flow in Pb+Pb collisions at $\sqrt{s_{_\text {NN}}} = 5.02$ Tev with the ATLAS detector,''
  Eur.\ Phys.\ J.\ C {\bf 78} (2018) no.9,  784
  doi:10.1140/epjc/s10052-018-6243-9
  [arXiv:1807.05198 [nucl-ex]].
  %%CITATION = doi:10.1140/epjc/s10052-018-6243-9;%%
  %11 citations counted in INSPIRE as of 01 Feb 2020

%\cite{Abelev:2012rv}
\bibitem{Abelev:2012rv} 
  B.~Abelev {\it et al.}  [ALICE Collaboration],
  %``J/psi production at low transverse momentum in Pb-Pb collisions at sqrt(sNN) = 2.76 TeV,''
  [arXiv:1202.1383 [hep-ex]].
  %%CITATION = ARXIV:1202.1383;%%

%\cite{ALICE:2013xna}
\bibitem{ALICE:2013xna}
  E.~Abbas {\it et al.} [ALICE Collaboration],
  %``J/Psi Elliptic Flow in Pb-Pb Collisions at $\sqrt{s_{\rm NN}}$ = 2.76 TeV,''
  Phys.\ Rev.\ Lett.\  {\bf 111} (2013) 162301
  doi:10.1103/PhysRevLett.111.162301
  [arXiv:1303.5880 [nucl-ex]].
  %%CITATION = doi:10.1103/PhysRevLett.111.162301;%%
  %110 citations counted in INSPIRE as of 01 Feb 2020

%\cite{Abelev:2013ila}
\bibitem{Abelev:2013ila} 
  B.~B.~Abelev {\it et al.} [ALICE Collaboration],
  %``Centrality, rapidity and transverse momentum dependence of $J/\psi$ suppression in Pb-Pb collisions at $\sqrt{s_{\rm NN}}$=2.76 TeV,''
  Phys.\ Lett.\ B {\bf 734}, 314 (2014)
  doi:10.1016/j.physletb.2014.05.064
  [arXiv:1311.0214 [nucl-ex]].
  %%CITATION = doi:10.1016/j.physletb.2014.05.064;%%
  %136 citations counted in INSPIRE as of 18 Aug 2017

%\cite{Adam:2015rba}
\bibitem{Adam:2015rba} 
  J.~Adam {\it et al.} [ALICE Collaboration],
  %``Inclusive, prompt and non-prompt J/$\psi$ production at mid-rapidity in Pb-Pb collisions at $\sqrt{s_{\rm NN}}$ = 2.76 TeV,''
  JHEP {\bf 1507}, 051 (2015)
  doi:10.1007/JHEP07(2015)051
  [arXiv:1504.07151 [nucl-ex]].
  %%CITATION = doi:10.1007/JHEP07(2015)051;%%
  %44 citations counted in INSPIRE as of 18 Aug 2017

%\cite{Adam:2015isa}
\bibitem{Adam:2015isa}
  J.~Adam {\it et al.} [ALICE Collaboration],
  %``Differential studies of inclusive J/ψ and ψ(2S) production at forward rapidity in Pb-Pb collisions at $ \sqrt{s_{\mathrm{NN}}}=2.76 $ TeV,''
  JHEP {\bf 1605} (2016) 179
  doi:10.1007/JHEP05(2016)179
  [arXiv:1506.08804 [nucl-ex]].
  %%CITATION = doi:10.1007/JHEP05(2016)179;%%
  %90 citations counted in INSPIRE as of 01 Feb 2020

%\cite{Adam:2015gba}
\bibitem{Adam:2015gba}
  J.~Adam {\it et al.} [ALICE Collaboration],
  %``Measurement of an excess in the yield of $J/\psi$ at very low $p_{\rm T}$ in Pb-Pb collisions at $\sqrt{s_{\rm NN}}$ = 2.76 TeV,''
  Phys.\ Rev.\ Lett.\  {\bf 116} (2016) no.22,  222301
  doi:10.1103/PhysRevLett.116.222301
  [arXiv:1509.08802 [nucl-ex]].
  %%CITATION = doi:10.1103/PhysRevLett.116.222301;%%
  %84 citations counted in INSPIRE as of 01 Feb 2020

%\cite{Adam:2016rdg}
\bibitem{Adam:2016rdg}
  J.~Adam {\it et al.} [ALICE Collaboration],
  %``J/$\psi$ suppression at forward rapidity in Pb-Pb collisions at $\mathbf{\sqrt{s_{{\rm NN}}} = 5.02}$ TeV,''
  Phys.\ Lett.\ B {\bf 766} (2017) 212
  doi:10.1016/j.physletb.2016.12.064
  [arXiv:1606.08197 [nucl-ex]].
  %%CITATION = doi:10.1016/j.physletb.2016.12.064;%%
  %98 citations counted in INSPIRE as of 01 Feb 2020

%\cite{Acharya:2017tgv}
\bibitem{Acharya:2017tgv}
  S.~Acharya {\it et al.} [ALICE Collaboration],
  %``J/$\psi$ elliptic flow in Pb-Pb collisions at $\sqrt{s_\mathrm{NN}}=5.02$ TeV,''
  Phys.\ Rev.\ Lett.\  {\bf 119} (2017) no.24,  242301
  doi:10.1103/PhysRevLett.119.242301
  [arXiv:1709.05260 [nucl-ex]].
  %%CITATION = doi:10.1103/PhysRevLett.119.242301;%%
  %44 citations counted in INSPIRE as of 01 Feb 2020

%\cite{Acharya:2018jvc}
\bibitem{Acharya:2018jvc}
  S.~Acharya {\it et al.} [ALICE Collaboration],
  %``Inclusive J/$\psi$ production in Xe–Xe collisions at $\sqrt{s_{\rm NN}}$ = 5.44 TeV,''
  Phys.\ Lett.\ B {\bf 785} (2018) 419
  doi:10.1016/j.physletb.2018.08.047
  [arXiv:1805.04383 [nucl-ex]].
  %%CITATION = doi:10.1016/j.physletb.2018.08.047;%%
  %10 citations counted in INSPIRE as of 01 Feb 2020

%\cite{Acharya:2018pjd}
\bibitem{Acharya:2018pjd}
  S.~Acharya {\it et al.} [ALICE Collaboration],
  %``Study of J/$\psi$ azimuthal anisotropy at forward rapidity in Pb-Pb collisions at $ \sqrt{s_{\mathrm{NN}}}=5.02 $ TeV,''
  JHEP {\bf 1902} (2019) 012
  doi:10.1007/JHEP02(2019)012
  [arXiv:1811.12727 [nucl-ex]].
  %%CITATION = doi:10.1007/JHEP02(2019)012;%%
  %9 citations counted in INSPIRE as of 01 Feb 2020

%\cite{Acharya:2019iur}
\bibitem{Acharya:2019iur}
  S.~Acharya {\it et al.} [ALICE Collaboration],
  %``Studies of J/$\psi$ production at forward rapidity in Pb-Pb collisions at $\sqrt{s_{\rm{NN}}}$ = 5.02 TeV,''
  arXiv:1909.03158 [nucl-ex].
  %%CITATION = ARXIV:1909.03158;%%
  %1 citations counted in INSPIRE as of 01 Feb 2020

%\cite{Acharya:2019lkh}
\bibitem{Acharya:2019lkh}
  S.~Acharya {\it et al.} [ALICE Collaboration],
  %``Centrality and transverse momentum dependence of inclusive J/$\psi$ production at midrapidity in Pb-Pb collisions at $\sqrt{s_{\rm NN}}$ = 5.02 TeV,''
  arXiv:1910.14404 [nucl-ex].
  %%CITATION = ARXIV:1910.14404;%%


%-- LHC Upsilon
%\cite{Chatrchyan:2011pe}
\bibitem{Chatrchyan:2011pe}
  S.~Chatrchyan {\it et al.}  [CMS Collaboration],
  %``Suppression of Upsilon excited states in PbPb collisions at a
  %nucleon-nucleon centre-of-mass energy of 2.76 TeV,''
  Phys.\ Rev.\ Lett.\  {\bf 107}, 052302 (2011).
  %%CITATION = PRLTA,107,052302;%%

%\cite{Khachatryan:2016xxp}
\bibitem{Khachatryan:2016xxp} 
  V.~Khachatryan {\it et al.} [CMS Collaboration],
  %``Suppression of $\Upsilon(1S), \Upsilon(2S)$ and $\Upsilon(3S)$ production in PbPb collisions at $\sqrt{s_{\rm NN}}$ = 2.76 TeV,''
  Phys.\ Lett.\ B {\bf 770}, 357 (2017)
  doi:10.1016/j.physletb.2017.04.031
  [arXiv:1611.01510 [nucl-ex]].
  %%CITATION = doi:10.1016/j.physletb.2017.04.031;%%
  %17 citations counted in INSPIRE as of 18 Aug 2017

%\cite{Sirunyan:2017lzi}
\bibitem{Sirunyan:2017lzi} 
  A.~M.~Sirunyan {\it et al.} [CMS Collaboration],
  %``Suppression of excited Upsilon states relative to the ground state in PbPb collisions at sqrt(sNN) = 5.02 TeV,''
  arXiv:1706.05984 [hep-ex].
  %%CITATION = ARXIV:1706.05984;%%
  %2 citations counted in INSPIRE as of 29 Aug 2017

%--- CNM experiments
%\cite{Adare:2007gn}
\bibitem{Adare:2007gn}
  A.~Adare {\it et al.} [PHENIX Collaboration],
  %``Cold Nuclear Matter Effects on J/Psi as Constrained by Deuteron-Gold Measurements at s(NN)**(1/2) = 200-GeV,''
  Phys.\ Rev.\ C {\bf 77} (2008) 024912
   Erratum: [Phys.\ Rev.\ C {\bf 79} (2009) 059901]
  doi:10.1103/PhysRevC.77.024912, 10.1103/PhysRevC.79.059901
  [arXiv:0903.4845 [nucl-ex], arXiv:0711.3917 [nucl-ex]].
  %%CITATION = doi:10.1103/PhysRevC.77.024912, 10.1103/PhysRevC.79.059901;%%
  %158 citations counted in INSPIRE as of 01 Feb 2020

%\cite{Adler:2005ph}
\bibitem{Adler:2005ph}
  S.~S.~Adler {\it et al.} [PHENIX Collaboration],
  %``J/psi production and nuclear effects for d+Au and p+p collisions at s(NN)**(1/2) = 200-GeV,''
  Phys.\ Rev.\ Lett.\  {\bf 96} (2006) 012304
  doi:10.1103/PhysRevLett.96.012304
  [nucl-ex/0507032].
  %%CITATION = doi:10.1103/PhysRevLett.96.012304;%%
  %173 citations counted in INSPIRE as of 01 Feb 2020

%\cite{Adare:2010fn}
\bibitem{Adare:2010fn}
  A.~Adare {\it et al.} [PHENIX Collaboration],
  %``Cold Nuclear Matter Effects on $J/\psi$ Yields as a Function of Rapidity and Nuclear Geometry in Deuteron-Gold Collisions at $\sqrt{s_{NN}}=200$ GeV,''
  Phys.\ Rev.\ Lett.\  {\bf 107} (2011) 142301
  doi:10.1103/PhysRevLett.107.142301
  [arXiv:1010.1246 [nucl-ex]].
  %%CITATION = doi:10.1103/PhysRevLett.107.142301;%%
  %145 citations counted in INSPIRE as of 01 Feb 2020

%\cite{Adamczyk:2016dhc}
\bibitem{Adamczyk:2016dhc}
  L.~Adamczyk {\it et al.} [STAR Collaboration],
  %``$\rm{J}/\psi$ production at low transverse momentum in p+p and d+Au collisions at $\sqrt{s_{NN}}$ = 200 GeV,''
  Phys.\ Rev.\ C {\bf 93} (2016) no.6,  064904
  doi:10.1103/PhysRevC.93.064904
  [arXiv:1602.02212 [nucl-ex]].
  %%CITATION = doi:10.1103/PhysRevC.93.064904;%%
  %8 citations counted in INSPIRE as of 01 Feb 2020

%\cite{Abelev:2013yxa}
\bibitem{Abelev:2013yxa} 
  B.~B.~Abelev {\it et al.} [ALICE Collaboration],
  %``$J/\psi$ production and nuclear effects in p-Pb collisions at $\sqrt{S_{NN}}$ = 5.02 TeV,''
  JHEP {\bf 1402}, 073 (2014)
  doi:10.1007/JHEP02(2014)073
  [arXiv:1308.6726 [nucl-ex]].
  %%CITATION = doi:10.1007/JHEP02(2014)073;%%
  %148 citations counted in INSPIRE as of 29 Aug 2017

%\cite{Abelev:2014zpa}
\bibitem{Abelev:2014zpa} 
  B.~B.~Abelev {\it et al.} [ALICE Collaboration],
  %``Suppression of $\psi$(2S) production in p-Pb collisions at $\sqrt{s_{\rm NN}}$ = 5.02 TeV,''
  JHEP {\bf 1412}, 073 (2014)
  doi:10.1007/JHEP12(2014)073
  [arXiv:1405.3796 [nucl-ex]].
  %%CITATION = doi:10.1007/JHEP12(2014)073;%%
  %78 citations counted in INSPIRE as of 29 Aug 2017

%\cite{Abelev:2014oea}
\bibitem{Abelev:2014oea} 
  B.~B.~Abelev {\it et al.} [ALICE Collaboration],
  %``Production of inclusive $\Upsilon$(1S) and $\Upsilon$(2S) in p-Pb collisions at $\mathbf{\sqrt{s_{{\rm NN}}} = 5.02}$ TeV,''
  Phys.\ Lett.\ B {\bf 740}, 105 (2015)
  doi:10.1016/j.physletb.2014.11.041
  [arXiv:1410.2234 [nucl-ex]].
  %%CITATION = doi:10.1016/j.physletb.2014.11.041;%%
  %51 citations counted in INSPIRE as of 29 Aug 2017

%\cite{Adam:2015iga}
\bibitem{Adam:2015iga} 
  J.~Adam {\it et al.} [ALICE Collaboration],
  %``Rapidity and transverse-momentum dependence of the inclusive J/$\psi$ nuclear modification factor in p-Pb collisions at $ \sqrt{s_{N\ N}} =$ 5.02 TeV,''
  JHEP {\bf 1506}, 055 (2015)
  doi:10.1007/JHEP06(2015)055
  [arXiv:1503.07179 [nucl-ex]].
  %%CITATION = doi:10.1007/JHEP06(2015)055;%%
  %53 citations counted in INSPIRE as of 29 Aug 2017

%\cite{Adam:2015jsa}
\bibitem{Adam:2015jsa} 
  J.~Adam {\it et al.} [ALICE Collaboration],
  %``Centrality dependence of inclusive J/ψ production in p-Pb collisions at $ \sqrt{s_{\mathrm{NN}}}=5.02 $ TeV,''
  JHEP {\bf 1511}, 127 (2015)
  doi:10.1007/JHEP11(2015)127
  [arXiv:1506.08808 [nucl-ex]].
  %%CITATION = doi:10.1007/JHEP11(2015)127;%%
  %39 citations counted in INSPIRE as of 29 Aug 2017

%\cite{Adamova:2017uhu}
\bibitem{Adamova:2017uhu}
  D.~Adamová {\it et al.} [ALICE Collaboration],
  %``J/$\psi$ production as a function of charged-particle pseudorapidity density in p-Pb collisions at $\sqrt{s_{\rm NN}} = 5.02$ TeV,''
  Phys.\ Lett.\ B {\bf 776} (2018) 91
  doi:10.1016/j.physletb.2017.11.008
  [arXiv:1704.00274 [nucl-ex]].
  %%CITATION = doi:10.1016/j.physletb.2017.11.008;%%
  %17 citations counted in INSPIRE as of 01 Feb 2020

%\cite{Acharya:2018yud}
\bibitem{Acharya:2018yud}
  S.~Acharya {\it et al.} [ALICE Collaboration],
  %``Prompt and non-prompt $\hbox {J}/\psi $ production and nuclear modification at mid-rapidity in p–Pb collisions at $\mathbf{\sqrt{{ s}_{\text {NN}}}= 5.02}$  TeV,''
  Eur.\ Phys.\ J.\ C {\bf 78} (2018) no.6,  466
  doi:10.1140/epjc/s10052-018-5881-2
  [arXiv:1802.00765 [nucl-ex]].
  %%CITATION = doi:10.1140/epjc/s10052-018-5881-2;%%
  %4 citations counted in INSPIRE as of 01 Feb 2020

%\cite{Acharya:2018kxc}
\bibitem{Acharya:2018kxc}
  S.~Acharya {\it et al.} [ALICE Collaboration],
  %``Inclusive J/$\psi$ production at forward and backward rapidity in p-Pb collisions at $\sqrt{s_{\rm NN}}$ = 8.16 TeV,''
  JHEP {\bf 1807} (2018) 160
  doi:10.1007/JHEP07(2018)160
  [arXiv:1805.04381 [nucl-ex]].
  %%CITATION = doi:10.1007/JHEP07(2018)160;%%
  %10 citations counted in INSPIRE as of 01 Feb 2020

%\cite{Sirunyan:2017mzd}
\bibitem{Sirunyan:2017mzd} 
  A.~M.~Sirunyan {\it et al.} [CMS Collaboration],
  %``Measurement of prompt and nonprompt $\mathrm{J}/{\psi }$ production in $\mathrm {p}\mathrm {p}$ and $\mathrm {p}\mathrm {Pb}$ collisions at $\sqrt{s_{\mathrm {NN}}} =5.02\,\text {TeV} $,''
  Eur.\ Phys.\ J.\ C {\bf 77}, no. 4, 269 (2017)
  doi:10.1140/epjc/s10052-017-4828-3
  [arXiv:1702.01462 [nucl-ex]].
  %%CITATION = doi:10.1140/epjc/s10052-017-4828-3;%%
  %2 citations counted in INSPIRE as of 29 Aug 2017

%\cite{Aad:2015ddl}
\bibitem{Aad:2015ddl}
  G.~Aad {\it et al.} [ATLAS Collaboration],
  %``Measurement of differential $J/\psi$ production cross sections and forward-backward ratios in p + Pb collisions with the ATLAS detector,''
  Phys.\ Rev.\ C {\bf 92} (2015) no.3,  034904
  doi:10.1103/PhysRevC.92.034904
  [arXiv:1505.08141 [hep-ex]].
  %%CITATION = doi:10.1103/PhysRevC.92.034904;%%
  %43 citations counted in INSPIRE as of 01 Feb 2020

%\cite{TheATLAScollaboration:2015zdl}
\bibitem{TheATLAScollaboration:2015zdl} 
  The ATLAS collaboration,
  %``Measurement of $\Upsilon(\mathrm{nS})$ production with $p$+Pb collisions at $\sqrt{s_{\rm NN}} = 5.02~\mathrm {TeV}$ and $pp$ collisions at $\sqrt{s} = 2.76~\mathrm {TeV}$,''
  ATLAS-CONF-2015-050.
  %%CITATION = ATLAS-CONF-2015-050;%%
  %8 citations counted in INSPIRE as of 29 Aug 2017

%\cite{Aaij:2013zxa}
\bibitem{Aaij:2013zxa} 
  R.~Aaij {\it et al.} [LHCb Collaboration],
  %``Study of $J/\psi$ production and cold nuclear matter effects in $pPb$ collisions at $\sqrt{s_{NN}} = 5$ TeV,''
  JHEP {\bf 1402}, 072 (2014)
  doi:10.1007/JHEP02(2014)072
  [arXiv:1308.6729 [nucl-ex]].
  %%CITATION = doi:10.1007/JHEP02(2014)072;%%
  %122 citations counted in INSPIRE as of 29 Aug 2017


%%--  D Meson Experiment, Experimental Observations
%%\cite{Knospe:2012qf}
%\bibitem{Knospe:2012qf}
%  A.~Knospe,
%  %``Yield and suppression of electrons from open heavy-flavor decays in
%  %heavy-ion collisions,''.
%  [arXiv:1201.0242]
%  %%CITATION = ARXIV:1201.0242;%%
%
%%\cite{Adare:2006nq}
%\bibitem{Adare:2006nq}
%  A.~Adare {\it et al.}  [PHENIX Collaboration],
%  %``Energy Loss and Flow of Heavy Quarks in Au$+$Au Collisions at \sqrt{s_{NN}} =
%  %200 GeV,''
%  Phys.\ Rev.\ Lett.\  {\bf 98}, 172301 (2007).
%  %%CITATION = PRLTA,98,172301;%%
%
%%\cite{Abelev:2006db}
%\bibitem{Abelev:2006db}
%  B.~I.~Abelev {\it et al.}  [STAR Collaboration],
%  %``Erratum: Transverse momentum and centrality dependence of high-\pt\
%  %non-photonic electron suppression in Au$+$Au collisions at \sqrtsNN\ = 200
%  %GeV,''
%  Phys.\ Rev.\ Lett.\  {\bf 98}, 192301 (2007)
%  [Erratum-ibid.\  {\bf 106}, 159902 (2011)].
%  %%CITATION = PRLTA,98,192301;%%
%
%%\cite{Dainese:2011vb}
%\bibitem{Dainese:2011vb}
%  A.~Dainese,
%  %``Heavy-flavour production in Pb-Pb collisions at the LHC, measured with the ALICE detector,''
%  J.\ Phys.\ G G {\bf 38}, 124032 (2011).
%  %%CITATION = ARXIV:1106.4042;%%

%--Other calculations
%-- Blaizot
%%\cite{Beraudo:2007ky}
%\bibitem{Beraudo:2007ky}
%  A.~Beraudo, J.-P.~Blaizot and C.~Ratti,
%  %``Real and imaginary-time Q anti-Q correlators in a thermal medium,''
%  Nucl.\ Phys.\ A {\bf 806} (2008) 312
%  doi:10.1016/j.nuclphysa.2008.03.001
%  [arXiv:0712.4394 [nucl-th]].
%  %%CITATION = doi:10.1016/j.nuclphysa.2008.03.001;%%
%  %197 citations counted in INSPIRE as of 28 Dec 2019

%%\cite{Beraudo:2008fx}
%\bibitem{Beraudo:2008fx}
%  A.~Beraudo, J.~P.~Blaizot and C.~Ratti,
%  %``Real and imaginary-time quarkonium correlators in a hot plasma,''
%  PoS CONFINEMENT {\bf 8} (2008) 117
%  doi:10.22323/1.077.0117
%  [arXiv:0812.1130 [hep-ph]].
%  %%CITATION = doi:10.22323/1.077.0117;%%
%  %8 citations counted in INSPIRE as of 27 Jan 2020

%%\cite{Beraudo:2009zz}
%\bibitem{Beraudo:2009zz}
%  A.~Beraudo, J.~P.~Blaizot, G.~Garberoglio and P.~Faccioli,
%  %``Heavy-quarks in the QGP: Study of medium effects through euclidean propagators and spectral functions,''
%  Nucl.\ Phys.\ A {\bf 830} (2009) 319C
%  doi:10.1016/j.nuclphysa.2009.10.027
%  Quark Matter 2009,
%  [arXiv:0907.1797 [hep-ph]].
%  %%CITATION = doi:10.1016/j.nuclphysa.2009.10.027;%%
%  %5 citations counted in INSPIRE as of 28 Dec 2019

%%\cite{Beraudo:2010tw}
%\bibitem{Beraudo:2010tw}
%  A.~Beraudo, J.~P.~Blaizot, P.~Faccioli and G.~Garberoglio,
%  %``A path integral for heavy-quarks in a hot plasma,''
%  Nucl.\ Phys.\ A {\bf 846} (2010) 104
%  doi:10.1016/j.nuclphysa.2010.06.007
%  [arXiv:1005.1245 [hep-ph]].
%  %%CITATION = doi:10.1016/j.nuclphysa.2010.06.007;%%
%  %19 citations counted in INSPIRE as of 27 Jan 2020

%--Grandchamp Rapp
%\cite{Grandchamp:2001pf}
\bibitem{Grandchamp:2001pf}
  L.~Grandchamp and R.~Rapp,
  %``Thermal versus direct J / Psi production in ultrarelativistic heavy ion collisions,''
  Phys.\ Lett.\ B {\bf 523} (2001) 60
  doi:10.1016/S0370-2693(01)01311-9
  [hep-ph/0103124].
  %%CITATION = doi:10.1016/S0370-2693(01)01311-9;%%
  %198 citations counted in INSPIRE as of 28 Jan 2020

%\cite{Grandchamp:2002wp}
\bibitem{Grandchamp:2002wp}
  L.~Grandchamp and R.~Rapp,
  %``Charmonium suppression and regeneration from SPS to RHIC,''
  Nucl.\ Phys.\ A {\bf 709} (2002) 415
  doi:10.1016/S0375-9474(02)01027-8
  [hep-ph/0205305].
  %%CITATION = doi:10.1016/S0375-9474(02)01027-8;%%
  %202 citations counted in INSPIRE as of 28 Jan 2020

%\cite{Grandchamp:2003}
\bibitem{Grandchamp:2003}
  L.~Grandchamp, R.~Rapp and G.~E.~Brown,
  %``In medium effects on charmonium production in heavy ion collisions,''
  Phys.\ Rev.\ Lett.\  {\bf 92} (2004) 212301
  doi:10.1103/PhysRevLett.92.212301
  [hep-ph/0306077].
  %%CITATION = doi:10.1103/PhysRevLett.92.212301;%%
  %182 citations counted in INSPIRE as of 28 Jan 2020

%\cite{Grandchamp:2005yw}
\bibitem{Grandchamp:2005yw}
  L.~Grandchamp, S.~Lumpkins, D.~Sun, H.~van Hees and R.~Rapp,
  %``Bottomonium production at RHIC and CERN LHC,''
  Phys.\ Rev.\ C {\bf 73} (2006) 064906
  doi:10.1103/PhysRevC.73.064906
  [hep-ph/0507314].
  %%CITATION = doi:10.1103/PhysRevC.73.064906;%%
  %73 citations counted in INSPIRE as of 28 Jan 2020
% Rapp T-matrix and later
%\cite{Mannarelli:2005pz}
\bibitem{Mannarelli:2005pz}
  M.~Mannarelli and R.~Rapp,
  %``Hadronic modes and quark properties in the quark-gluon plasma,''
  Phys.\ Rev.\ C {\bf 72} (2005) 064905
  doi:10.1103/PhysRevC.72.064905
  [hep-ph/0505080].
  %%CITATION = doi:10.1103/PhysRevC.72.064905;%%
  %77 citations counted in INSPIRE as of 28 Jan 2020

%%\bibitem{Mannarelli:2005} 
%\bibitem{Mannarelli:2005}
%  M.~Mannarelli and R.~Rapp,
%  %``Scattering of quark-quasiparticles in the quark-gluon plasma,''
%  Nucl.\ Phys.\ A {\bf 774} (2006) 761
%  doi:10.1016/j.nuclphysa.2006.06.016
%  [hep-ph/0509310].
%  %%CITATION = doi:10.1016/j.nuclphysa.2006.06.016;%%
%  %8 citations counted in INSPIRE as of 28 Jan 2020
%
%%\bibitem{Cabrera:2006}
%\bibitem{Cabrera:2006}
%  D.~Cabrera and R.~Rapp,
%  %``Q anti-Q modes in the quark-gluon plasma,''
%  Eur.\ Phys.\ J.\ A {\bf 31} (2007) 858
%  doi:10.1140/epja/i2006-10237-y
%  [hep-ph/0610254].
%  %%CITATION = doi:10.1140/epja/i2006-10237-y;%%
%  %15 citations counted in INSPIRE as of 28 Jan 2020
 
%\cite{Cabrera:2006wh}
\bibitem{Cabrera:2006wh}
  D.~Cabrera and R.~Rapp,
  %``T-Matrix Approach to Quarkonium Correlation Functions in the QGP,''
  Phys.\ Rev.\ D {\bf 76} (2007) 114506
  doi:10.1103/PhysRevD.76.114506
  [hep-ph/0611134].
  %%CITATION = doi:10.1103/PhysRevD.76.114506;%%
  %105 citations counted in INSPIRE as of 28 Jan 2020

%\cite{vanHees:2007me}
\bibitem{vanHees:2007me}
  H.~van Hees, M.~Mannarelli, V.~Greco and R.~Rapp,
  %``Nonperturbative heavy-quark diffusion in the quark-gluon plasma,''
  Phys.\ Rev.\ Lett.\  {\bf 100} (2008) 192301
  doi:10.1103/PhysRevLett.100.192301
  [arXiv:0709.2884 [hep-ph]].
  %%CITATION = doi:10.1103/PhysRevLett.100.192301;%%
  %214 citations counted in INSPIRE as of 28 Jan 2020

%\cite{Zhao:2007}
\bibitem{Zhao:2007}
  X.~Zhao and R.~Rapp,
  %``Transverse Momentum Spectra of $J/\psi$ in Heavy-Ion Collisions,''
  Phys.\ Lett.\ B {\bf 664} (2008) 253
  doi:10.1016/j.physletb.2008.03.068
  [arXiv:0712.2407 [hep-ph]].
  %%CITATION = doi:10.1016/j.physletb.2008.03.068;%%
  %137 citations counted in INSPIRE as of 28 Jan 2020
  
%\cite{Zhao:2008vu}
\bibitem{Zhao:2008vu}
  X.~Zhao and R.~Rapp,
  %``Charmonium Production at High $p_t$ at RHIC,''
  [arXiv:0806.1239 [nucl-th]].
  %%CITATION = ARXIV:0806.1239;%%

%\cite{Riek:2010fk}
\bibitem{Riek:2010fk}
  F.~Riek and R.~Rapp,
  %``Quarkonia and Heavy-Quark Relaxation Times in the Quark-Gluon Plasma,''
  Phys.\ Rev.\ C {\bf 82} (2010) 035201
  doi:10.1103/PhysRevC.82.035201
  [arXiv:1005.0769 [hep-ph]].
  %%CITATION = doi:10.1103/PhysRevC.82.035201;%%
  %103 citations counted in INSPIRE as of 28 Jan 2020

%\cite{Zhao:2010nk}
\bibitem{Zhao:2010nk}
  X.~Zhao and R.~Rapp,
  %``Charmonium in Medium: From Correlators to Experiment,''
  Phys.\ Rev.\ C {\bf 82} (2010) 064905
  doi:10.1103/PhysRevC.82.064905
  [arXiv:1008.5328 [hep-ph]].
  %%CITATION = doi:10.1103/PhysRevC.82.064905;%%
  %139 citations counted in INSPIRE as of 28 Jan 2020

%%\cite{Rapp:2011dz}
%\bibitem{Rapp:2011dz}
%  R.~Rapp,
%  %``Theory of Heavy Flavor in the Quark-Gluon Plasma,''
%  Nucl.\ Phys.\ A {\bf 862-863} (2011) 153
%  doi:10.1016/j.nuclphysa.2011.05.034
%  [arXiv:1101.5194 [nucl-th]].
%  %%CITATION = doi:10.1016/j.nuclphysa.2011.05.034;%%

%\cite{Emerick:2011xu}
\bibitem{Emerick:2011xu}
  A.~Emerick, X.~Zhao and R.~Rapp,
  %``Bottomonia in the Quark-Gluon Plasma and their Production at RHIC and LHC,''
  Eur.\ Phys.\ J.\ A {\bf 48} (2012) 72
  doi:10.1140/epja/i2012-12072-y
  [arXiv:1111.6537 [hep-ph]].
  %%CITATION = doi:10.1140/epja/i2012-12072-y;%%
  %134 citations counted in INSPIRE as of 28 Jan 2020

%\cite{Zhao:2012gc}
\bibitem{Zhao:2012gc} 
  X.~Zhao, A.~Emerick and R.~Rapp,
  %``In-Medium Quarkonia at SPS, RHIC and LHC,''
  Nucl.\ Phys.\ A {\bf 904-905}, 611c (2013)
  doi:10.1016/j.nuclphysa.2013.02.088
  [arXiv:1210.6583 [hep-ph]].
  %%CITATION = doi:10.1016/j.nuclphysa.2013.02.088;%%
  %23 citations counted in INSPIRE as of 27 Aug 2017

%\cite{Liu:2015ypa}
\bibitem{Liu:2015ypa}
  S.~Y.~F.~Liu and R.~Rapp,
  %``An in-medium heavy-quark potential from the $Q\bar{Q}$ free energy,''
  Nucl.\ Phys.\ A {\bf 941} (2015) 179
  doi:10.1016/j.nuclphysa.2015.07.001
  [arXiv:1501.07892 [hep-ph]].
  %%CITATION = doi:10.1016/j.nuclphysa.2015.07.001;%%
  %18 citations counted in INSPIRE as of 28 Jan 2020

%\cite{Du:2017qkv}
\bibitem{Du:2017qkv} 
  X.~Du, R.~Rapp and M.~He,
  %``Color Screening and Regeneration of Bottomonia at RHIC and the LHC,''
  arXiv:1706.08670 [hep-ph].
  %%CITATION = ARXIV:1706.08670;%%
  %1 citations counted in INSPIRE as of 25 Aug 2017

%\cite{Liu:2017qah}
\bibitem{Liu:2017qah}
  S.~Y.~F.~Liu and R.~Rapp,
  %``$T$-matrix Approach to Quark-Gluon Plasma,''
  Phys.\ Rev.\ C {\bf 97} (2018) no.3,  034918
  doi:10.1103/PhysRevC.97.034918
  [arXiv:1711.03282 [nucl-th]].
  %%CITATION = doi:10.1103/PhysRevC.97.034918;%%
  %21 citations counted in INSPIRE as of 02 Feb 2020

%%\cite{Liu:2018syc}
%\bibitem{Liu:2018syc}
%  S.~Y.~F.~Liu, M.~He and R.~Rapp,
%  %``Probing the in-Medium QCD Force by Open Heavy-Flavor Observables,''
%  Phys.\ Rev.\ C {\bf 99} (2019) no.5,  055201
%  doi:10.1103/PhysRevC.99.055201
%  [arXiv:1806.05669 [nucl-th]].
%  %%CITATION = doi:10.1103/PhysRevC.99.055201;%%
%  %5 citations counted in INSPIRE as of 02 Feb 2020

%%\cite{Du:2018wsj}
%\bibitem{Du:2018wsj}
%  X.~Du and R.~Rapp,
%  %``In-Medium Charmonium Production in Proton-Nucleus Collisions,''
%  JHEP {\bf 1903} (2019) 015
%  doi:10.1007/JHEP03(2019)015
%  [arXiv:1808.10014 [nucl-th]].
%  %%CITATION = doi:10.1007/JHEP03(2019)015;%%
%  %18 citations counted in INSPIRE as of 02 Feb 2020

%\cite{Du:2019tjf}
\bibitem{Du:2019tjf}
  X.~Du, S.~Y.~F.~Liu and R.~Rapp,
  %``Extraction of the Heavy-Quark Potential from Bottomonium Observables in Heavy-Ion Collisions,''
  Phys.\ Lett.\ B {\bf 796} (2019) 20
  doi:10.1016/j.physletb.2019.07.032
  [arXiv:1904.00113 [nucl-th]].
  %%CITATION = doi:10.1016/j.physletb.2019.07.032;%%
  %3 citations counted in INSPIRE as of 02 Feb 2020

%%\cite{He:2019vgs}
%\bibitem{He:2019vgs}
%  M.~He and R.~Rapp,
%  %``Hadronization and Charm-Hadron Ratios in Heavy-Ion Collisions,''
%  Phys.\ Rev.\ Lett.\  {\bf 124} (2020) 042301
%  doi:10.1103/PhysRevLett.124.042301
%  [arXiv:1905.09216 [nucl-th]].
%  %%CITATION = doi:10.1103/PhysRevLett.124.042301;%%
%  %5 citations counted in INSPIRE as of 02 Feb 2020

%-- Recombination, other
%\cite{Greco:2003vf}
\bibitem{Greco:2003vf}
  V.~Greco, C.~M.~Ko and R.~Rapp,
  %``Quark coalescence for charmed mesons in ultrarelativistic heavy ion collisions,''
  Phys.\ Lett.\ B {\bf 595} (2004) 202
  doi:10.1016/j.physletb.2004.06.064
  [nucl-th/0312100].
  %%CITATION = doi:10.1016/j.physletb.2004.06.064;%%
  %281 citations counted in INSPIRE as of 02 Feb 2020

%\cite{Zhang:2002ug}
\bibitem{Zhang:2002ug}
  B.~Zhang, C.~M.~Ko, B.~A.~Li, Z.~W.~Lin and S.~Pal,
  %``J / psi production in relativistic heavy ion collisions from a multiphase transport model,''
  Phys.\ Rev.\ C {\bf 65} (2002) 054909
  doi:10.1103/PhysRevC.65.054909
  [nucl-th/0201038].
  %%CITATION = doi:10.1103/PhysRevC.65.054909;%%
  %72 citations counted in INSPIRE as of 02 Feb 2020

%\cite{Thews:2006ia}
\bibitem{Thews:2006ia} 
  R.~L.~Thews,
  %``Quarkonium production via recombination,''
  Nucl.\ Phys.\ A {\bf 783}, 301 (2007)
  doi:10.1016/j.nuclphysa.2006.11.084
  [hep-ph/0609121].
  %%CITATION = doi:10.1016/j.nuclphysa.2006.11.084;%%
  %15 citations counted in INSPIRE as of 29 Aug 2017

%\cite{Bass:2006vu}
\bibitem{Bass:2006vu}
  S.~A.~Bass,
  %``Review of parton recombination models,''
  J.\ Phys.\ Conf.\ Ser.\  {\bf 50} 279, (2006).
  %%CITATION = 00462,50,279;%%

%\cite{Yan:2006ve}
\bibitem{Yan:2006ve}
  L.~Yan, P.~Zhuang and N.~Xu,
  %``Competition between J / psi suppression and regeneration in quark-gluon plasma,''
  Phys.\ Rev.\ Lett.\  {\bf 97} (2006) 232301
  doi:10.1103/PhysRevLett.97.232301
  [nucl-th/0608010].
  %%CITATION = doi:10.1103/PhysRevLett.97.232301;%%
  %159 citations counted in INSPIRE as of 02 Feb 2020

%\cite{Capella:2007jv}
\bibitem{Capella:2007jv}
  A.~Capella, L.~Bravina, E.~G.~Ferreiro, A.~B.~Kaidalov, K.~Tywoniuk and E.~Zabrodin,
  %``Charmonium dissociation and recombination at RHIC and LHC,''
  Eur.\ Phys.\ J.\ C {\bf 58} (2008) 437
  doi:10.1140/epjc/s10052-008-0772-6
  [arXiv:0712.4331 [hep-ph]].
  %%CITATION = doi:10.1140/epjc/s10052-008-0772-6;%%
  %72 citations counted in INSPIRE as of 02 Feb 2020

%\cite{Bravina:2008su}
\bibitem{Bravina:2008su}
  L.~Bravina, A.~Capella, E.~G.~Ferreiro, A.~B.~Kaidalov, K.~Tywoniuk and E.~Zabrodin,
  %``Can the RHIC J/psi puzzle(s) be settled at LHC?,''
  Eur.\ Phys.\ J.\ C {\bf 61} (2009) 865
  doi:10.1140/epjc/s10052-009-0906-5
  [arXiv:0811.0790 [hep-ph]].
  %%CITATION = doi:10.1140/epjc/s10052-009-0906-5;%%
  %4 citations counted in INSPIRE as of 02 Feb 2020

%\cite{Peng:2010zza}
\bibitem{Peng:2010zza}
  R.~Peng and C.~B.~Yang,
  %``J/psi production in Au + Au collisions at s(NN)**(1/2) = 200-GeV in the recombination model,''
  Nucl.\ Phys.\ A {\bf 837} (2010) 54.
  doi:10.1016/j.nuclphysa.2010.02.006
  %%CITATION = doi:10.1016/j.nuclphysa.2010.02.006;%%
  %6 citations counted in INSPIRE as of 02 Feb 2020

%\cite{Ferreiro:2012rq}
\bibitem{Ferreiro:2012rq}
  E.~G.~Ferreiro,
  %``Charmonium dissociation and recombination at LHC: Revisiting comovers,''
  Phys.\ Lett.\ B {\bf 731} (2014) 57
  doi:10.1016/j.physletb.2014.02.011
  [arXiv:1210.3209 [hep-ph]].
  %%CITATION = doi:10.1016/j.physletb.2014.02.011;%%
  %87 citations counted in INSPIRE as of 02 Feb 2020

%%\cite{Chaudhuri:2008if}
%\bibitem{Chaudhuri:2008if} 
%  A.~K.~Chaudhuri,
%  %``J / psi suppression and p(T) spectra in RHIC and LHC energy collisions,''
%  Eur.\ Phys.\ J.\ C {\bf 61}, 331 (2009)
%  doi:10.1140/epjc/s10052-009-0963-9
%  [arXiv:0808.2702 [nucl-th]].
%  %%CITATION = doi:10.1140/epjc/s10052-009-0963-9;%%
%  %4 citations counted in INSPIRE as of 29 Aug 2017

%-- Thermal
%\cite{BraunMunzinger:2000ep}
\bibitem{BraunMunzinger:2000ep}
  P.~Braun-Munzinger and J.~Stachel,
  %``On charm production near the phase boundary,''
  Nucl.\ Phys.\ A {\bf 690} (2001) 119
  doi:10.1016/S0375-9474(01)00936-8
  [nucl-th/0012064].
  %%CITATION = doi:10.1016/S0375-9474(01)00936-8;%%
  %110 citations counted in INSPIRE as of 07 Feb 2020

%\cite{Thews:2000rj}
\bibitem{Thews:2000rj}
  R.~L.~Thews, M.~Schroedter and J.~Rafelski,
  %``Enhanced $J/\psi$ production in deconfined quark matter,''
  Phys.\ Rev.\ C {\bf 63} (2001) 054905
  doi:10.1103/PhysRevC.63.054905
  [hep-ph/0007323].
  %%CITATION = doi:10.1103/PhysRevC.63.054905;%%
  %504 citations counted in INSPIRE as of 02 Feb 2020

\bibitem{Yan:2006}
Yan, L., Zhuang, P., and Xu, N. . 
%``J/ψ production in quark-gluon plasma.''
Physical Review letters, {\bf{97}} 23,(2006) 232301.

%\cite{Kostyuk:2003kt}
\bibitem{Kostyuk:2003kt}
  A.~P.~Kostyuk, M.~I.~Gorenstein, H.~Stoecker and W.~Greiner,
  %``Charm coalescence at RHIC,''
  Phys.\ Rev.\ C {\bf 68} (2003) 041902
  doi:10.1103/PhysRevC.68.041902
  [hep-ph/0305277].
  %%CITATION = doi:10.1103/PhysRevC.68.041902;%%
  %52 citations counted in INSPIRE as of 02 Feb 2020

%\cite{Andronic:2011yq}
\bibitem{Andronic:2011yq}
  A.~Andronic, P.~Braun-Munzinger, K.~Redlich and J.~Stachel,
  %``The thermal model on the verge of the ultimate test: particle production in Pb-Pb collisions at the LHC,''
  J.\ Phys.\ G {\bf 38} (2011) 124081
  doi:10.1088/0954-3899/38/12/124081
  [arXiv:1106.6321 [nucl-th]].
  %%CITATION = doi:10.1088/0954-3899/38/12/124081;%%
  %165 citations counted in INSPIRE as of 02 Feb 2020

%\cite{Gupta:2014ova}
\bibitem{Gupta:2014ova} 
  S.~Gupta and R.~Sharma,
  %``Thermalization of quarkonia at energies available at the CERN Large Hadron Collider,''
  Phys.\ Rev.\ C {\bf 89}, no. 5, 057901 (2014)
  doi:10.1103/PhysRevC.89.057901
  [arXiv:1401.2930 [nucl-th]].
  %%CITATION = doi:10.1103/PhysRevC.89.057901;%%
  %3 citations counted in INSPIRE as of 29 Aug 2017

%-- Wolschin
%\cite{Hoelck:2016tqf}
\bibitem{Hoelck:2016tqf}
  J.~Hoelck, F.~Nendzig and G.~Wolschin,
  %``In-medium $\Upsilon$ suppression and feed-down in UU and PbPb collisions,''
  Phys.\ Rev.\ C {\bf 95} (2017) no.2,  024905
  doi:10.1103/PhysRevC.95.024905
  [arXiv:1602.00019 [hep-ph]].
  %%CITATION = doi:10.1103/PhysRevC.95.024905;%%
  %20 citations counted in INSPIRE as of 28 Dec 2019

%\cite{Hoelck:2017dby}
\bibitem{Hoelck:2017dby}
  J.~Hoelck and G.~Wolschin,
  %``Electromagnetic field effects on $\Upsilon$ -meson dissociation in PbPb collisions at LHC energies,''
  Eur.\ Phys.\ J.\ A {\bf 53} (2017) no.12,  241
  doi:10.1140/epja/i2017-12441-0
  [arXiv:1712.06871 [hep-ph]].
  %%CITATION = doi:10.1140/epja/i2017-12441-0;%%
  %7 citations counted in INSPIRE as of 28 Dec 2019

%%\cite{Wolschin:2018lpd}
%\bibitem{Wolschin:2018lpd}
%  G.~Wolschin,
%  %``Spectroscopy in the quark-gluon plasma with Bottomonia,''
%  PoS HardProbes {\bf 2018} (2018) 140.
%  doi:10.22323/1.345.0140
%  %%CITATION = doi:10.22323/1.345.0140;%%
%  %1 citations counted in INSPIRE as of 27 Jan 2020

%%\cite{Wolschin:2019ovv}
%\bibitem{Wolschin:2019ovv}
%  G.~Wolschin,
%  %``Bottomonia physics at RHIC and LHC,''
%  PoS CORFU {\bf 2018} (2019) 165.
%  doi:10.22323/1.347.0165
%  %%CITATION = doi:10.22323/1.347.0165;%%

%%\cite{Dinh:2019ajl}
%\bibitem{Dinh:2019ajl}
%  V.~H.~Dinh, J.~Hoelck and G.~Wolschin,
%  %``Hot-medium effects on $\Upsilon$ yields in pPb collisions at $\sqrt{s_\text{NN}}=8.16$ TeV,''
%  Phys.\ Rev.\ C {\bf 100} (2019) no.2,  024906
%  doi:10.1103/PhysRevC.100.024906
%  [arXiv:1903.12594 [hep-ph]].
%  %%CITATION = doi:10.1103/PhysRevC.100.024906;%%
%  %1 citations counted in INSPIRE as of 27 Jan 2020


%--Strickland
%\cite{Dumitru:2007hy}
\bibitem{Dumitru:2007hy}
  A.~Dumitru, Y.~Guo and M.~Strickland,
  %``The Heavy-quark potential in an anisotropic (viscous) plasma,''
  Phys.\ Lett.\ B {\bf 662} (2008) 37
  doi:10.1016/j.physletb.2008.02.048
  [arXiv:0711.4722 [hep-ph]].
  %%CITATION = doi:10.1016/j.physletb.2008.02.048;%%
  %98 citations counted in INSPIRE as of 28 Dec 2019

%\cite{Dumitru:2009ni}
\bibitem{Dumitru:2009ni}
  A.~Dumitru, Y.~Guo, A.~Mocsy and M.~Strickland,
  %``Quarkonium states in an anisotropic QCD plasma,''
  Phys.\ Rev.\ D {\bf 79} (2009) 054019
  doi:10.1103/PhysRevD.79.054019
  [arXiv:0901.1998 [hep-ph]].
  %%CITATION = doi:10.1103/PhysRevD.79.054019;%%
  %68 citations counted in INSPIRE as of 26 Jan 2020

%\cite{Dumitru:2009fy}
\bibitem{Dumitru:2009fy}
  A.~Dumitru, Y.~Guo and M.~Strickland,
  %``The Imaginary part of the static gluon propagator in an anisotropic (viscous) QCD plasma,''
  Phys.\ Rev.\ D {\bf 79} (2009) 114003
  doi:10.1103/PhysRevD.79.114003
  [arXiv:0903.4703 [hep-ph]].
  %%CITATION = doi:10.1103/PhysRevD.79.114003;%%
  %87 citations counted in INSPIRE as of 26 Jan 2020

%\cite{Strickland:2011mw}
\bibitem{Strickland:2011mw}
  M.~Strickland,
  %``Thermal $\upsilon_{1s}$ and chi_b1 suppression in $\sqrt{s_{NN}}=2.76$ TeV Pb-Pb collisions at the LHC,''
  Phys.\ Rev.\ Lett.\  {\bf 107} (2011) 132301
  doi:10.1103/PhysRevLett.107.132301
  [arXiv:1106.2571 [hep-ph]].
  %%CITATION = doi:10.1103/PhysRevLett.107.132301;%%
  %102 citations counted in INSPIRE as of 28 Dec 2019

%\cite{Strickland:2011aa}
\bibitem{Strickland:2011aa}
  M.~Strickland and D.~Bazow,
  %``Thermal Bottomonium Suppression at RHIC and LHC,''
  Nucl.\ Phys.\ A {\bf 879} (2012) 25
  doi:10.1016/j.nuclphysa.2012.02.003
  [arXiv:1112.2761 [nucl-th]].
  %%CITATION = doi:10.1016/j.nuclphysa.2012.02.003;%%
  %145 citations counted in INSPIRE as of 28 Dec 2019

%\cite{Margotta:2011ta}
\bibitem{Margotta:2011ta}
  M.~Margotta, K.~McCarty, C.~McGahan, M.~Strickland and D.~Yager-Elorriaga,
  %``Quarkonium states in a complex-valued potential,''
  Phys.\ Rev.\ D {\bf 83}, 105019 (2011)
   [Erratum-ibid.\ D {\bf 84}, 069902 (2011)].
  %%CITATION = ARXIV:1101.4651;%%

%%\cite{Strickland:2012cq}
%\bibitem{Strickland:2012cq}
%  M.~Strickland,
%  %``Thermal Bottomonium Suppression,''
%  AIP Conf.\ Proc.\  {\bf 1520} (2013) no.1,  179
%  doi:10.1063/1.4795953
%  [arXiv:1207.5327 [hep-ph]].
%  %%CITATION = doi:10.1063/1.4795953;%%
%  %26 citations counted in INSPIRE as of 26 Jan 2020

%%\cite{Strickland:2012as}
%\bibitem{Strickland:2012as} 
%  M.~Strickland,
%  %``Bottomonia in the Quark Gluon Plasma,''
%  J.\ Phys.\ Conf.\ Ser.\  {\bf 432}, 012015 (2013)
%  doi:10.1088/1742-6596/432/1/012015
%  [arXiv:1210.7512 [nucl-th]].
%  %%CITATION = doi:10.1088/1742-6596/432/1/012015;%%
%  %6 citations counted in INSPIRE as of 26 Jan 2020

%\cite{Machado:2013rta}
\bibitem{Machado:2013rta}
  C.~S.~Machado, F.~S.~Navarra, E.~G.~de Oliveira, J.~Noronha and M.~Strickland,
  %``Heavy quarkonium production in a strong magnetic field,''
  Phys.\ Rev.\ D {\bf 88} (2013) 034009
  doi:10.1103/PhysRevD.88.034009
  [arXiv:1305.3308 [hep-ph]].
  %%CITATION = doi:10.1103/PhysRevD.88.034009;%%
  %43 citations counted in INSPIRE as of 26 Jan 2020

%\cite{Alford:2013jva}
\bibitem{Alford:2013jva}
  J.~Alford and M.~Strickland,
  %``Charmonia and Bottomonia in a Magnetic Field,''
  Phys.\ Rev.\ D {\bf 88} (2013) 105017
  doi:10.1103/PhysRevD.88.105017
  [arXiv:1309.3003 [hep-ph]].
  %%CITATION = doi:10.1103/PhysRevD.88.105017;%%
  %62 citations counted in INSPIRE as of 26 Jan 2020

%\cite{Krouppa:2015yoa}
\bibitem{Krouppa:2015yoa}
  B.~Krouppa, R.~Ryblewski and M.~Strickland,
  %``Bottomonia suppression in 2.76 TeV Pb-Pb collisions,''
  Phys.\ Rev.\ C {\bf 92} (2015) no.6,  061901
  doi:10.1103/PhysRevC.92.061901
  [arXiv:1507.03951 [hep-ph]].
  %%CITATION = doi:10.1103/PhysRevC.92.061901;%%
  %53 citations counted in INSPIRE as of 28 Dec 2019

%\cite{Krouppa:2016jcl}
\bibitem{Krouppa:2016jcl}
  B.~Krouppa and M.~Strickland,
  %``Predictions for Bottomonia suppression in 5.023 TeV Pb-Pb collisions,''
  Universe {\bf 2} (2016) no.3,  16
  doi:10.3390/universe2030016
  [arXiv:1605.03561 [hep-ph]].
  %%CITATION = doi:10.3390/universe2030016;%%
  %45 citations counted in INSPIRE as of 26 Jan 2020

%%\cite{Krouppa:2017lsw}
%\bibitem{Krouppa:2017lsw}
%  B.~Krouppa, R.~Ryblewski and M.~Strickland,
%  %``Bottomonium suppression in heavy-ion collisions,''
%  Nucl.\ Phys.\ A {\bf 967} (2017) 604
%  doi:10.1016/j.nuclphysa.2017.05.073
%  [arXiv:1704.02361 [nucl-th]].
%  %%CITATION = doi:10.1016/j.nuclphysa.2017.05.073;%%
%  %10 citations counted in INSPIRE as of 28 Dec 2019

%\cite{Krouppa:2017jlg}
\bibitem{Krouppa:2017jlg}
  B.~Krouppa, A.~Rothkopf and M.~Strickland,
  %``Bottomonium suppression using a lattice QCD vetted potential,''
  Phys.\ Rev.\ D {\bf 97} (2018) no.1,  016017
  doi:10.1103/PhysRevD.97.016017
  [arXiv:1710.02319 [hep-ph]].
  %%CITATION = doi:10.1103/PhysRevD.97.016017;%%
  %36 citations counted in INSPIRE as of 28 Dec 2019

%%\cite{Krouppa:2018lkt}
%\bibitem{Krouppa:2018lkt}
%  B.~Krouppa, A.~Rothkopf and M.~Strickland,
%  %``Bottomonium suppression at RHIC and LHC,''
%  Nucl.\ Phys.\ A {\bf 982} (2019) 727
%  doi:10.1016/j.nuclphysa.2018.09.034
%  [arXiv:1807.07452 [hep-ph]].
%  %%CITATION = doi:10.1016/j.nuclphysa.2018.09.034;%%
%  %4 citations counted in INSPIRE as of 26 Jan 2020

%\cite{Bhaduri:2018iwr}
\bibitem{Bhaduri:2018iwr} 
  P.~P.~Bhaduri, N.~Borghini, A.~Jaiswal and M.~Strickland,
  %``Anisotropic escape mechanism and elliptic flow of Bottomonia,''
  Phys.\ Rev.\ C {\bf 100}, no. 5, 051901 (2019)
  doi:10.1103/PhysRevC.100.051901
  [arXiv:1809.06235 [hep-ph]].
  %%CITATION = doi:10.1103/PhysRevC.100.051901;%%
  %8 citations counted in INSPIRE as of 28 Dec 2019

%\cite{Boyd:2019arx}
\bibitem{Boyd:2019arx} 
  J.~Boyd, T.~Cook, A.~Islam and M.~Strickland,
  %``Heavy quarkonium suppression beyond the adiabatic limit,''
  Phys.\ Rev.\ D {\bf 100}, no. 7, 076019 (2019)
  doi:10.1103/PhysRevD.100.076019
  [arXiv:1905.05676 [hep-ph]].
  %%CITATION = doi:10.1103/PhysRevD.100.076019;%%
  %2 citations counted in INSPIRE as of 28 Dec 2019

%-- Song
%\cite{Song:2005yd}
\bibitem{Song:2005yd}
  T.~Song and S.~H.~Lee,
  %``Quarkonium-hadron interactions in perturbative QCD,''
  Phys.\ Rev.\ D {\bf 72} (2005) 034002
  doi:10.1103/PhysRevD.72.034002
  [hep-ph/0501252].
  %%CITATION = doi:10.1103/PhysRevD.72.034002;%%
  %35 citations counted in INSPIRE as of 28 Jan 2020

%\cite{Park:2007zza}
\bibitem{Park:2007zza}
  Y.~Park, K.~I.~Kim, T.~Song, S.~H.~Lee and C.~Y.~Wong,
  %``Widths of quarkonia in quark gluon plasma,''
  Phys.\ Rev.\ C {\bf 76} (2007) 044907
  doi:10.1103/PhysRevC.76.044907
  [arXiv:0704.3770 [hep-ph]].
  %%CITATION = doi:10.1103/PhysRevC.76.044907;%%
  %54 citations counted in INSPIRE as of 28 Jan 2020

%\cite{Song:2007gm}
\bibitem{Song:2007gm}
  T.~Song, Y.~Park, S.~H.~Lee and C.~Y.~Wong,
  %``The Thermal width of heavy quarkonia moving in quark gluon plasma,''
  Phys.\ Lett.\ B {\bf 659} (2008) 621
  doi:10.1016/j.physletb.2007.11.084
  [arXiv:0709.0794 [hep-ph]].
  %%CITATION = doi:10.1016/j.physletb.2007.11.084;%%
  %27 citations counted in INSPIRE as of 28 Jan 2020

%\cite{Song:2010ix}
\bibitem{Song:2010ix}
  T.~Song, W.~Park and S.~H.~Lee,
  %``$R_{AA}$ of $J/\psi$ near mid-rapidity in heavy ion collisions at $\sqrt{s_{NN}}=200$ GeV,''
  Phys.\ Rev.\ C {\bf 81} (2010) 034914
  doi:10.1103/PhysRevC.81.034914
  [arXiv:1002.1884 [nucl-th]].
  %%CITATION = doi:10.1103/PhysRevC.81.034914;%%
  %25 citations counted in INSPIRE as of 28 Jan 2020

%\cite{Song:2010er}
\bibitem{Song:2010er}
  T.~Song, C.~M.~Ko, S.~H.~Lee and J.~Xu,
  %``$J/\psi$ production and elliptic flow in relativistic heavy-ion collisions,''
  Phys.\ Rev.\ C {\bf 83} (2011) 014914
  doi:10.1103/PhysRevC.83.014914
  [arXiv:1008.2730 [hep-ph]].
  %%CITATION = doi:10.1103/PhysRevC.83.014914;%%
  %25 citations counted in INSPIRE as of 28 Jan 2020

%\cite{Song:2011xi}
\bibitem{Song:2011xi}
  T.~Song, K.~C.~Han and C.~M.~Ko,
  %``Charmonium production in relativistic heavy-ion collisions,''
  Phys.\ Rev.\ C {\bf 84} (2011) 034907
  doi:10.1103/PhysRevC.84.034907, 10.1103/PhysRevC.84.039902
  [arXiv:1103.6197 [nucl-th]].
  %%CITATION = doi:10.1103/PhysRevC.84.034907, 10.1103/PhysRevC.84.039902;%%
  %28 citations counted in INSPIRE as of 28 Jan 2020

%\cite{Song:2011nu}
\bibitem{Song:2011nu}
  T.~Song, K.~C.~Han and C.~M.~Ko,
  %``Bottomonia suppression in heavy-ion collisions,''.
  %%CITATION = ARXIV:1109.6691;%%
  [arXiv:1109.6691 [nucl-th]].
  Phys.\ Rev.\ C {\bf{85}}, 014902 (2012).

%%\cite{Song:2011ev}
%\bibitem{Song:2011ev}
%  T.~Song, K.~C.~Han and C.~M.~Ko,
%  %``The effect of initial fluctuations on Bottomonia suppression in relativistic heavy-ion collisions,''
%  arXiv:1112.0613 [nucl-th].
%  %%CITATION = ARXIV:1112.0613;%%
%  %10 citations counted in INSPIRE as of 28 Jan 2020
%
%%\cite{Song:2012at}
%\bibitem{Song:2012at}
%  T.~Song, K.~C.~Han and C.~M.~Ko,
%  %``Charmonium production from nonequilibrium charm and anticharm quarks in quark-gluon plasma,''
%  Phys.\ Rev.\ C {\bf 85} (2012) 054905
%  doi:10.1103/PhysRevC.85.054905
%  [arXiv:1203.2964 [nucl-th]].
%  %%CITATION = doi:10.1103/PhysRevC.85.054905;%%
%  %15 citations counted in INSPIRE as of 28 Jan 2020

%\cite{Song:2013lov}
\bibitem{Song:2013lov}
  T.~Song, C.~M.~Ko and S.~H.~Lee,
  %``Quarkonium formation time in quark-gluon plasma,''
  Phys.\ Rev.\ C {\bf 87} (2013) no.3,  034910
  doi:10.1103/PhysRevC.87.034910
  [arXiv:1302.4395 [nucl-th]].
  %%CITATION = doi:10.1103/PhysRevC.87.034910;%%
  %11 citations counted in INSPIRE as of 28 Jan 2020

%\cite{Lee:2013dca}
\bibitem{Lee:2013dca}
  S.~H.~Lee, K.~Morita, T.~Song and C.~M.~Ko,
  %``Free energy versus internal energy potential for heavy quark systems at finite temperature,''
  Phys.\ Rev.\ D {\bf 89} (2014) no.9,  094015
  doi:10.1103/PhysRevD.89.094015
  [arXiv:1304.4092 [nucl-th]].
  %%CITATION = doi:10.1103/PhysRevD.89.094015;%%
  %16 citations counted in INSPIRE as of 28 Jan 2020

%\cite{Liu:2013kkg}
\bibitem{Liu:2013kkg}
  Y.~Liu, C.~M.~Ko and T.~Song,
  %``Gluon dissociation of J/ψ beyond the dipole approximation,''
  Phys.\ Rev.\ C {\bf 88} (2013) no.6,  064902
  doi:10.1103/PhysRevC.88.064902
  [arXiv:1307.4427 [hep-ph]].
  %%CITATION = doi:10.1103/PhysRevC.88.064902;%%
  %6 citations counted in INSPIRE as of 28 Jan 2020

%\cite{Song:2014qoa}
\bibitem{Song:2014qoa} 
  T.~Song,
  %``Charmonia formation in quark-gluon plasma,''
  Phys.\ Rev.\ C {\bf 89}, no. 4, 044903 (2014)
  doi:10.1103/PhysRevC.89.044903
  [arXiv:1402.3451 [nucl-th]].
  %%CITATION = doi:10.1103/PhysRevC.89.044903;%%
  %5 citations counted in INSPIRE as of 25 Aug 2017

%\cite{Song:2015bja}
\bibitem{Song:2015bja}
  T.~Song, C.~M.~Ko and S.~H.~Lee,
  %``Quarkonium formation time in relativistic heavy-ion collisions,''
  Phys.\ Rev.\ C {\bf 91} (2015) no.4,  044909
  doi:10.1103/PhysRevC.91.044909
  [arXiv:1502.05734 [nucl-th]].
  %%CITATION = doi:10.1103/PhysRevC.91.044909;%%
  %10 citations counted in INSPIRE as of 28 Jan 2020

%ADS/CFT
%%\cite{Ejaz:2007hg}
%\bibitem{Ejaz:2007hg}
%  Q.~J.~Ejaz, T.~Faulkner, H.~Liu, K.~Rajagopal and U.~A.~Wiedemann,
%  %``A limiting velocity for quarkonium propagation in a strongly coupled plasma
%  %via AdS/CFT,''
%  JHEP {\bf 0804} (2008) 089.
%  %%CITATION = JHEPA,0804,089;%%

%--color-singlet production
%\cite{Baier:1983}
\bibitem{Baier:1983}
{
    {R.~Baier and  R.~R{\"u}ckl},
    %{Hadronic collisions: A quarkonium factory},
    {Zeitschrift f{\"u}r Physik C Particles and Fields},
    %{Springer Berlin / Heidelberg},
    {0170-9739},
    %{Physics and Astronomy},
    {\bf{19}},
    {3},
    {251-266}
    {(1983)}.
}

%\cite{Humpert:1987}
\bibitem{Humpert:1987}
    B.~Humpert, 
    %Narrow heavy resonance production by gluons, 
    Phys. Lett. B, {\bf{184}}, Issue 1, 105-107 (1987).
    %ISSN 0370-2693, 
    %10.1016/0370-2693(87)90496-5.
    %(http://www.sciencedirect.com/science/article/pii/0370269387904965)

%--Leibovich.
%\cite{Cho:1995ce}
\bibitem{Cho:1995ce}
  P.~L.~Cho and A.~K.~Leibovich,
  %``Color-octet quarkonia production II,''
  Phys.\ Rev.\  D {\bf 53}, 6203 (1996).
  %%CITATION = PHRVA,D53,6203;%%

%\cite{Cho:1995vh}
\bibitem{Cho:1995vh}
  P.~L.~Cho and A.~K.~Leibovich,
  %``color-octet quarkonia production,''
  Phys.\ Rev.\  D {\bf 53}, 150 (1996).
  %%CITATION = PHRVA,D53,150;%%

%\cite{Braaten:2000cm}
\bibitem{Braaten:2000cm}
  E.~Braaten, S.~Fleming and A.~K.~Leibovich,
  %``NRQCD analysis of bottomonium production at the Tevatron,''
  Phys.\ Rev.\  D {\bf 63}, 094006 (2001).
  %%CITATION = PHRVA,D63,094006;%%

%\cite{Sridhar:1996vd}
\bibitem{Sridhar:1996vd}
  K.~Sridhar,
  %``P wave singlet charmonium production at the Tevatron,''
  Phys.\ Rev.\ Lett.\  {\bf 77} (1996) 4880
  doi:10.1103/PhysRevLett.77.4880
  [hep-ph/9609285].
  %%CITATION = doi:10.1103/PhysRevLett.77.4880;%%
  %18 citations counted in INSPIRE as of 25 Dec 2019


%--NNLO production
%\cite{Butenschoen:2010rq}
\bibitem{Butenschoen:2010rq} 
  M.~Butenschon and B.~A.~Kniehl,
  %``Reconciling $J/\psi$ production at HERA, RHIC, Tevatron, and LHC with NRQCD factorization at next-to-leading order,''
  Phys.\ Rev.\ Lett.\  {\bf 106}, 022003 (2011).
  %%CITATION = ARXIV:1009.5662;%%

%\cite{Butenschoen:Long}
\bibitem{Butenschoen:Long}
  M.~Butenschoen and B.~A.~Kniehl,
  %``World data of J/psi production consolidate NRQCD factorization at NLO,''
  Phys.\ Rev.\  D {\bf 84}, 051501 (2011).
  %%CITATION = PHRVA,D84,051501;%%

%\cite{Butenschoen:polarised}
\bibitem{Butenschoen:polarised}
  M.~Butenschoen and B.~A.~Kniehl,
  %``Probing nonrelativistic QCD factorization in polarized J/psi
  %photoproduction at next-to-leading order,''
  Phys.\ Rev.\ Lett.\  {\bf 107}, 232001 (2011).
  %%CITATION = PRLTA,107,232001;%%

%\cite{Wang:2012is}
\bibitem{Wang:2012is} 
  K.~Wang, Y.~Q.~Ma and K.~T.~Chao,
  %``$\Upsilon(1S)$ prompt production at the Tevatron and LHC in nonrelativistic QCD,''
  Phys.\ Rev.\ D {\bf 85}, 114003 (2012)
  doi:10.1103/PhysRevD.85.114003
  [arXiv:1202.6012 [hep-ph]].
  %%CITATION = doi:10.1103/PhysRevD.85.114003;%%
  %34 citations counted in INSPIRE as of 27 Aug 2017

%\cite{Shao:2014yta}
\bibitem{Shao:2014yta} 
  H.~S.~Shao, H.~Han, Y.~Q.~Ma, C.~Meng, Y.~J.~Zhang and K.~T.~Chao,
  %``Yields and polarizations of prompt $J/\psi$ and $\psi(2S)$ production in hadronic collisions,''
  JHEP {\bf 1505}, 103 (2015)
  doi:10.1007/JHEP05(2015)103
  [arXiv:1411.3300 [hep-ph]].
  %%CITATION = doi:10.1007/JHEP05(2015)103;%%
  %34 citations counted in INSPIRE as of 27 Aug 2017

%-- Eloss
%\cite{Gyulassy:1993hr}
\bibitem{Gyulassy:1993hr}
  M.~Gyulassy and X.~N.~Wang,
  %``Multiple collisions and induced gluon Bremsstrahlung in QCD,''
  Nucl.\ Phys.\ B {\bf 420} (1994) 583
  doi:10.1016/0550-3213(94)90079-5
  [nucl-th/9306003].
  %%CITATION = doi:10.1016/0550-3213(94)90079-5;%%
  %713 citations counted in INSPIRE as of 22 Dec 2019

%\cite{Baier:1996sk}
\bibitem{Baier:1996sk}
  R.~Baier, Y.~L.~Dokshitzer, A.~H.~Mueller, S.~Peigne and D.~Schiff,
  %``Radiative energy loss and p(T) broadening of high-energy partons in nuclei,''
  Nucl.\ Phys.\ B {\bf 484} (1997) 265
  doi:10.1016/S0550-3213(96)00581-0
  [hep-ph/9608322].
  %%CITATION = doi:10.1016/S0550-3213(96)00581-0;%%
  %1090 citations counted in INSPIRE as of 24 Dec 2019

%\cite{Zakharov:1996fv}
\bibitem{Zakharov:1996fv}
  B.~G.~Zakharov,
  %``Fully quantum treatment of the Landau-Pomeranchuk-Migdal effect in QED and QCD,''
  JETP Lett.\  {\bf 63} (1996) 952
  doi:10.1134/1.567126
  [hep-ph/9607440].
  %%CITATION = doi:10.1134/1.567126;%%
  %554 citations counted in INSPIRE as of 24 Dec 2019

%\cite{Baier:1998kq}
\bibitem{Baier:1998kq}
  R.~Baier, Y.~L.~Dokshitzer, A.~H.~Mueller and D.~Schiff,
  %``Medium induced radiative energy loss: Equivalence between the BDMPS and Zakharov formalisms,''
  Nucl.\ Phys.\ B {\bf 531} (1998) 403
  doi:10.1016/S0550-3213(98)00546-X
  [hep-ph/9804212].
  %%CITATION = doi:10.1016/S0550-3213(98)00546-X;%%
  %274 citations counted in INSPIRE as of 22 Dec 2019

%\cite{Gyulassy:1999zd}
\bibitem{Gyulassy:1999zd}
  M.~Gyulassy, P.~Levai and I.~Vitev,
  %``Jet quenching in thin quark gluon plasmas. 1. Formalism,''
  Nucl.\ Phys.\ B {\bf 571} (2000) 197
  doi:10.1016/S0550-3213(99)00713-0
  [hep-ph/9907461].
  %%CITATION = doi:10.1016/S0550-3213(99)00713-0;%%
  %285 citations counted in INSPIRE as of 22 Dec 2019

%\cite{Gyulassy:2000er}
\bibitem{Gyulassy:2000er}
  M.~Gyulassy, P.~Levai and I.~Vitev,
  %``Reaction operator approach to nonAbelian energy loss,''
  Nucl.\ Phys.\ B {\bf 594} (2001) 371
  doi:10.1016/S0550-3213(00)00652-0
  [nucl-th/0006010].
  %%CITATION = doi:10.1016/S0550-3213(00)00652-0;%%
  %711 citations counted in INSPIRE as of 22 Dec 2019

%\cite{Gyulassy:2002yv}
\bibitem{Gyulassy:2002yv}
  M.~Gyulassy, P.~Levai and I.~Vitev,
  %``Reaction operator approach to multiple elastic scatterings,''
  Phys.\ Rev.\ D {\bf 66} (2002) 014005
  doi:10.1103/PhysRevD.66.014005
  [nucl-th/0201078].
  %%CITATION = doi:10.1103/PhysRevD.66.014005;%%
  %78 citations counted in INSPIRE as of 26 Dec 2019

%\cite{Wiedemann:2000za}
\bibitem{Wiedemann:2000za}
  U.~A.~Wiedemann,
  %``Gluon radiation off hard quarks in a nuclear environment: Opacity expansion,''
  Nucl.\ Phys.\ B {\bf 588} (2000) 303
  doi:10.1016/S0550-3213(00)00457-0
  [hep-ph/0005129].
  %%CITATION = doi:10.1016/S0550-3213(00)00457-0;%%
  %516 citations counted in INSPIRE as of 24 Dec 2019

%\cite{Salgado:2003gb}
\bibitem{Salgado:2003gb}
  C.~A.~Salgado and U.~A.~Wiedemann,
  %``Calculating quenching weights,''
  Phys.\ Rev.\ D {\bf 68} (2003) 014008
  doi:10.1103/PhysRevD.68.014008
  [hep-ph/0302184].
  %%CITATION = doi:10.1103/PhysRevD.68.014008;%%
  %418 citations counted in INSPIRE as of 22 Dec 2019

%%---Dead cone effect
%%\cite{Dokshitzer:2001}
%\bibitem{Dokshitzer:2001} 
% Dokshitzer, Y. L., and D. E. Kharzeev. 
% %"Heavy-quark colorimetry of QCD matter." 
% Physics Letters B {\bf{519}} (2001), {199-206}.
%
%%\cite{Djordjevic:2003zk}
%\bibitem{Djordjevic:2003zk} 
%  M.~Djordjevic and M.~Gyulassy,
%  %``Heavy quark radiative energy loss in QCD matter,''
%  Nucl.\ Phys.\ A {\bf 733}, 265 (2004)
%  %%CITATION = NUCL-TH/0310076;%%

%%--CNM for open heavy quarks
%%\cite{Vitev:2008vk}
%\bibitem{Vitev:2008vk}
% I.~Vitev and B.~W.~Zhang,
% %``A systematic study of direct photon production in heavy ion collisions,''
% Phys.\ Lett.\  B {\bf 669} (2008), 337.
% %%CITATION = PHLTA,B669,337;%%
%
%%\cite{Vitev:2006bi}
%\bibitem{Vitev:2006bi}
%  I.~Vitev, T.~Goldman, M.~B.~Johnson and J.~W.~Qiu,
%  %``Open charm tomography of cold nuclear matter,''
%  Phys.\ Rev.\ D {\bf 74}, 054010 (2006) [hep-ph/0605200].
%  %%CITATION = HEP-PH 0605200;%%
%
%%\cite{Kang:2011rt}
%\bibitem{Kang:2011rt}
%  Z.~B.~Kang and I.~Vitev,
%  %``Photon-tagged heavy meson production in high energy nuclear collisions,''
%  Phys.\ Rev.\  D {\bf 84}, 014034 (2011).
%  %%CITATION = PHRVA,D84,014034;%%
%
%%--Cronin      
%%\cite{Accardi:2002ik}
%\bibitem{Accardi:2002ik}
% A.~Accardi,
% %``Cronin effect in proton nucleus collisions: A survey of theoretical
% %models,''
% [arXiv:hep-ph/0212148] and references therein.
% %%CITATION = HEP-PH/0212148;%%

%-- Vitev
%\cite{Adil:2006ra}
\bibitem{Adil:2006ra}
  A.~Adil and I.~Vitev,
  %``Collisional dissociation of heavy mesons in dense QCD matter,''
  Phys.\ Lett.\  B {\bf 649}, 139 (2007).
  %%CITATION = PHLTA,B649,139;%%

%\cite{Sharma:2009hn}
\bibitem{Sharma:2009hn}
  R.~Sharma, I.~Vitev and B.~W.~Zhang,
  %``Light-cone wave function approach to open heavy flavor dynamics in QCD matter,''
  Phys.\ Rev.\ C {\bf 80} (2009) 054902
  doi:10.1103/PhysRevC.80.054902
  [arXiv:0904.0032 [hep-ph]].
  %%CITATION = doi:10.1103/PhysRevC.80.054902;%%
  %234 citations counted in INSPIRE as of 24 Dec 2019

%\cite{Sharma:2012dy}
\bibitem{Sharma:2012dy}
  R.~Sharma and I.~Vitev,
  %``High transverse momentum quarkonium production and dissociation in heavy ion collisions,''
  Phys.\ Rev.\ C {\bf 87} (2013) no.4,  044905
  doi:10.1103/PhysRevC.87.044905
  [arXiv:1203.0329 [hep-ph]].
  %%CITATION = doi:10.1103/PhysRevC.87.044905;%%
  %100 citations counted in INSPIRE as of 24 Dec 2019
%%---Collisional Dissociation------
%%\cite{Dominguez:2008be}
%\bibitem{Dominguez:2008be}
%  F.~Dominguez and B.~Wu,
%  %``On dissociation of heavy mesons in a hot quark-gluon plasma,''
%  Nucl.\ Phys.\ A {\bf 818}, 246 (2009).
%  %%CITATION = NUPHA,A818,246;%%
%
%%\cite{Dominguez:2008aa}
%\bibitem{Dominguez:2008aa}
%  F.~Dominguez, C.~Marquet and B.~Wu,
%  %``On multiple scatterings of mesons in hot and cold QCD matter,''.
%  Nucl.\ Phys.\ A {\bf{823}} 99-119(2009).
%  %[arXiv:0812.3878]
%  %%CITATION = ARXIV:0812.3878;%%

% EFT for quarkonia in the QGP
%\cite{Makris:2019ttx}
\bibitem{Makris:2019ttx}
  Y.~Makris and I.~Vitev,
  %``An Effective Theory of Quarkonia in QCD Matter,''
  JHEP {\bf 1910} (2019) 111
  doi:10.1007/JHEP10(2019)111
  [arXiv:1906.04186 [hep-ph]].
  %%CITATION = doi:10.1007/JHEP10(2019)111;%%
  %3 citations counted in INSPIRE as of 25 Dec 2019
  
%\cite{Makris:2019kap}
\bibitem{Makris:2019kap}
  Y.~Makris and I.~Vitev,
  %``An Effective Theory of Quarkonia in QCD Matter,''
  arXiv:1912.08008 [hep-ph].
  %%CITATION = ARXIV:1912.08008;%%

%-- Adiabatic
%\cite{Dutta:2012nw}
\bibitem{Dutta:2012nw}
  N.~Dutta and N.~Borghini,
  %``Sequential suppression of quarkonia and high-energy nucleus–nucleus collisions,''
  Mod.\ Phys.\ Lett.\ A {\bf 30} (2015) no.37,  1550205
  doi:10.1142/S0217732315502053
  [arXiv:1206.2149 [nucl-th]].
  %%CITATION = doi:10.1142/S0217732315502053;%%
  %15 citations counted in INSPIRE as of 28 Dec 2019

%-- Open quantum systems
%\cite{Breuer:2002pc}
\bibitem{Breuer:2002pc}
  H.~P.~Breuer and F.~Petruccione,
  %``The theory of open quantum systems,''
  Oxford, UK: Univ. Pr. (2002) 625 p
  %92 citations counted in INSPIRE as of 20 Dec 2019

%-- Lindblad
%\cite{Lindblad:1975ef}
\bibitem{Lindblad:1975ef}
  G.~Lindblad,
  %``On the Generators of Quantum Dynamical Semigroups,''
  Commun.\ Math.\ Phys.\  {\bf 48} (1976) 119.
  doi:10.1007/BF01608499
  %%CITATION = doi:10.1007/BF01608499;%%
  %652 citations counted in INSPIRE as of 29 Jan 2020

%%-- J/psi jets
%%\cite{Aaij:2017fak}
%\bibitem{Aaij:2017fak} 
%  R.~Aaij {\it et al.} [LHCb Collaboration],
%  %``Study of J/? Production in Jets,''
%  Phys.\ Rev.\ Lett.\  {\bf 118}, no. 19, 192001 (2017)
%  doi:10.1103/PhysRevLett.118.192001
%  [arXiv:1701.05116 [hep-ex]].
%  %%CITATION = doi:10.1103/PhysRevLett.118.192001;%%
%  %8 citations counted in INSPIRE as of 15 Aug 2017

%%-- HQ diffusion from AdS/CFT
%\cite{CasalderreySolana:2006rq}
\bibitem{CasalderreySolana:2006rq}
  J.~Casalderrey-Solana and D.~Teaney,
  %``Heavy quark diffusion in strongly coupled N=4 Yang-Mills,''
  Phys.\ Rev.\ D {\bf 74} (2006) 085012
  doi:10.1103/PhysRevD.74.085012
  [hep-ph/0605199].
  %%CITATION = doi:10.1103/PhysRevD.74.085012;%%
  %395 citations counted in INSPIRE as of 04 Feb 2020
\end{thebibliography}
\end{document}